\documentstyle[twoside,fleqn,epsfig]{article}

\def\be{\begin{equation}}
\def\ee{\end{equation}}
\def\bea{\begin{eqnarray}}
\def\eea{\end{eqnarray}}
\newcommand{\lsim}{\mathrel{\mathop{\kern 0pt \rlap
  {\raise.2ex\hbox{$<$}}}
  \lower.9ex\hbox{\kern-.190em $\sim$}}}
\newcommand{\gsim}{\mathrel{\mathop{\kern 0pt \rlap
  {\raise.2ex\hbox{$>$}}}
  \lower.9ex\hbox{\kern-.190em $\sim$}}}

\newcommand{\AmS}{{\protect\the\textfont2
  A\kern-.1667em\lower.5ex\hbox{M}\kern-.125emS}}
\normalsize
\begin{document}
\begin{flushright}
\large
{\bf ROM2F/2008/07 \\}
{\bf April 2008 \\}
\end{flushright}

\baselineskip=0.65cm

\begin{center}
\Large 
{\bf First results from DAMA/LIBRA and the combined results with DAMA/NaI } \\
\rm
\end{center}

\vspace{-0.2cm}
\large

\begin{center}
R.\,Bernabei $^{a,b}$,~P.\,Belli $^{b}$, ~F.\,Cappella $^{c,d}$, R.\,Cerulli $^{e}$, C.J.\,Dai $^{f}$,
\vspace{1mm}

A. d'Angelo $^{c,d}$, ~H.L.\,He $^{f}$, ~A.\,Incicchitti $^{d}$, ~H.H.\,Kuang $^{f}$, 
\vspace{1mm}

~J.M.\,Ma $^{f}$, ~F.\,Montecchia $^{a,b}$,~F.\,Nozzoli $^{a,b}$, 
\vspace{1mm}

~D.\,Prosperi $^{c,d}$, ~X.D.\,Sheng $^{f}$, ~Z.P.\,Ye $^{f,g}$
\vspace{1mm}

\normalsize

\vspace{0.2cm}

$^{a}${\it Dip. di Fisica, Universit\`a di Roma ``Tor Vergata'', I-00133 Rome, Italy}
\vspace{1mm}

$^{b}${\it INFN, sez. Roma ``Tor Vergata'', I-00133 Rome, Italy}
\vspace{1mm}

$^{c}${\it Dip. di Fisica, Universit\`a di Roma ``La Sapienza'', I-00185 Rome, Italy}
\vspace{1mm}

$^{d}${\it INFN, sez. Roma, I-00185 Rome, Italy}
\vspace{1mm}

$^{e}${\it Laboratori Nazionali del Gran Sasso, I.N.F.N., Assergi, Italy}
\vspace{1mm}

$^{f}${\it IHEP, Chinese Academy, P.O. Box 918/3, Beijing 100039, China}
\vspace{1mm}

$^{g}${\it University of Jing Gangshan, Jiangxi, China}
\vspace{1mm}

\end{center}

\vspace{-0.5cm}

\normalsize

\begin{abstract}

The highly radiopure $\simeq$ 250 kg NaI(Tl) DAMA/LIBRA set-up is running at the Gran Sasso
National Laboratory of the I.N.F.N.. In this paper the first result obtained by exploiting
the model independent annual modulation signature for Dark Matter (DM) particles is presented. 
It refers to an exposure of 0.53 ton$\times$yr.
The collected DAMA/LIBRA data satisfy all the many peculiarities of the DM annual modulation signature.
Neither systematic effects nor side reactions can account for the observed modulation amplitude
and contemporaneously satisfy all the several requirements of this DM 
signature. Thus, the presence of Dark Matter particles in the galactic halo is supported also by 
DAMA/LIBRA and,
considering the former DAMA/NaI and the present DAMA/LIBRA data all together (total exposure 
0.82 ton$\times$yr), the presence of Dark Matter particles in the galactic halo is supported 
at 8.2 $\sigma$ C.L..

\end{abstract}

\vspace{2.0mm}


\vspace{2.0mm}


\vspace{-2.0mm}

\section{Introduction}

DAMA/LIBRA is part of the DAMA project, which is mainly based on the development and use of low 
background scintillators 
\cite{allDM,Nim98,Sist,RNC,ijmd,ijma,epj06,ijma07,chan,wimpele,ldm,allRare,supclu,IDM96,ref1}.  

In particular, the former DAMA/NaI and the present DAMA/LIBRA set-ups have 
the main aim to perform a direct detection of Dark Matter (DM) particles in 
the galactic halo
through the model independent annual modulation signature (originally suggested in ref. \cite{Freese}). 

The former DAMA/NaI experiment 
\cite{allDM,Nim98,Sist,RNC,ijmd,ijma,epj06,ijma07,chan,wimpele,ldm,allRare,supclu,IDM96}
has achieved many competitive results on rare processes and, in particular,  
has pointed out a model independent evidence for the presence of DM particles in the galactic halo
with high C.L..

In 1996 -- while running the DAMA/NaI set-up 
-- DAMA proposed to INFN to develop and build a one ton set-up \cite{Prop} to further investigate 
Dark Matter particles and other rare processes.  
Thus, a second generation R\&D project was funded to develop new  highly 
radiopure NaI(Tl) detectors towards 
the achievement of an intermediate step: the  $\simeq$ 250 kg highly radiopure NaI(Tl) DAMA/LIBRA 
(Large sodium Iodide Bulk for RAre processes) set-up, which is now in data taking. 

The exploitation of the annual modulation DM signature 
with highly radiopure NaI(Tl) as target material 
can permit to answer -- by direct detection and in a way largely 
independent on the nature of the candidate and 
on the astrophysical, nuclear and particle Physics assumptions -- 
the main question: ``Are there Dark Matter (DM) particles in the galactic halo?''

In particular, the use of the highly radiopure DAMA/LIBRA (and, previously, DAMA/NaI) 
NaI(Tl) scintillators as target-detectors offers many specific advantages
thanks e.g. to the intrinsic radiopurity, to the large sensitivity 
to many of the DM candidates, of the interactions and of astrophysical, nuclear and particle 
Physics scenarios, to the granularity of the set-up, to the data taking up to the MeV scale 
(even though the optimization is made for the lowest energy region), to the full controll of the running 
conditions,
etc.. 

Phenomenological properties of some basic interaction mechanisms induced by DM particles
are discussed, for instance, in ref. \cite{Freese,sem1,Wei01,sem2,RNC,ijmd,ijma,ijma07,chan,wimpele,ldm}.
The DM annual modulation signature exploits the effect of the Earth revolution 
around the Sun on the number of events induced by DM particles in a suitable low background 
set-up placed deep underground. 
In particular, as a consequence of its annual revolution, the Earth should be crossed 
by a larger flux of DM particles around roughly June 2$^{nd}$ (when its rotational velocity is summed to 
the one of 
the solar system with respect to the Galaxy) and by a smaller one around roughly December 2$^{nd}$ (when 
the two 
velocities are subtracted) \cite{RNC}. 
Thus, the contribution of the signal to the counting rate in the $k$--th energy interval can be written as
(see e.g. ref. \cite{RNC,ijmd}): 
\begin{equation}
 S_k = S_{0,k} + S_{m,k} \cos \omega(t-t_0)  \; , 
\label{eq1}
\end{equation}
where: i) $S_{0,k}$ is the constant part of the signal; ii) $S_{m,k}$ is the modulation amplitude; 
iii) $\omega = \frac{2\pi}{T}$ with period $T$; iv) $t_0$ is the phase.

This annual modulation signature is very distinctive since a seasonal effect induced by DM 
particles must simultaneously satisfy all the following requirements: 
1) the rate must contain a component modulated according to a cosine function; 
2) with one year period; 
3) with a phase roughly around June 2$^{nd}$ in case of usually adopted halo models (slight variations 
may occur in case of presence of 
   non thermalized DM components in the halo);
4) this modulation must be present only in a well-defined low energy range, where DM particles can induce signals; 
5) it must be present only in those events where just a single detector, among all the available ones in the used 
      set-up, actually ``fires'' ({\it single-hit} events), since the probability that DM particles experience multiple 
      interactions is negligible; 
6) the modulation amplitude in the region of maximal sensitivity has to be $\lsim$ 7\% 
      in case of usually adopted halo distributions, but it may be significantly larger in case of 
      some particular scenarios such as $e.g.$ those of ref. \cite{Wei01}. 
To mimic such a signature spurious effects or side reactions should be able not only to account for the observed 
modulation amplitude but also to contemporaneously satisfy all the requirements of the signature; 
none of these has been found or suggested by anyone over more than a decade 
(see e.g. ref. \cite{RNC,ijmd}, the references therein and later). 

The corollary question: ``Which are exactly the nature of the DM
particle(s) detected 
by the annual modulation signature
and the related astrophysical, nuclear and particle Physics scenarios?''
requires subsequent model dependent corollary analyses
as those performed e.g. in refs. \cite{RNC,ijmd,ijma,epj06,ijma07,chan,wimpele,ldm}.
One should stress that it does not exist 
any approach to investigate the nature of the candidate in the direct and indirect DM searches which can 
offer these information independently on assumed astrophysical, nuclear and particle Physics
scenarios. 

As regards complementary information from accelerators, and most noticeably from LHC, 
the existence of some of the possible candidates could be tested at some extent; this will be 
very useful. However, it is worth noting that interesting DM candidates and scenarios for them  
exist, which are beyond the reach of that class of experiments (but potentially capable of 
determining the annual modulation effect).

\vspace{0.3cm}

The main goal of the DAMA/LIBRA experiment is to further study the presence of DM particles in the 
galactic halo pointed out by the former DAMA/NaI experiment \cite{RNC,ijmd} 
exploiting the annual modulation signature and to get improved
information on the corollary quests on the nature of the candidate particle(s) and on the related astrophysical, 
nuclear and particle physics models. 
Moreover, second order effects are planned to be investigated (see e.g. ref. \cite{ijmd,epj06}), 
and dedicated data takings will also allow the study of many other rare 
processes (as e.g. already performed with DAMA/NaI \cite{allRare,supclu,IDM96})
thanks to the peculiarity of the experimental set-up.

DAMA/LIBRA and the former DAMA/NaI are the only experiments effectively exploiting in all the
aspects the DM annual modulation signature, and with highly radiopure NaI(Tl) detectors.
Note that approaches based on many selections and handling procedures to
``reject'' the electromagntic component of the counting rate
cannot offer any signature for Dark Matter particles 
even under the assumption of an ``ideal'' electromagnetic
component rejection, since e.g. the neutrons and the internal end-range $\alpha$'s
induce signals indistinguishable from recoils (they are looking for) which 
cannot be estimated and subtracted in any reliable manner at the
needed precision, and since part or all the signal can have electromagnetic nature.
Moreover, in a safe investigation of the DM annual modulation signature
those data handlings cannot be applied e.g. because of their -- always -- statistical nature 
which would affect the annual modulation analysis and restrict the sensitivity to many 
kinds of candidates (including also the WIMPs).
On the other hand, as known, the exploitation of the DM annual modulation signature acts itself 
as an effective background rejection.

\vspace{0.3cm}

In this paper some main features of the DAMA/LIBRA set-up \cite{perflibra}
will be shortly summarized in sect. \ref{c1}. 
The model independent experimental results obtained 
by DAMA/LIBRA (exposure of 0.53 ton$\times$yr collected over 4 annual cycles)
and the combined ones with 
DAMA/NaI (exposure of 0.29 ton$\times$yr collected over 7 annual cycles) are presented (total exposure 
of 0.82 ton$\times$yr) in sect. \ref{sc:evi}. In sect. \ref{sistse} the quantitative investigation on 
possible 
systematic effects and side processes is discussed. The corollary model dependent analyses on the 
candidate particle(s) and astrophysical, nuclear and particle physics scenarios will be presented 
elsewhere in a dedicated publication; here in Appendix A just few arguments are mentioned for 
some illustrative purposes.  

\section{The experimental set-up}
\label{c1}

The DAMA/LIBRA set-up, its main features and radiopurity have been discussed in the 
devoted ref. \cite{perflibra}. Here we just shortly summarize some information.

The installation of DAMA/LIBRA started in July 2002 after the dismounting of the former DAMA/NaI.
The experimental site as well as many components of the installation itself have been implemented.
All the procedures performed during the dismounting of DAMA/NaI and the installation of DAMA/LIBRA detectors have 
been carried out in high purity (HP) Nitrogen atmosphere.

The sensitive part of DAMA/LIBRA is made of 25 highly radiopure NaI(Tl) crystal scintillators 
in a 5-rows 5-columns matrix. Each NaI(Tl) detector has  
9.70 kg mass and a size 
of ($10.2 \times 10.2 \times 25.4$) cm$^{3}$. 
The bare crystals are enveloped in Tetratec-teflon foils and encapsulated in radiopure OFHC Cu housing; 
10 cm long special quartz light guides act also as 
optical windows on the two end faces of the crystals and are coupled to two low background photomultipliers (PMT). 
The threshold of each one of the two PMTs on a detector is set at single photoelectron level; their coincidence provides 
the trigger of the detector. 
The software energy threshold has been cautiously taken at 2 keV electron equivalent (hereafter keV). The 
measured light response is 5.5—-7.5 photoelectrons/keV depending on the detector.
The detectors are housed in a low radioactivity sealed 
copper box installed in the center of a low-radioactivity Cu/Pb/Cd-foils/polyethylene/paraffin shield;
moreover, about 1 m concrete (made from the Gran Sasso rock material) almost fully surrounds (mostly outside the 
barrack) this passive shield, acting as a further neutron moderator.
The copper box is maintained in HP Nitrogen atmosphere in slightly overpressure with respect to the 
external environment; it is part of the 3-levels sealing system which excludes the detectors from environmental air. 
The whole installation is air-conditioned and the temperature is continuously monitored and recorded; 
moreover, it is worth noting that the detectors have copper housings in direct contact with the multi-tons passive 
shield and its huge heat capacity ($\approx 10^6$ cal$/^{\circ}$C ) further assures a relevant stability of the 
detectors operating temperature (see also later).

Following the same strategy as DAMA/NaI, on the top of the shield a glove-box (also continuously maintained in the HP Nitrogen
atmosphere) is directly connected to the inner Cu box, housing the detectors, through Cu pipes. 
The pipes are filled with low radioactivity Cu bars (covered by 10 cm of low radioactive Cu and 15 cm of low
radioactive Pb) which can be removed to allow the insertion of radioactive sources for calibrating the detectors 
in the same running condition, without any contact with external air. The glove-box is also equipped with a compensation chamber.

A hardware/software system to monitor the running conditions is operative and self-controlled 
computer processes automatically control several parameters and manage alarms.
For the electronic chain, the data acquisition system and for all other details see ref. \cite{perflibra}.

The DAMA/LIBRA set-up, as the former DAMA/NaI, allows the recording both of the {\it single-hit} events 
(those events where just one detector of many actually fires) and of the {\it multiple-hit} events 
(those events where more than one detector fire).
 
The experiment take data up to the MeV scale despite the optimization is made for the lowest energy region. 
The linearity and the energy resolution of the detectors at low and high energy have been investigated using 
several sources as discussed in ref. \cite{perflibra}.
In particular, as regards the low energy region, calibrations down to the 3.2 keV X-ray have been carried out .
During the production runs periodical calibrations (every $\simeq$ 10 days) are carried out with $^{241}$Am 
sources, introduced in the proximity of the detectors by source holders inserted in the Cu pipes mentioned above; 
the latter one is also continuously maintained in the HP Nitrogen atmosphere.

The energy threshold, the PMT gain, the electronic line stability are continuously verified and monitored during 
the data taking by the routine calibrations, by the position and energy resolution of internal lines \cite{perflibra}
and by the study of the hardware rate behaviours with time.

The main procedures of the DAMA data taking for the investigation of DM particles 
annual modulation signature are:
1) the data taking of each annual cycle starts from autumn/winter (when $\cos \omega (t-t_0) \simeq 0$) 
towards summer (maximum expected);
2) the routine calibrations with radioactive sources are performed about each 10 days 
(collecting typically $\simeq 10^4 - 10^5$ events per keV), moreover regularly intrinsic 
calibration are carried out, etc. \cite{perflibra};
3) the on-line monitoring of all the running parameters is continuously carried out with automatic alarm to operator 
if any would go out of allowed range.

\section{The model-independent experimental results}
\label{sc:evi}

As mentioned, DAMA/LIBRA started the first operations in March 2003.
However, in order to allow the decay of medium half-life isotopes \cite{perflibra},
the data taking for the investigation 
of the annual modulation signature, reported here, has been started on September 9, 2003. 
Moreover, it is worth noting that one of the more external detectors has been put out of operation few months after 
installation because of a PMT break; since the disinstallation and reinstallation of this detector would require the 
opening of the set-up, the installation of the room for doing it in HP Nitrogen atmosphere and some time of stop of 
the experiment, we have delayed this. The related procedure are planned to occur in 2008 when also an upgrade of
the electronics will occur. 
Therefore, the exposed mass in the four annual cycles, presented here, is 232.8 kg for a total 
exposure of 0.53 ton$\times$yr.

The only data treatment, which is performed on the raw data, is to eliminate obvious noise pulses 
(mainly PMT noise, Cherenkov light in the light guides and in the PMT windows, and afterglows)
near the energy threshold in the {\it single-hit} events \cite{perflibra}; the number of such pulses 
sharply decreases when increasing the number of available photoelectrons.
In particular, as mentioned, the DAMA/LIBRA detectors are seen by two PMTs working in coincidence and 
this already strongly reduces the noise near the energy threshold for the {\it single-hit} events 
(of interest for the detection of DM particles), 
while obviously noise is practically absent in the {\it multiple-hit} events, since the probability to 
have random coincidences is negligible ($<3\times 10^{-6}$). This rejection of the noise near energy 
threshold is based on the different time structures of the pulse profile of noise pulses 
(time decay of order of tens ns) and of the scintillation pulses
(time decay of order of hundreds ns). The high number of photoelectrons/keV assures a very good 
separation between the two populations,
nevertheless stringent acceptance windows are used in order to assure full noise rejection near energy 
threshold; related acceptance window efficiencies are measured by devoted source calibrations. 
For a description of the used procedure and details see ref. \cite{perflibra}. 

Detailed information about the four annual cycles by DAMA/LIBRA, considered here, is given in Table \ref{tb:years}. 
\begin{table}[ht]
\caption{DAMA/LIBRA annual cycles. There $\alpha=\langle cos^2\omega (t-t_0) \rangle$ 
is the mean value of the squared cosine and $\beta=\langle cos \omega (t-t_0) \rangle$ 
is the mean value of the cosine (the averages are taken over the live time of the data taking
and $t_0=152.5$ day, i.e. June 2$^{nd}$); 
thus, $\alpha - \beta^2$ indicates the variance of the cosine
(i.e. it is 0.5 for a full year of data taking). 
The exposed mass in these four annual cycles is 232.8 kg; see text.}
\begin{center}
\begin{tabular}{|c|c|c|c|}
\hline
\hline
 Period & &Exposure (kg$\times$day) & $\alpha - \beta^2$ \\
\hline
 & & & \\
 DAMA/LIBRA-1 & Sept. 9, 2003 - July 21, 2004 &51405  & 0.562 \\ 
   &        &       & \\ 
 DAMA/LIBRA-2 & July 21, 2004 - Oct. 28, 2005 &52597  & 0.467 \\ 
   &        &       & \\ 
 DAMA/LIBRA-3 & Oct. 28, 2005 - July 18, 2006 &39445  & 0.591 \\ 
   &        &       & \\ 
 DAMA/LIBRA-4 & July 19, 2006 - July 17, 2007 &49377  & 0.541 \\ 
   &        &       & \\ 
\hline
Total         &                               &192824 & 0.537 \\
              &  &     $\simeq 0.53$ ton$\times$yr & \\
\hline
\hline
\end{tabular}
\end{center}
\label{tb:years}
\end{table}
%
In these annual cycles about $4.4 \times 10^7$ events have also been collected for energy calibrations 
and about $6.0 \times 10^7$ events for the evaluation of the acceptance windows efficiency for noise rejection near energy 
threshold \cite{perflibra}.
The periodical calibrations and, in particular, those related with the acceptance windows efficiency
mainly affect the duty cycle of the experiment; 
in the present data taking it is of the same order as the one of DAMA/NaI, despite the larger number of 
involved detectors, thanks also to the improvements in the electronics and in the data acquisition (DAQ) system.

\begin{figure}[!t]
\vspace{-0.4cm}
\centering
\includegraphics[width=10.cm] {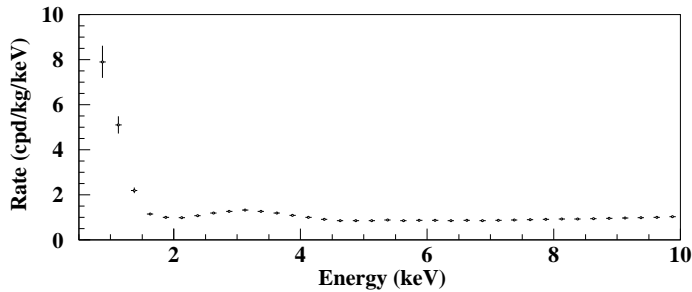}
\vspace{-0.6cm}
\caption{Cumulative low-energy distribution 
of the {\it single-hit} scintillation events (that is each detector has all the others as veto),
as measured 
by the DAMA/LIBRA detectors in an exposure of 0.53 ton $\times$ yr.  The energy threshold of the 
experiment is 2 keV and corrections for efficiencies are already applied.}
\label{fg:0}
\end{figure}

Figure \ref{fg:0} shows the cumulative low-energy distribution 
of the {\it single-hit} scintillation events  (of interest for the DM particles since DM particle multiple interaction  
probability is negligible), as measured 
by the DAMA/LIBRA detectors in the 0.53 ton$\times$yr exposure. 

In order to further investigate the presence of DM particles in the galactic halo, 
a model-independent investigation of the annual modulation signature has been carried out by
exploiting the time behaviour of the residual rates of the {\it single-hit} events in the
lowest energy regions of the DAMA/LIBRA data.
These residual rates are calculated from the measured rate of the {\it single-hit} events (obviously 
corrections for the overall efficiency and for the acquisition dead time are already applied)
after subtracting the constant part: $<r_{ijk}-flat_{jk}>_{jk}$.
Here $r_{ijk}$ is the rate in the considered $i$-th time interval for the $j$-th detector in the 
$k$-th energy bin, while $flat_{jk}$ is the rate of the $j$-th detector in the $k$-th energy bin 
averaged over the cycles. 
The average is made on all the detectors ($j$ index) and on all the 1 keV bins ($k$ index) which 
constitute the considered energy interval. The weighted mean of the residuals must 
obviously be zero over one cycle. 

\begin{figure}[p]
\begin{center}
\vspace{-0.7cm}
\includegraphics[width=13.cm] {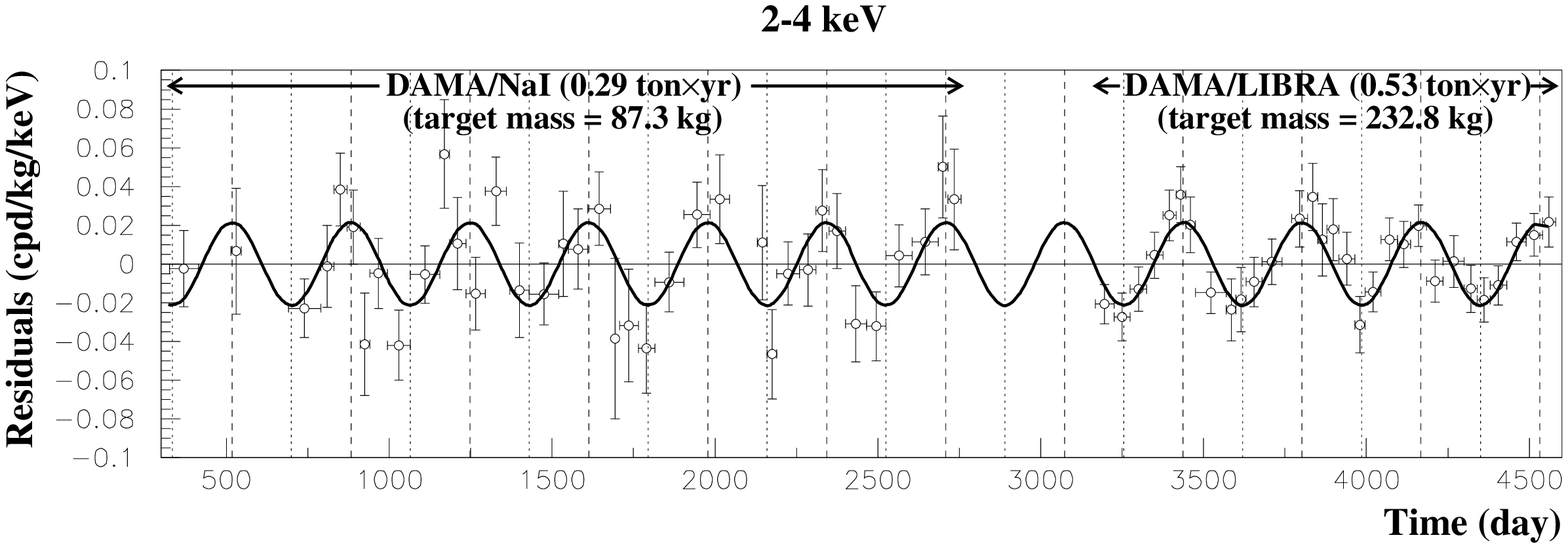}
\includegraphics[width=13.cm] {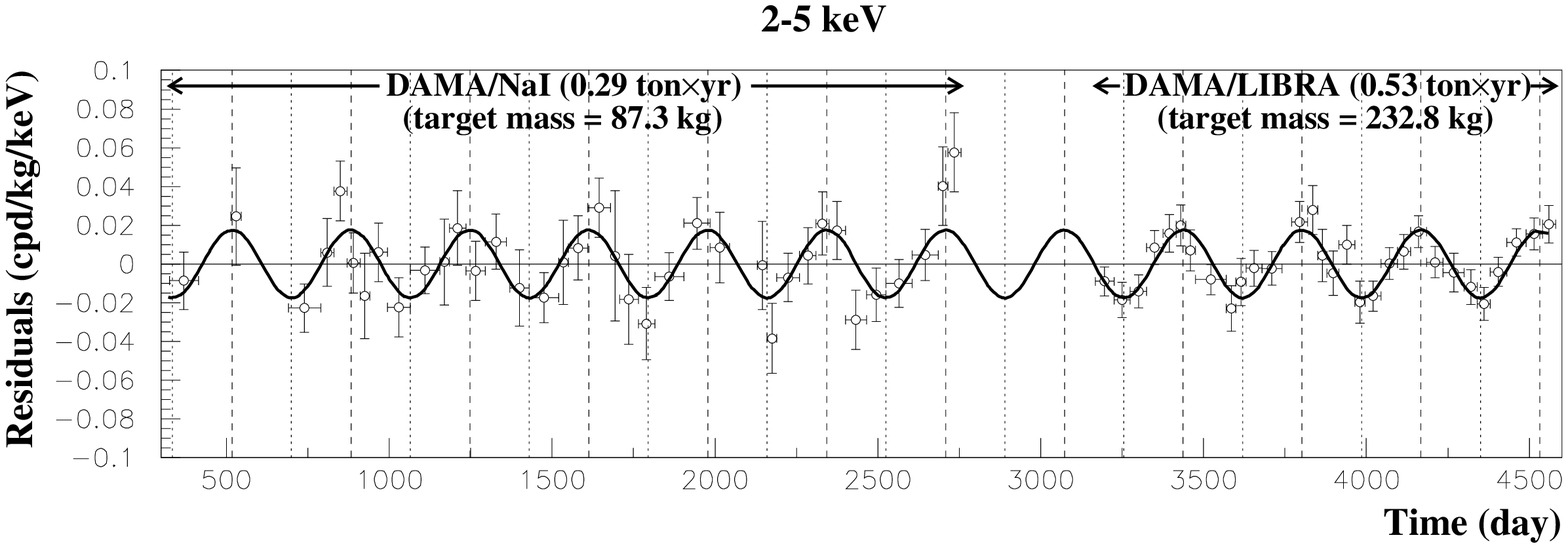}
\includegraphics[width=13.cm] {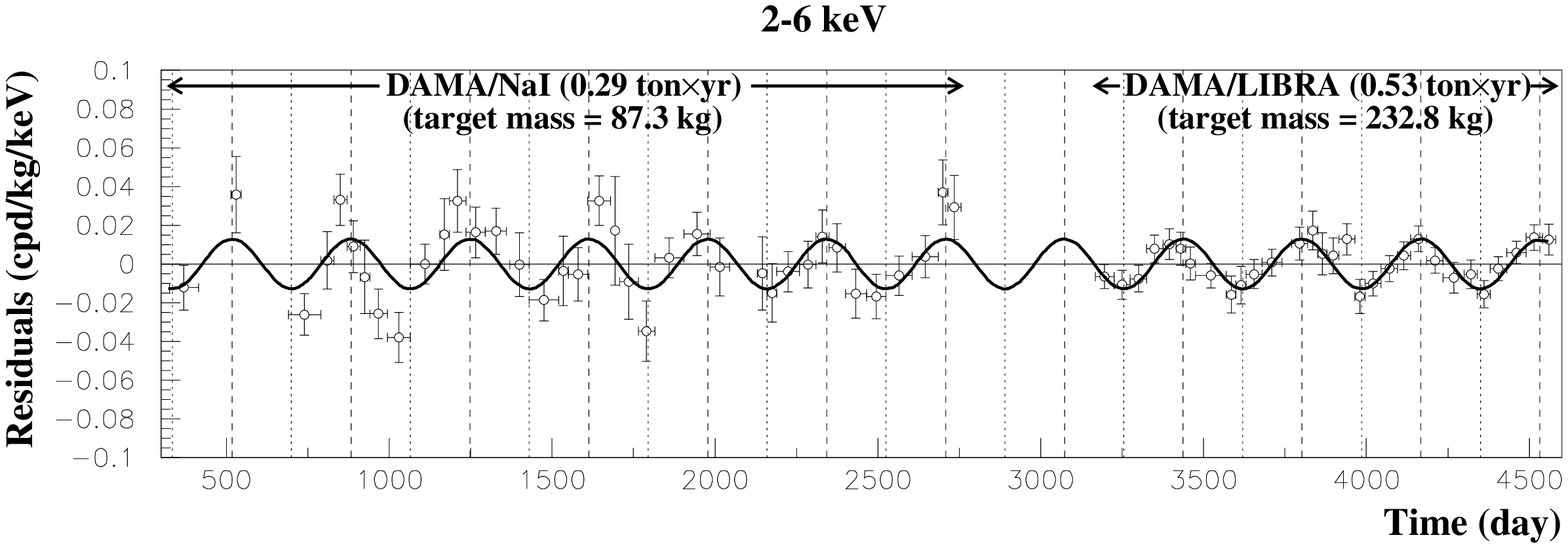}
\end{center}
\vspace{-0.6cm}
\caption{Model-independent residual rate of the {\it single-hit} scintillation events, measured by the new DAMA/LIBRA 
experiment in the (2 -- 4), (2 -- 5) and (2 -- 6) keV energy intervals as a function of the time. The 
residuals measured by DAMA/NaI and 
already published in ref. \cite{RNC,ijmd} are also shown. The zero of the time scale is January 1$^{st}$ 
of the first year of data taking of the former DAMA/NaI experiment. 
The experimental points present the errors as vertical bars and the associated time bin width as horizontal bars. 
The superimposed curves represent the cosinusoidal functions behaviours $A \cos \omega(t-t_0)$  
with a period $T = \frac{2\pi}{\omega} =  1$ yr, with a phase $t_0 = 152.5$ day (June 2$^{nd}$) and with
modulation amplitudes, $A$, equal to the central values obtained by best fit over the whole data, that is: 
$(0.0215 \pm 0.0026)$ cpd/kg/keV, 
$(0.0176 \pm 0.0020)$ cpd/kg/keV and 
$(0.0129 \pm 0.0016)$ cpd/kg/keV 
for the (2 -- 4) keV, for the (2 -- 5) keV and for the (2 -- 6) keV energy intervals, respectively. See 
text. 
The dashed vertical lines correspond to the maximum of the signal (June 2$^{nd}$), while the dotted vertical lines correspond to the minimum.
The total exposure is 0.82 ton$\times$yr.}
\label{fig1}
\vspace{-0.4cm}
\end{figure}

Figure \ref{fig1} shows the time behaviour (over three energy intervals) of the model-independent 
residual rates for {\it single-hit} 
events collected by the new DAMA/LIBRA experiment over four annual cycles (0.53 ton$\times$yr). 
Those measured over seven annual cycles 
by the former DAMA/NaI experiment \cite{RNC,ijmd} (0.29 ton$\times$yr) are shown as well; the cumulative 
exposure of the two experiments is
0.82 ton$\times$yr. The advantage of the increased exposed mass and exposure is evident.

In particular, the residual rates in the (2 -- 4), (2 -- 5) and (2 -- 6) keV energy 
intervals are depicted in Fig. \ref{fig1};
the experimental points present the errors as vertical bars and the associated time bin width as horizontal bars. 
The superimposed curves represent the cosinusoidal functions behaviours $A \cos \omega(t-t_0)$  
with a period $T = \frac{2\pi}{\omega} =  1$ yr and with a phase $t_0 = 152.5$ day (June 2$^{nd}$) and 
modulation amplitudes, $A$, obtained by best fit over the whole data (DAMA/NaI \& DAMA/LIBRA).
The dashed vertical lines correspond to the maximum of the signal (June 2$^{nd}$), while the dotted vertical lines correspond to the minimum.
We note that, for simplicity, in Fig. \ref{fig1} the same time binning already considered e.g. in ref. \cite{RNC,ijmd} 
has been used. The result of this approach is similar by choosing other time binnings, as it is also evident
from the analysis on modulation amplitudes given in the following.

\vspace{0.2cm}

Table \ref{tb:libra2} summarizes the results obtained by fitting with the function 
$A \cos \omega(t-t_0)$: i) only the DAMA/NaI data \cite{RNC,ijmd}; 
ii) only the DAMA/LIBRA data; iii) 
all the data together.
A clear modulation is present in all the energy intervals and the periods and 
phases agree with those 
expected in the case of a DM particle induced effect.

\begin{table}[!ht]
\caption{Results obtained from the time behaviours of the residual rates of the {\it single-hit} scintillation events,
collected by DAMA/NaI, by DAMA/LIBRA and by the two experiments all together in the (2 -- 4), (2 -- 5) 
and (2 -- 6) keV energy intervals.
The data have been fitted with the function: $A \cos \omega(t-t_0)$.
The last column shows the C.L. obtained from the fitted modulation amplitudes.
See comments in the text.}
\vspace{0.6cm}
\centering
\resizebox{\textwidth}{!}{
\begin{tabular}{|l|c|c|c|c|} \hline
                 & $A$ (cpd/kg/keV)    & $T=\frac{2\pi}{\omega}$ (yr) & $t_0$ (day) & C.L. \\ \hline
 DAMA/NaI        &                     &                   &              &             \\
 (2--4) keV      & $0.0252 \pm 0.0050$ & $1.01 \pm 0.02$   & $125 \pm 30$ & $5.0\sigma$ \\ 
 (2--5) keV      & $0.0215 \pm 0.0039$ & $1.01 \pm 0.02$   & $140 \pm 30$ & $5.5\sigma$ \\
 (2--6) keV      & $0.0200 \pm 0.0032$ & $1.00 \pm 0.01$   & $140 \pm 22$ & $6.3\sigma$ \\ \hline
 DAMA/LIBRA      &                     &                   &              &             \\
 (2--4) keV      & $0.0213 \pm 0.0032$ & $0.997 \pm 0.002$ & $139 \pm 10$ & $6.7\sigma$ \\ 
 (2--5) keV      & $0.0165 \pm 0.0024$ & $0.998 \pm 0.002$ & $143 \pm  9$ & $6.9\sigma$ \\
 (2--6) keV      & $0.0107 \pm 0.0019$ & $0.998 \pm 0.003$ & $144 \pm 11$ & $5.6\sigma$ \\ \hline
 DAMA/NaI+ DAMA/LIBRA      &           &                   &              &             \\
 (2--4) keV      & $0.0223 \pm 0.0027$ & $0.996 \pm 0.002$ & $138 \pm  7$ & $8.3\sigma$ \\ 
 (2--5) keV      & $0.0178 \pm 0.0020$ & $0.998 \pm 0.002$ & $145 \pm  7$ & $8.9\sigma$ \\
 (2--6) keV      & $0.0131 \pm 0.0016$ & $0.998 \pm 0.003$ & $144 \pm  8$ & $8.2\sigma$ \\ \hline
\end{tabular}}
\vspace{0.3cm}
\label{tb:libra2}
\end{table}

It is worthwhile remarking how the larger exposed mass per annual cycle has improved the fit; for 
example, the $\chi^2$/{\it d.o.f.} of the best fit of the (2 -- 6) keV {\it single-hit} 
residual rate from DAMA/NaI 
plus DAMA/LIBRA given in Table \ref{tb:libra2} is 53.2/64. 
The period and phase substantially agree with 
$T=$1 yr and $t_0=$ 152.5 day.

\begin{figure}[!ht]
\begin{center}
\includegraphics[width=12.cm] {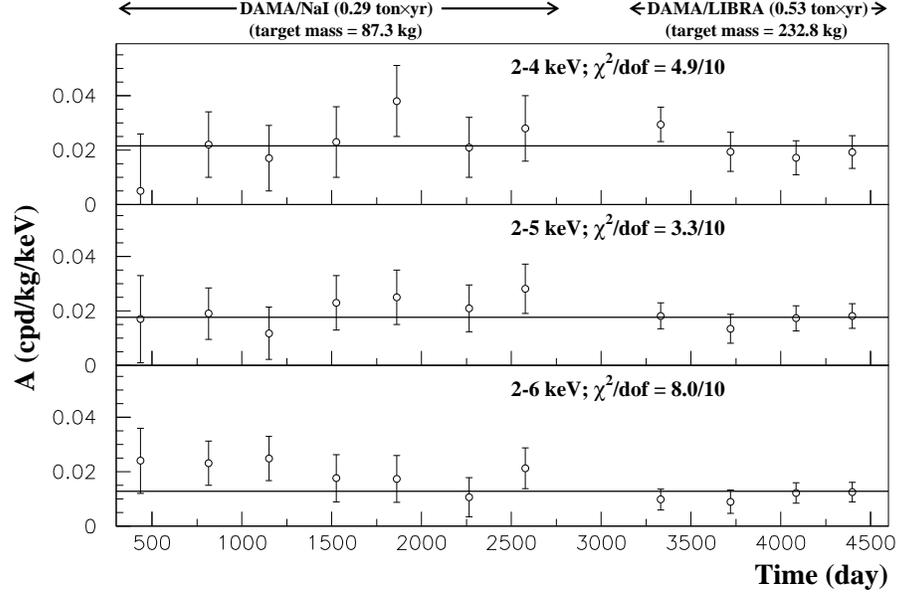}
\end{center}
\vspace{-0.4cm}
\caption{Modulation amplitudes of each single year of DAMA/NaI and DAMA/LIBRA experiments 
in the (2 -- 4), (2 -- 5) and (2 -- 6) keV 
energy intervals. The same time scale as in Fig. \ref{fig1} is adopted.
The solid horizontal lines shows the central values obtained by best fit over 
the whole data set (see Fig. \ref{fig1}).
The $\chi^2$ test and the {\it run test} accept the hypothesis at 90\% C.L. that the modulation amplitudes
are normally fluctuating around the best fit values. See text.}
\label{fg:cl}
\end{figure}

We note that the difference 
in the (2 -- 6) keV modulation amplitude between DAMA/NaI and DAMA/LIBRA 
depends mainly on the rate in the (5 -- 6) keV energy bin. 
In particular, the modulation amplitudes for the (2 -- 6) keV energy interval, obtained 
when fixing 
exactly the period at 1 yr and 
the phase at 152.5 days, are $(0.019\pm0.003)$ cpd/kg/keV and $(0.011\pm0.002)$ cpd/kg/keV 
for DAMA/NaI and DAMA/LIBRA, respectively; thus, their difference 
is about $\simeq 2 \sigma$ which correspond to a modest, but non negligible probability.
This is further supported by the analyses of the modulation amplitudes of each 
single year of DAMA/NaI and DAMA/LIBRA experiments,
as reported in Fig. \ref{fg:cl}. There the central values obtained by best fit over 
the whole data set (see Fig. \ref{fig1}) are also depicted.
The $\chi^2$ test 
($\chi^2 = 4.9$, 3.3 and 8.0 over 10 {\it d.o.f.} for the three energy intervals, respectively)
and the {\it run test} (lower tail probabilities 
of 74\%, 61\% and 11\% for the three energy intervals, respectively)
accept the hypothesis at 90\% C.L. 
that the modulation amplitudes are normally fluctuating around their best fit values.
Thus, the cumulative result from DAMA/NaI and DAMA/LIBRA can be adopted. 

In conclusion, the DAMA/LIBRA data are in substantial agreement with those of 
DAMA/NaI and the cumulative analysis 
favours the presence of a modulated cosine-like behaviour at 
8.2 $\sigma$ C.L. (see Table \ref{tb:libra2}). Moreover, the $\chi^2$ test on the residual rates disfavours 
the hypothesis of unmodulated behaviour ($A=0$)
giving probabilities of $1.3 \times 10^{-4}$ ($\chi^2$/{\it d.o.f.} = 117.7/67), 
$1.9 \times 10^{-4}$ ($\chi^2$/{\it d.o.f.} = 116.1/67)
and $1.8 \times 10^{-4}$ ($\chi^2$/{\it d.o.f.} = 116.4/67) for the three
energy intervals, respectively.

\begin{figure}[!tbh]
\centering
\vspace{-0.4cm}
\includegraphics[width=6.cm] {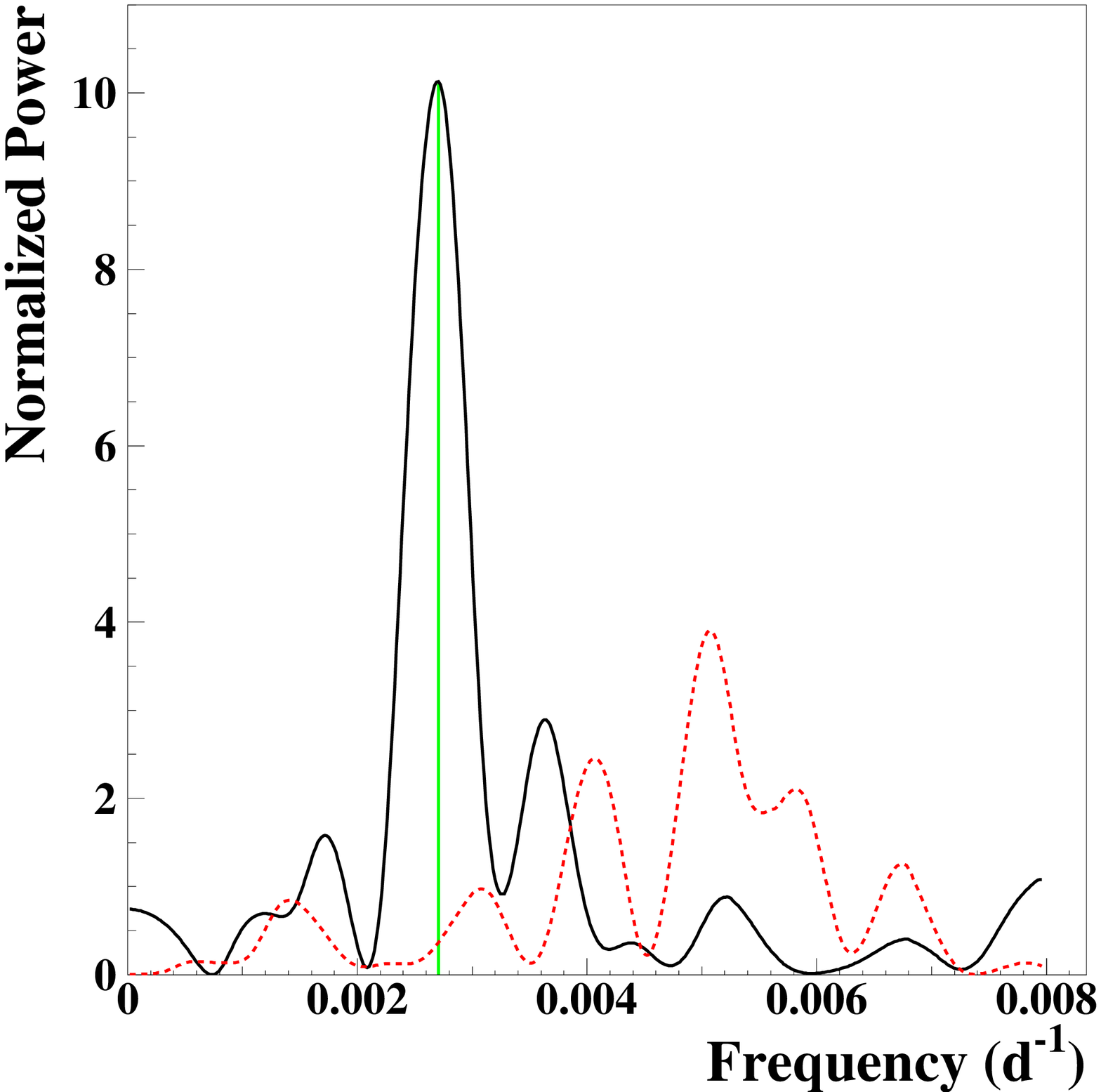}
\includegraphics[width=6.cm] {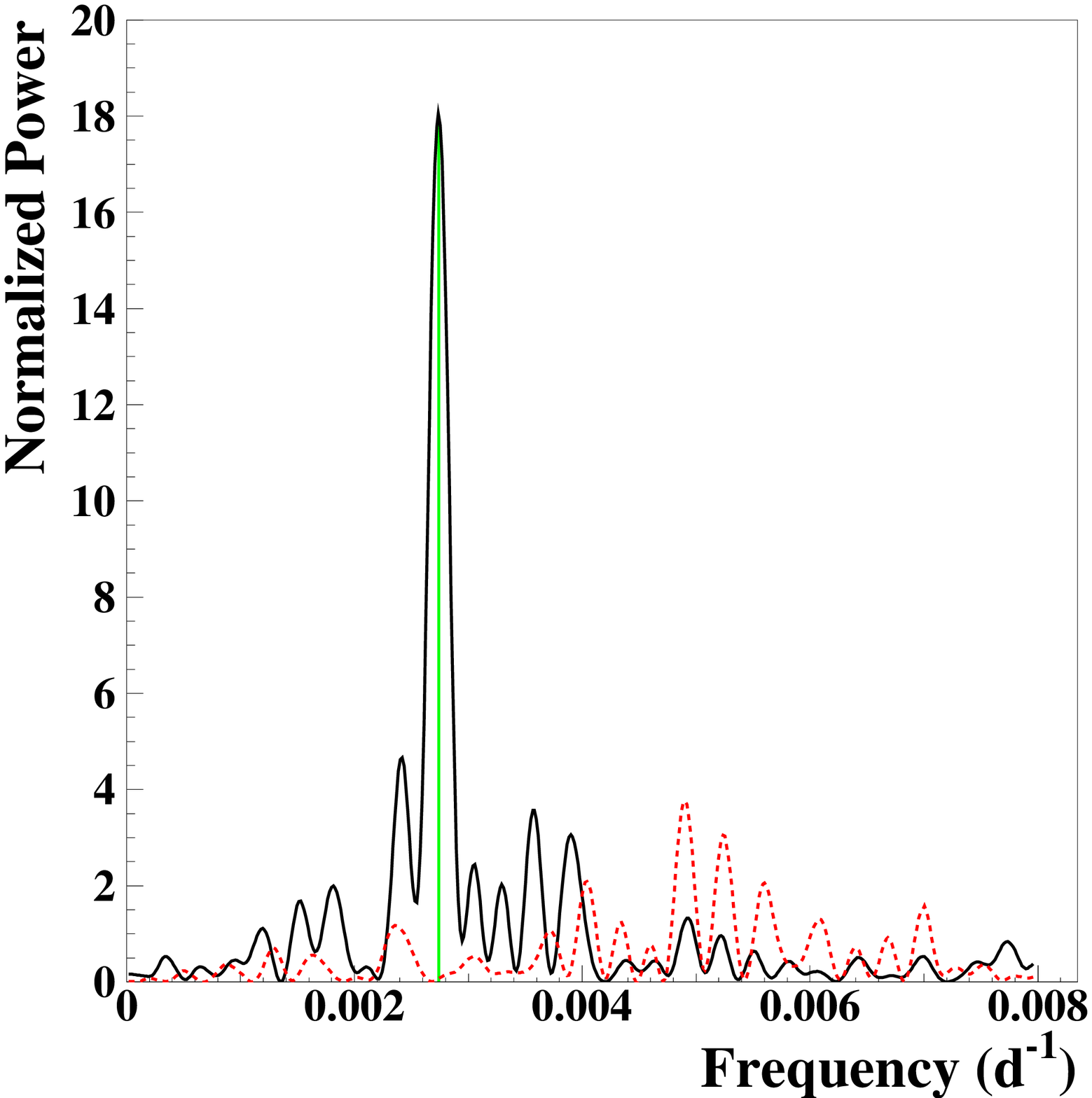}
\vspace{-0.3cm}
\caption{Power spectrum of the measured {\it single-hit} residuals for the (2--6) keV (solid lines) and 
(6--14) keV (dotted lines) energy intervals calculated according to ref. \cite{Lomb}, including also the 
treatment of the experimental errors and of the time binning. The data refer to: {\it left -} just to the DAMA/LIBRA data; 
{\it right -} to the cumulative DAMA/NaI and DAMA/LIBRA
data; the case of DAMA/NaI has been given in ref. \cite{RNC,ijmd}. 
As it can be seen, the principal mode present in the (2--6) keV energy interval corresponds to a frequency 
of $2.705 \times 10^{-3}$ d$^{-1}$ and $2.737 \times 10^{-3}$
d$^{-1}$, respectively (vertical lines); that is, they correspond to a period of $\simeq$ 1 year. 
A similar peak is not present in the (6--14) keV energy interval just above.}
\label{fig2}
\normalsize
\end{figure}

The same data of Fig. \ref{fig1} have also been investigated by a Fourier analysis 
(performed according to ref.
\cite{Lomb} including also the treatment of the experimental errors and of the time binning); in particular, 
Fig.~\ref{fig2} shows the result for the DAMA/LIBRA and for the cumulative exposure; 
the one for DAMA/NaI alone has been given in ref. \cite{RNC,ijmd}. 
Here a clear peak for a period of 1 year is evident in the lowest energy interval (2--6) keV.

\begin{figure}[!bth]
\vspace{-0.2cm}
\centering
\includegraphics[width=5.cm] {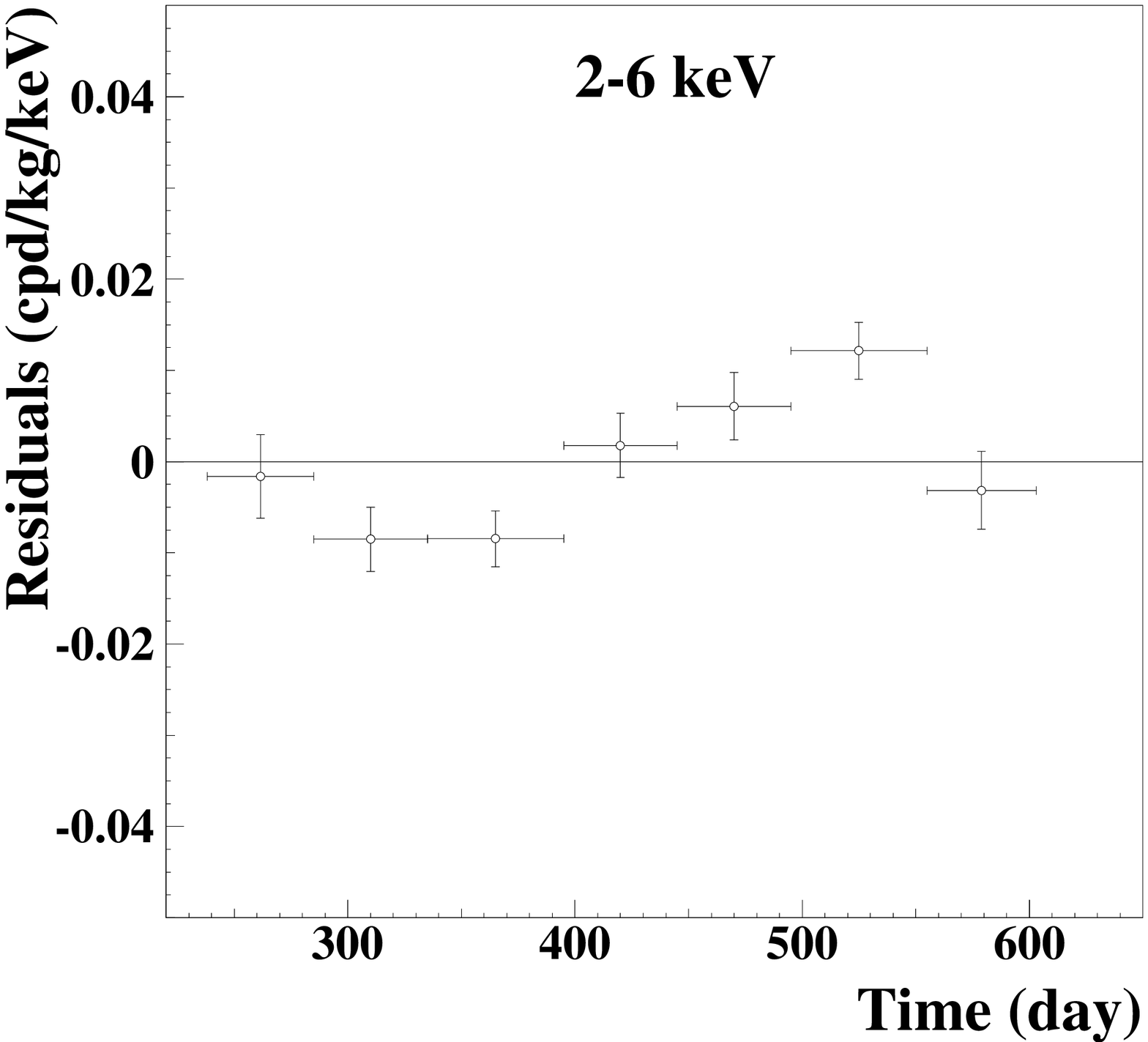}
\includegraphics[width=5.cm] {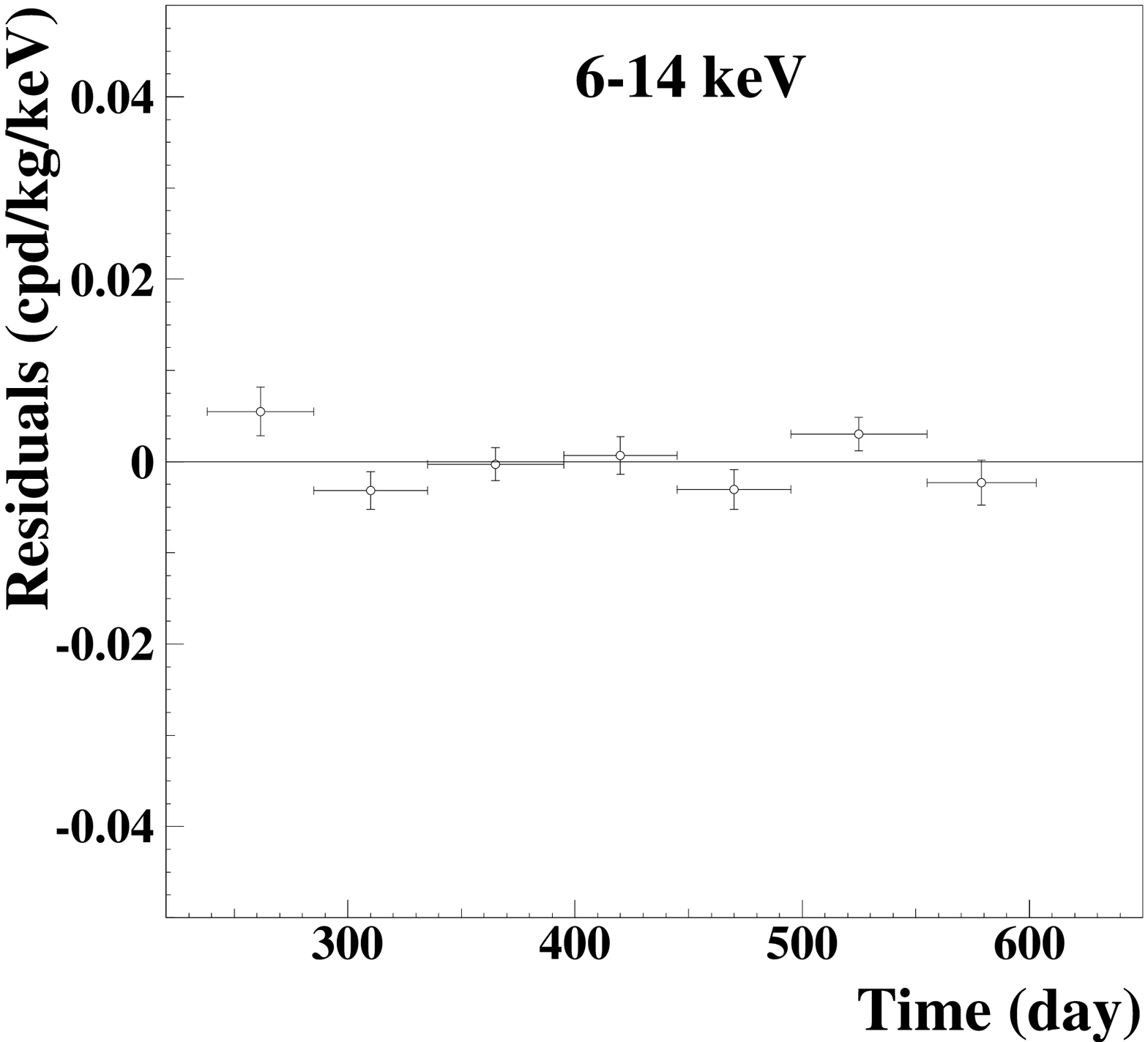}
\vspace{-0.2cm}
\caption{Experimental residuals in the (2 -- 6) keV region and those in the 
(6 -- 14) keV region just above 
for the DAMA/LIBRA data considered as collected in a
single annual cycle. The experimental points present the errors as vertical bars and the associated 
time bin width as horizontal bars. The initial time is taken at August 7$^{th}$.
The clear modulation is present in the lowest energy interval, 
while it is absent just above. See text.}
\label{fig2_1}  
\end{figure}

In the same Fig.~\ref{fig2}{\it --left} there is also shown 
the absence of modulation in the region (6 -- 14) keV just above the region 
where the modulation is present in the DAMA/LIBRA data; in 
Fig.~\ref{fig2}{\it --right} the same is shown for the cumulative DAMA/NaI 
and DAMA/LIBRA data (for the DAMA/NaI data alone 
see ref. \cite{RNC,ijmd}).

\vspace{0.3cm}

Fig.~\ref{fig2_1} compares the residuals in the (2 -- 6) keV region and those in 
the (6 -- 14) keV region just above 
for the DAMA/LIBRA data considered as collected in a
single annual cycle. A clear modulation is present in the lowest energy interval, 
while it is absent just above. In fact, the best fitted modulation amplitude 
in the (6 -- 14) keV energy region is well compatible with zero: 
(0.0009 $\pm$ 0.0011) cpd/kg/keV.

\begin{figure}[!ht]
\centering
a) \includegraphics[width=5.cm] {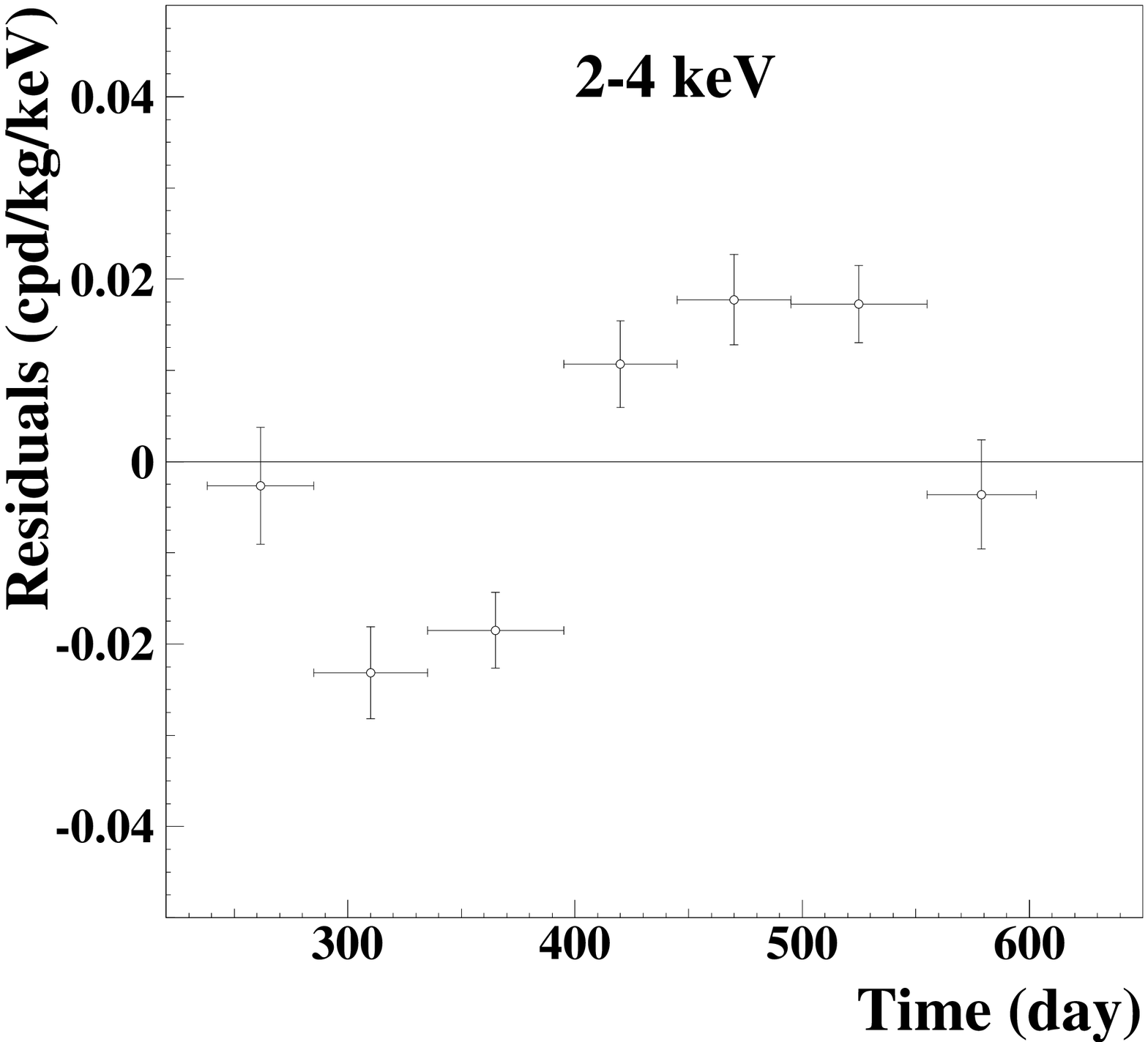}
b) \includegraphics[width=5.cm] {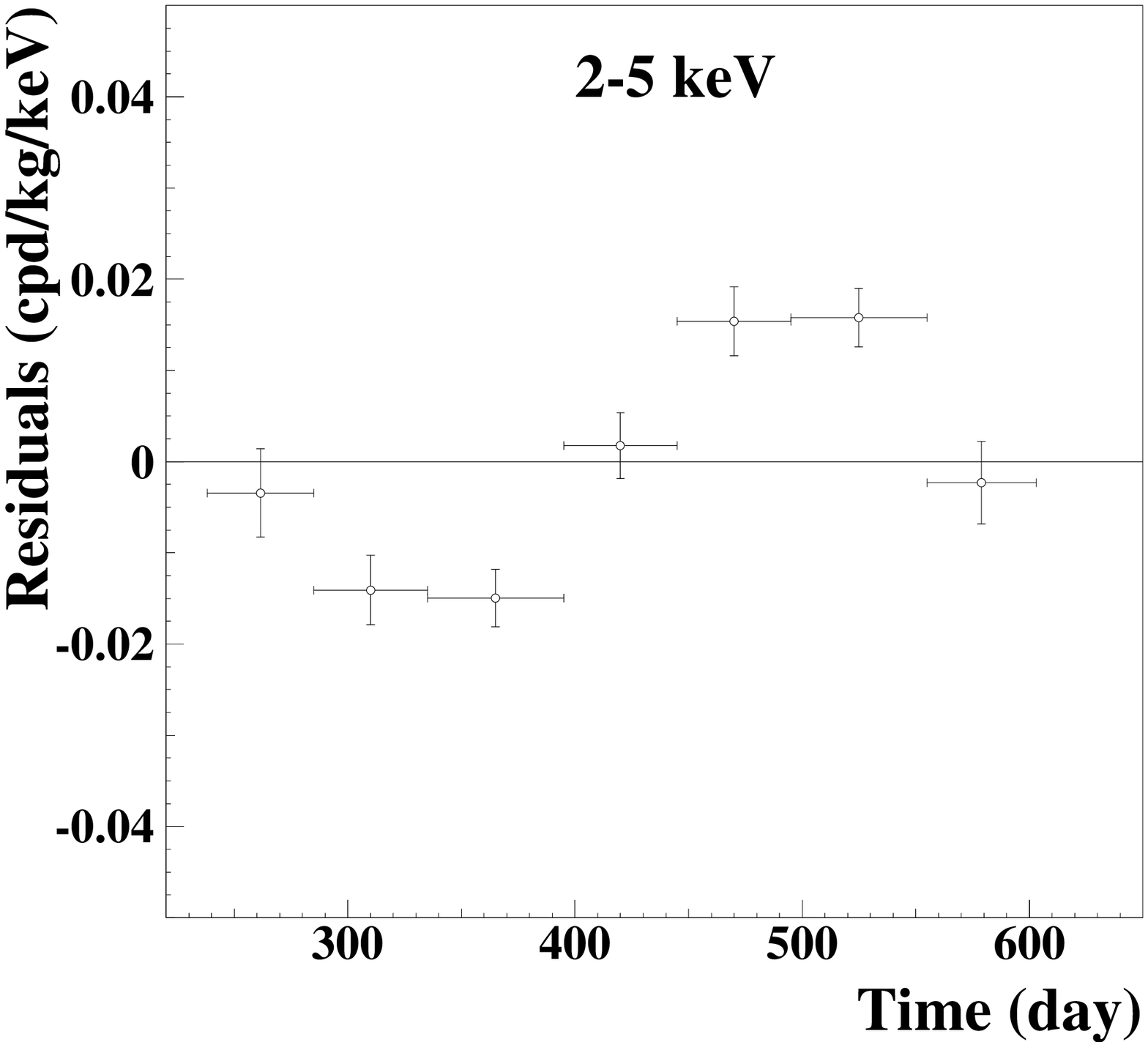} \\
c) \includegraphics[width=5.cm] {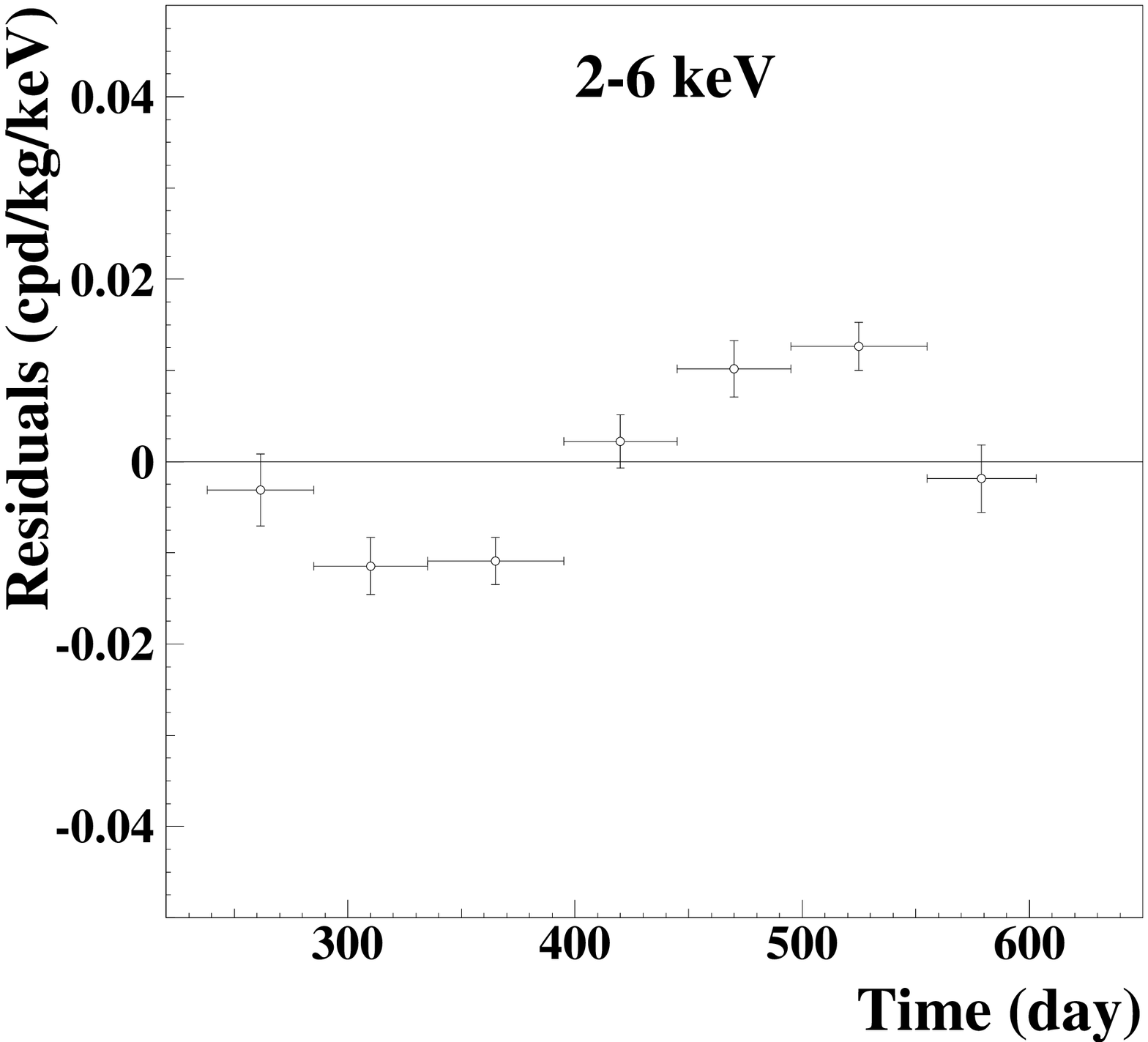}
d) \includegraphics[width=5.cm] {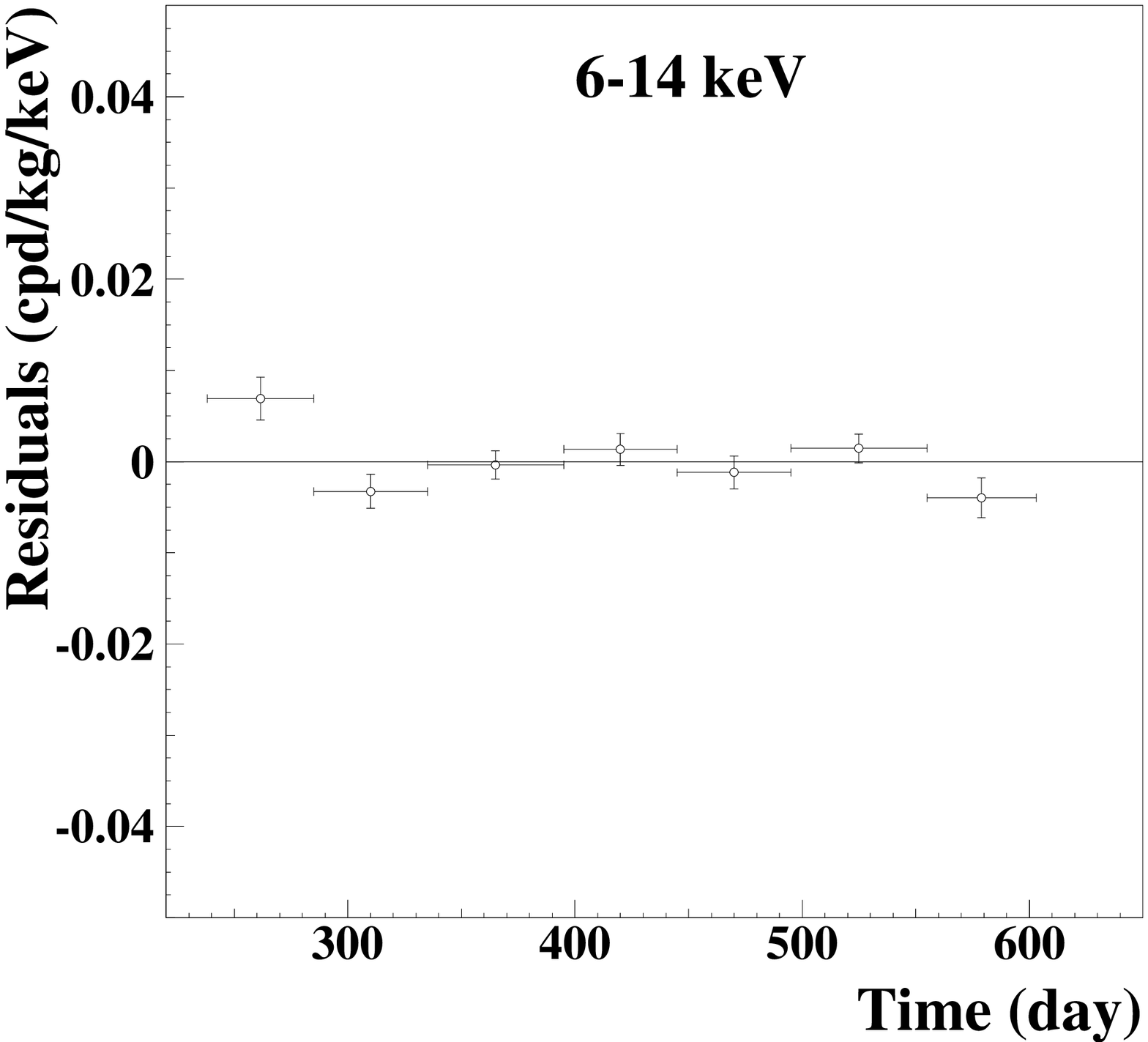} 
\vspace{-0.4cm}
\caption{Experimental {\it single-hit} residual rates, as collected in a single annual cycle,
for the total exposure of 0.82 ton$\times$yr (i.e. DAMA/NaI plus DAMA/LIBRA)
in the (2 -- 4) keV (a), (2 -- 5) keV (b), (2 -- 6) keV (c) and (6 -- 14) keV (d) energy intervals. 
The experimental points present the errors as vertical bars and the associated time bin width as horizontal 
bars. The initial time is taken at August 7$^{th}$.
Fitting the data with a cosinusoidal function when fixing exactly the period at 1 yr and the phase at 
152.5 days, the following 
amplitudes are obtained: 
a) $(0.0204 \pm 0.0026)$ cpd/kg/keV;
b) $(0.0166 \pm 0.0020)$ cpd/kg/keV;
c) $(0.0125 \pm 0.0016)$ cpd/kg/keV;
d) $(0.0004 \pm 0.0010)$ cpd/kg/keV. 
Thus, a clear modulation is present in the lowest energy regions, while it is absent just above.}
\label{fig3}  
\end{figure}

Finally, Fig.~\ref{fig3} shows -- for various energy intervals --
the experimental {\it single-hit} residual rates, as 
collected in a single annual cycle,
for the total exposure of 0.82 ton$\times$yr (i.e. DAMA/NaI \& DAMA/LIBRA).

To verify absence of annual modulation in other energy regions and, thus,  
to also verify the absence of any significant background modulation, 
the energy distribution measured during the data taking periods
in energy regions not of interest for DM detection 
have been investigated.
In fact, the background in the lowest energy region is
essentially due to ``Compton'' electrons, X-rays and/or Auger
electrons, muon induced events, etc., which are strictly correlated
with the events in the higher energy part of the spectrum.
Thus, if a modulation detected 
in the lowest energy region would be due to
a modulation of the background (rather than to a signal),
an equal or larger (sometimes much larger)
modulation in the higher energy regions should be present.
For this purpose, also in the present case we have investigated the rate 
integrated above 90 keV,  R$_{90}$, as a function of the time.
\begin{figure}[!ht]
\begin{center}
\includegraphics[width=4.cm] {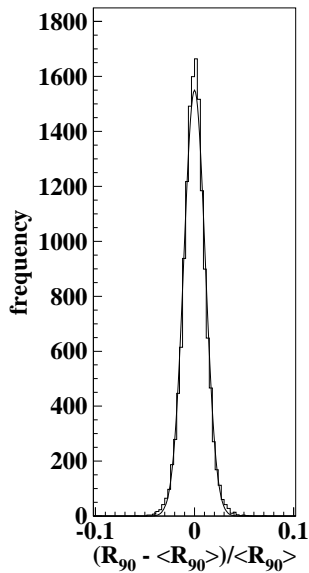}
\end{center}
\vspace{-0.3cm}
\caption{Distribution of the percentage variations
of R$_{90}$ with respect to the mean values for all the detectors in the DAMA/LIBRA-1 to -4 
annual cycles (histogram); the superimposed curve is a gaussian fit. See text and sect. \ref{sistse}.}
\label{fig_r90}
\vspace{-0.2cm}
\end{figure}
In Fig.~\ref{fig_r90} the distribution of the percentage variations
of R$_{90}$ with respect to the mean values for all the detectors
during the DAMA/LIBRA-1 to -4 annual cycles is given. 
It shows a cumulative gaussian behaviour
with $\sigma$ $\simeq$ 1\%, well accounted by the statistical 
spread expected from the used sampling time.
This result excludes any significant background variation (see also later).

Moreover, fitting
the time behaviour of R$_{90}$
with phase and period as for DM particles, a modulation amplitude compatible with zero 
is found in each running period:
$ -(0.05 \pm 0.19)$ cpd/kg, 
$ -(0.12 \pm 0.19)$ cpd/kg,
$ -(0.13 \pm 0.18)$ cpd/kg, and
$  (0.15 \pm 0.17)$ cpd/kg
for DAMA/LIBRA-1 to -4 annual cycles, respectively.
This excludes the presence of a background
modulation in the whole energy spectrum at a level much
lower than the effect found in the lowest energy region for the {\it single-hit} events; 
in fact, otherwise -- considering the R$_{90}$ mean values --
a modulation amplitude of order of tens 
cpd/kg, that $\simeq 100 \sigma$ far away from the measured value, would be present.
Other arguments are also given in sect. \ref{sistse}.

\begin{figure}[!hb]
\begin{center}
\includegraphics[width=12.cm] {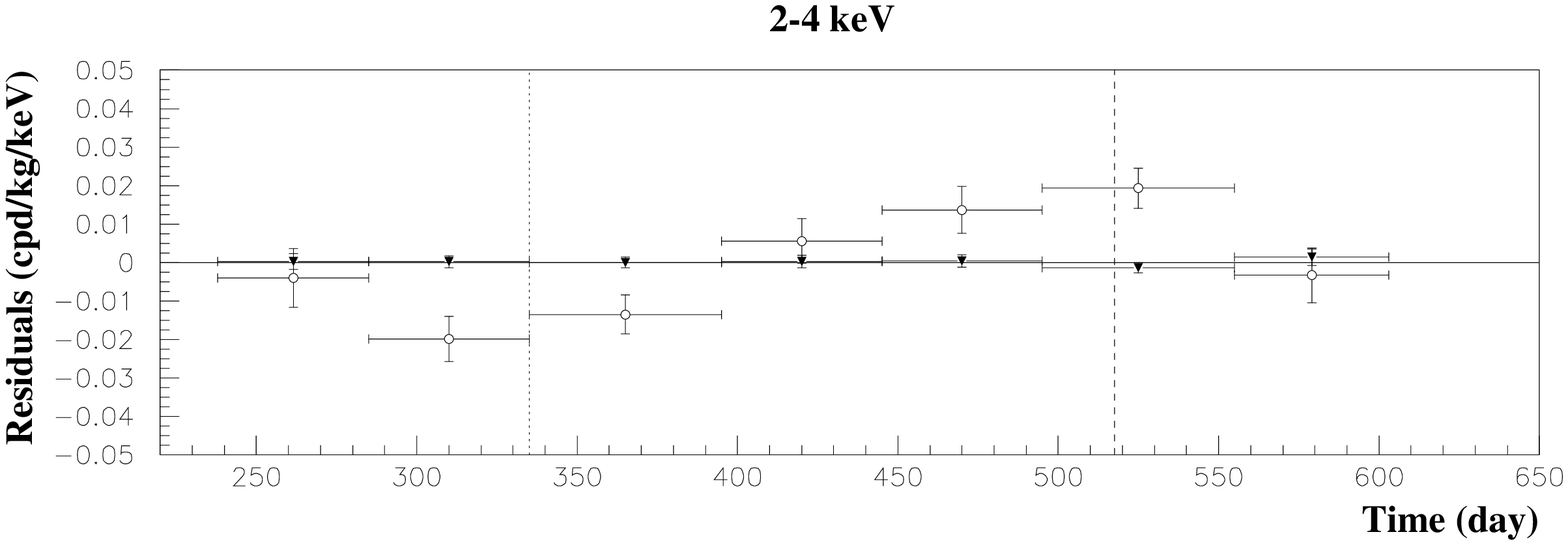}
\includegraphics[width=12.cm] {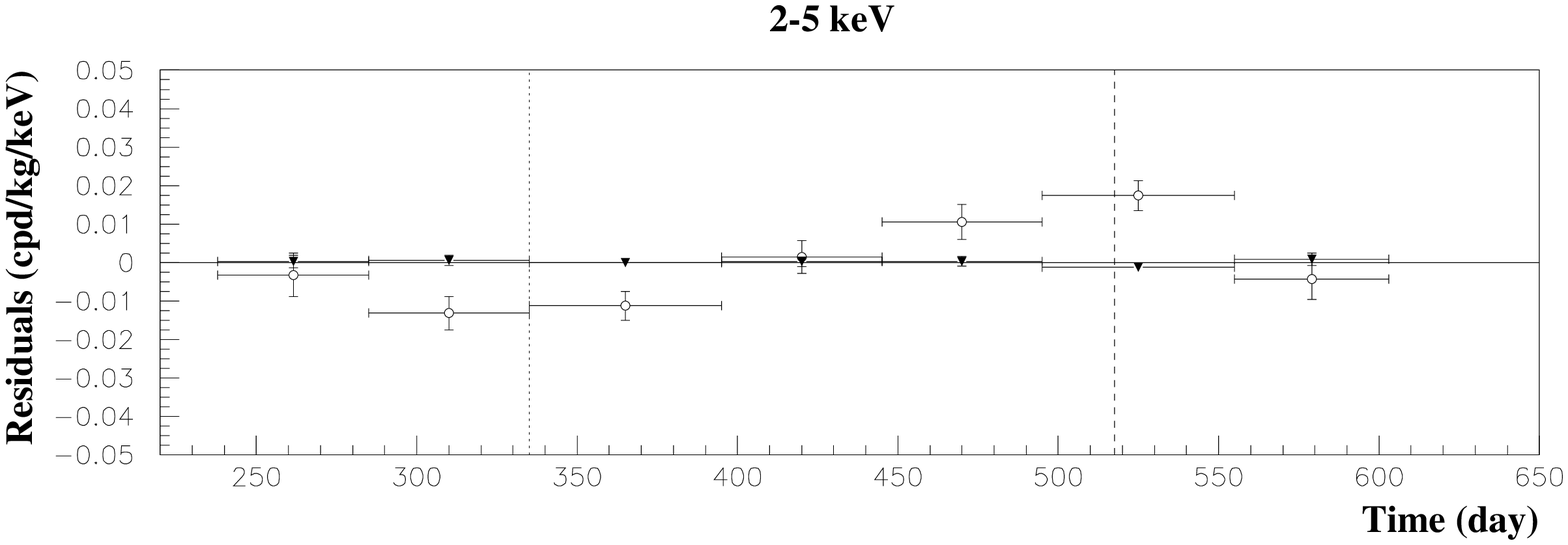}
\includegraphics[width=12.cm] {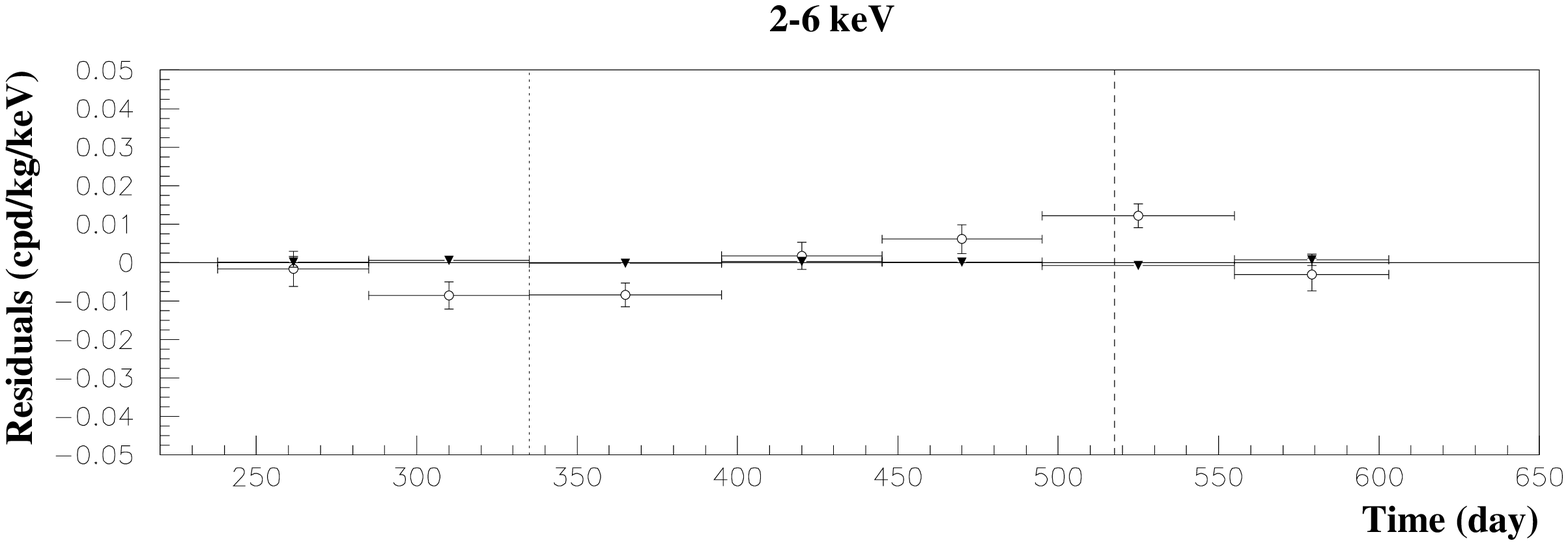}
\end{center}
\caption{Experimental residual rates over the four DAMA/LIBRA annual cycles for {\it single-hit} events (open 
circles) -- class of events to which DM events belong -- and for {\it multiple-hits} events (filled triangles)
-- class of events to which DM events do not belong -- in the (2--4), (2--5) and (2--6) keV energy intervals. 
They have been obtained by considering for each class of events the data as collected in a single annual cycle 
and by using in both cases the same identical hardware and the same identical software procedures.
The initial time of the scale is taken on August 7$^{th}$.
The experimental points present the errors as vertical bars and the associated time bin width as horizontal 
bars. See text.}
\label{fig_mul}
\end{figure}

Finally, a further relevant investigation has been done by applying the same hardware and software 
procedures, 
used to acquire and to analyse the {\it single-hit} residual rate, to the {\it multiple-hits} one. 
In fact, since the probability that a DM particle interacts in more than one detector 
is negligible, a DM signal can be present just in the {\it single-hit} residual rate.
Thus, this allows the test of the background behaviour in the same energy interval of the observed 
positive effect.
We remind that similar investigations have already been performed for the last two annual cycles of the
DAMA/NaI experiment \cite{ijmd}.
Thus, in Fig. \ref{fig_mul} the residual rates of the {\it single-hit} events measured over the four 
DAMA/LIBRA annual 
cycles are reported, as collected in a single annual cycle, together with the residual rates of the {\it 
multiple-hits} events, 
in the considered energy intervals. 

\vspace{0.3cm}

While, as already observed, a clear modulation is present in the {\it single-hit} events,
the fitted modulation amplitudes for the {\it multiple-hits} 
residual rate are well compatible with zero: 
$-(0.0004\pm0.0008)$ cpd/kg/keV,
$-(0.0005\pm0.0007)$ cpd/kg/keV,
and $-(0.0004\pm0.0006)$ cpd/kg/keV 
in the energy regions (2 -- 4), (2 -- 5) and (2 -- 6) keV, respectively.
Thus, again evidence of annual modulation with proper features as required by the DM annual modulation
signature is 
present in the {\it single-hit} residuals (events class to which the 
DM particle induced events belong), while it is absent in the {\it multiple-hits} residual rate (event class to 
which only background events belong).
Since the same identical hardware and the same identical software procedures have been used to analyse the 
two classes of events, the obtained result offers an additional strong support for the presence of a DM 
particle component in the galactic halo further excluding any side effect either from hardware or from software 
procedures or from background.

\vspace{0.4cm}

The annual modulation present at low energy can also be shown by depicting the
$S_{m,k}$ values as a function of the energy; the $S_{m,k}$ is the
modulation amplitude of the modulated part of the signal (see eq. \ref{eq1}) obtained
by maximum likelihood method over the data considering $T=$1 yr and $t_0=$ 152.5 day.
For such purpose the likelihood function of the {\it single-hit} experimental data
in the $k-$th energy bin is defined as:
\begin{equation}
{\it\bf L_k}  = {\bf \Pi}_{ij} e^{-\mu_{ijk}}
{\mu_{ijk}^{N_{ijk}} \over N_{ijk}!}
\label{eq:maxl}
\end{equation}
where $N_{ijk}$ is the number of events collected in the
$i$-th time interval (hereafter 1 day), by the $j$-th detector and in the
$k$-th energy bin. $N_{ijk}$ follows a Poissonian
distribution with expectation value
$\mu_{ijk} = \left[ b_{jk} + S_{0,k} + S_{m,k} \cdot \cos\omega(t_i-t_0)\right] M_j \Delta
t_i \Delta E \epsilon_{jk}$.
The b$_{jk}$ are the background contributions, $M_j$ is the mass of the $j-$th detector,
$\Delta t_i$ is the detector running time during the $i$-th time interval,
$\Delta E$ is the chosen energy bin,
$\epsilon_{jk}$ is the overall efficiency.
The usual procedure is to minimize the function $y_k=-2ln({\it\bf L_k}) - const$ for each energy bin;
the free parameters of the fit are the $(b_{jk} + S_{0,k})$ contributions and the $S_{m,k}$
parameter.

\begin{figure}[!thbp]
\vspace{0.2cm}
\begin{center}
\includegraphics[width=\textwidth] {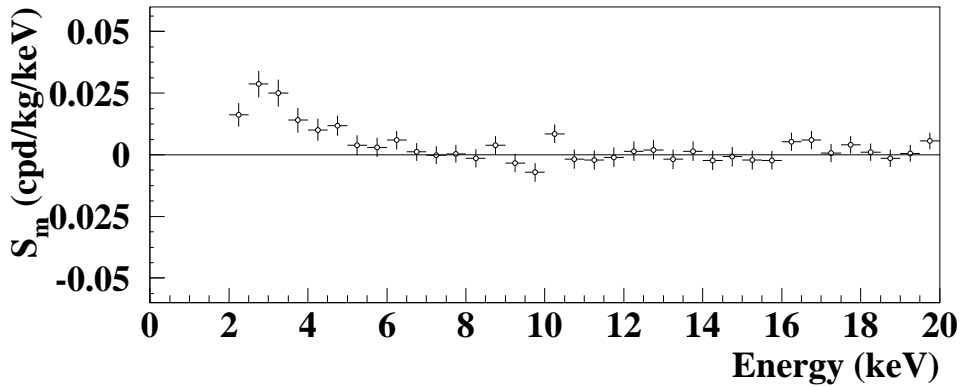}
\end{center}
\caption{Energy distribution of the $S_{m,k}$ variable for the
total exposure (0.82 ton$\times$yr, DAMA/NaI \& DAMA/LIBRA).
See text. A clear modulation is present in the lowest energy region, 
while $S_{m,k}$ values compatible with zero are present just above. In fact, the $S_{m,k}$ values
in the (6--20) keV energy interval have random fluctuations around zero with
$\chi^2$ equal to 24.4 for 28 degrees of freedom.
See also Appendix A.}
\label{sme}
\end{figure}

\vspace{0.3cm}

In Fig. \ref{sme} the $S_{m,k}$  are reported
for the seven annual cycles of DAMA/NaI and for the four annual cycles of DAMA/LIBRA
in each considered energy bin (here $\Delta E = 0.5$ keV).
It can be inferred that positive signal is present in the (2--6) keV energy interval, while $S_{m,k}$
values compatible with zero are present just above. In fact, the $S_{m,k}$ values
in the (6--20) keV energy interval have random fluctuations around zero with
$\chi^2$ equal to 24.4 for 28 degrees of freedom.
All this confirms the previous analyses.

\vspace{0.5cm}

The method also allows the extraction of the $S_m$ (hereafter the index $k$ 
is omitted) values
for each detector, for each annual cycle as well as for each considered energy bin.
The $S_m$ are expected to follow a normal distribution in absence of any systematic effects.
Therefore, in order to show if they are statistically well 
distributed 
in all the crystals, in all the annual cycles and in the energy bins, 
the variable $x = \frac {S_m - \langle S_m \rangle}{\sigma}$  is considered.
Here, $\sigma$ are the errors associated to $S_m$ and $\langle S_m \rangle$ 
are the mean values of the $S_m$ averaged over the detectors 
and the annual cycles for each considered energy bin (in the following $\Delta E = 0.25$ keV).
Similar investigations have already been performed also for DAMA/NaI \cite{RNC,ijmd}.

\begin{figure}[!ht]
\vspace{-0.3cm}
\begin{center}
\includegraphics[width=8.cm] {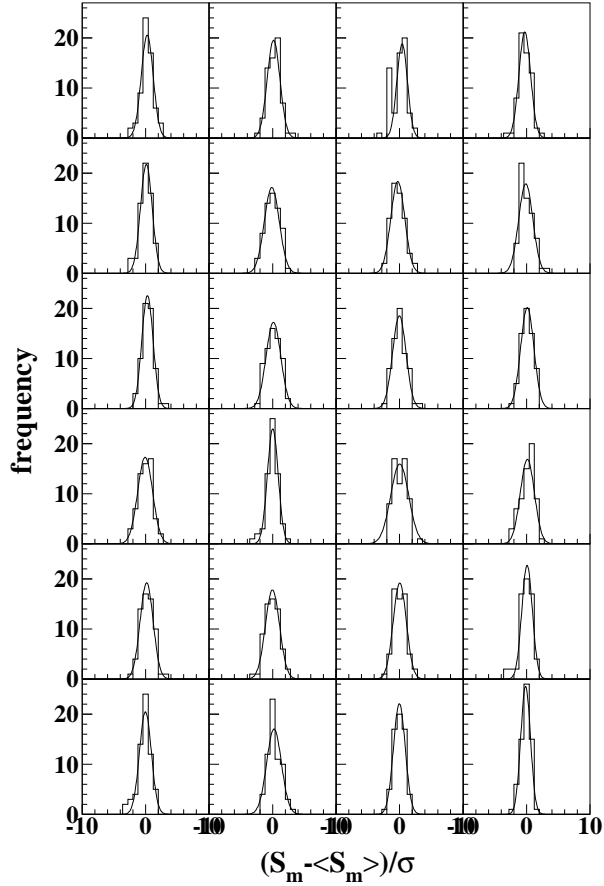}
\end{center}
\vspace{-0.3cm}
\caption{Distributions (histograms) of the variable $\frac {S_m - \langle S_m \rangle}{\sigma}$ (where 
$\sigma$ is the error
associated to the $S_m$ and $\langle S_m \rangle$ 
are the mean values of the modulation amplitudes averaged over the detectors 
and the annual cycles for each considered energy bin). The superimposed curves are gaussian 
fits. Each panel refers to a single DAMA/LIBRA detector
in the (2 -- 6) keV energy interval for the four DAMA/LIBRA annual cycles.}
\label{gaus2}
\vspace{-0.3cm}
\end{figure}

\begin{figure}[!ht]
\vspace{-0.4cm}
\begin{center}
\includegraphics[width=0.6\textwidth] {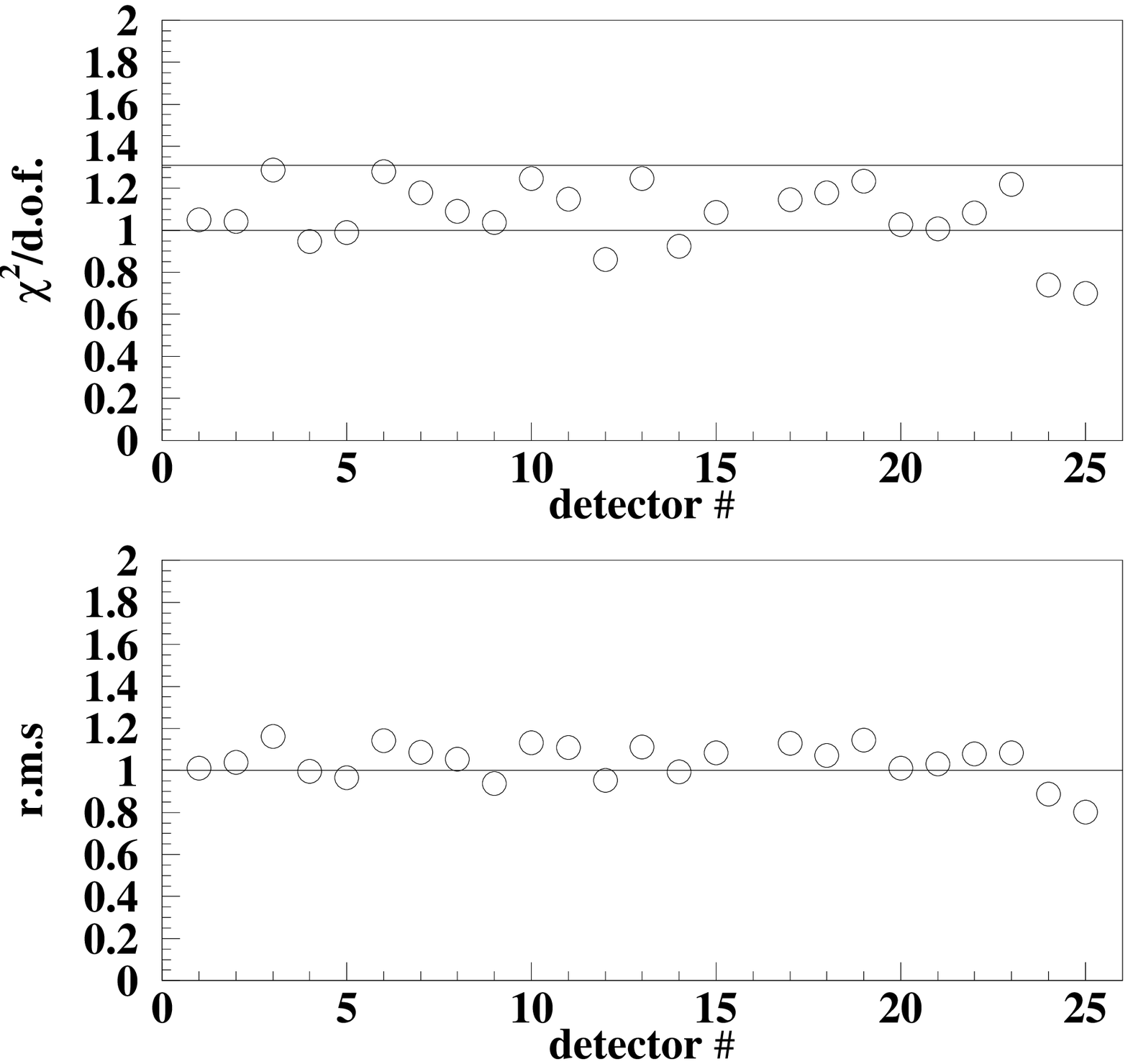}
\raisebox{7.6cm}{\includegraphics[width=0.225\textwidth,angle=270] {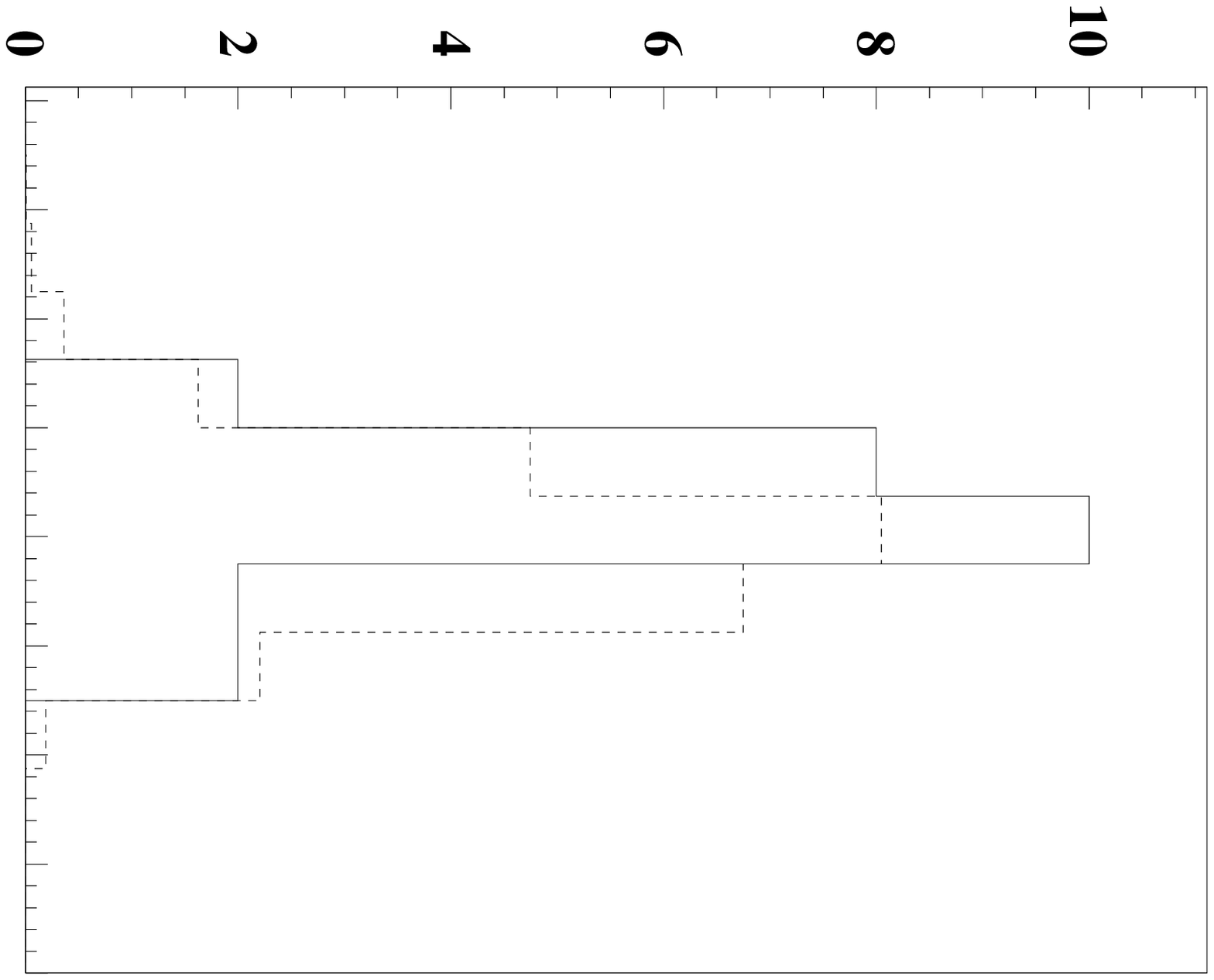}}
\end{center}
\vspace{-0.3cm}
\caption{{\it Top left:} $\chi^2/d.o.f.$ values of $S_m$ distributions around their mean value
for each DAMA/LIBRA detector in the (2--6) keV energy interval for the four annual cycles.
The line at $\chi^2/d.o.f. = 1.31$ corresponds to an upper tail probability of 5\%. 
All the $\chi^2/d.o.f.$ values are below this line and, thus, at 95\% C.L.
the observed annual modulation effect is well
distributed in all the detectors. {\it Top right:} distribution of the twenty-four points 
in the top left panel (solid histogram) compared with the $\chi^2$ distribution with 64 degrees of
freedom; the comparison gives a $\chi^2/d.o.f. = 8.1/7$. See text. 
{\it Bottom:} standard deviations of the $x$ variable for the DAMA/LIBRA detectors;
see also Fig. \ref{gaus2}. See text.}
\label{chi2}
\end{figure}

Figure \ref{gaus2} shows the distributions of the variable $x$
for the DAMA/LIBRA data in the (2--6) keV energy interval
plotted for each detector separately
(i.e. the entries of each histogram are the 
64 $x$ values, evaluated for the 16 energy bins in the considered
(2--6) keV energy interval and for the 4 DAMA/LIBRA annual cycles).
These distributions allow one to conclude that the observed annual modulation effect is well
distributed in all the detectors and annual cycles. In fact, the standard deviations of 
the $x$ variable for the DAMA/LIBRA detectors
range from 0.80 to 1.16 (see also Fig. \ref{chi2}{\it --bottom}). 
Defining $\chi^2 = \Sigma x^2$ -- where the sum is extended over 
all the 64 $x$ values -- $\chi^2/d.o.f.$ values ranging from 0.7 to 1.28 (see Fig. \ref{chi2}{\it --top}) are obtained.
The corresponding upper tail probabilities range from about 97\% to 6\%.
Therefore, the observed annual modulation effect is well
distributed in all the detectors at 95\% C.L..
The $\chi^2/d.o.f.$ values of the DAMA/LIBRA detectors show a distribution around their expectation value
(see Fig. \ref{chi2}{\it --top}). 
The twenty-four points follow a $\chi^2$ distribution with 64 degrees of freedom;
in fact, when compared with the expectation in Fig. \ref{chi2}{\it --top}-{\it right},
a $\chi^2/d.o.f. = 8.1/7$ is obtained.
The mean value of the twenty-four $\chi^2/d.o.f.$ values
is 1.072, slightly larger than expected.
Although this can be still ascribed to statistical fluctuations (see before),  
let us ascribe it to a possible systematics. In this case, one would 
have an additional error of $\leq 5 \times 10^{-4}$ cpd/kg/keV, if quadratically combined, or 
$\leq 7 \times 10^{-5}$ cpd/kg/keV, if linearly combined, to the modulation amplitude 
measured in the (2 -- 6) keV energy interval. 
This possible additional error -- $\leq 4.7\%$ or $\leq 0.7\%$, respectively, of the 
DAMA/LIBRA modulation amplitude --
can be considered as an upper limit of possible systematic effects (see also later the dedicated
section).

The above arguments demonstrate 
that the modulation amplitudes are statistically well distributed in all the 
crystals, in all the data taking periods and in the considered energy bins.

\begin{figure}[!htbp]
\begin{center}
\includegraphics[width=0.45\textwidth] {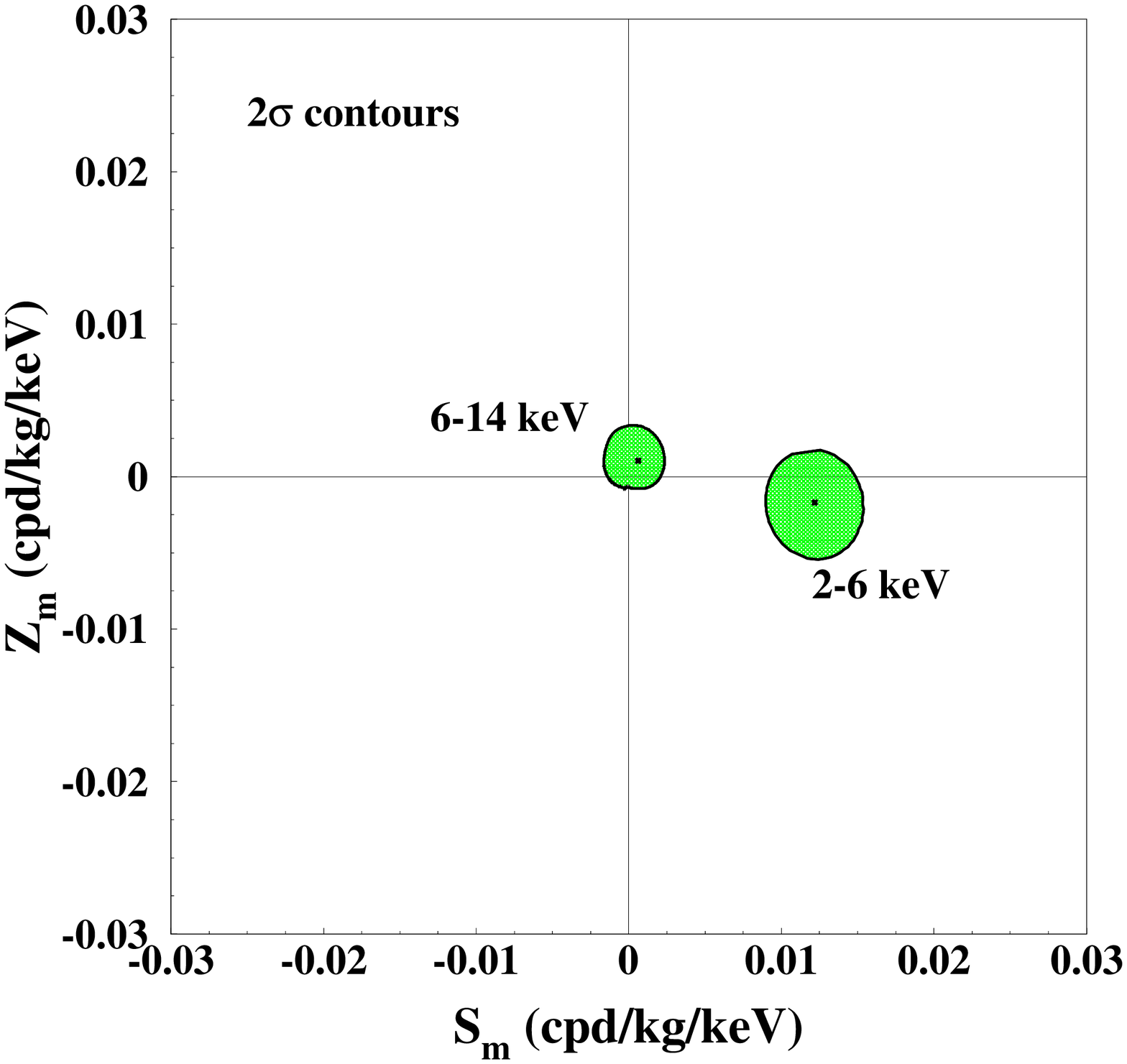}
\includegraphics[width=0.45\textwidth] {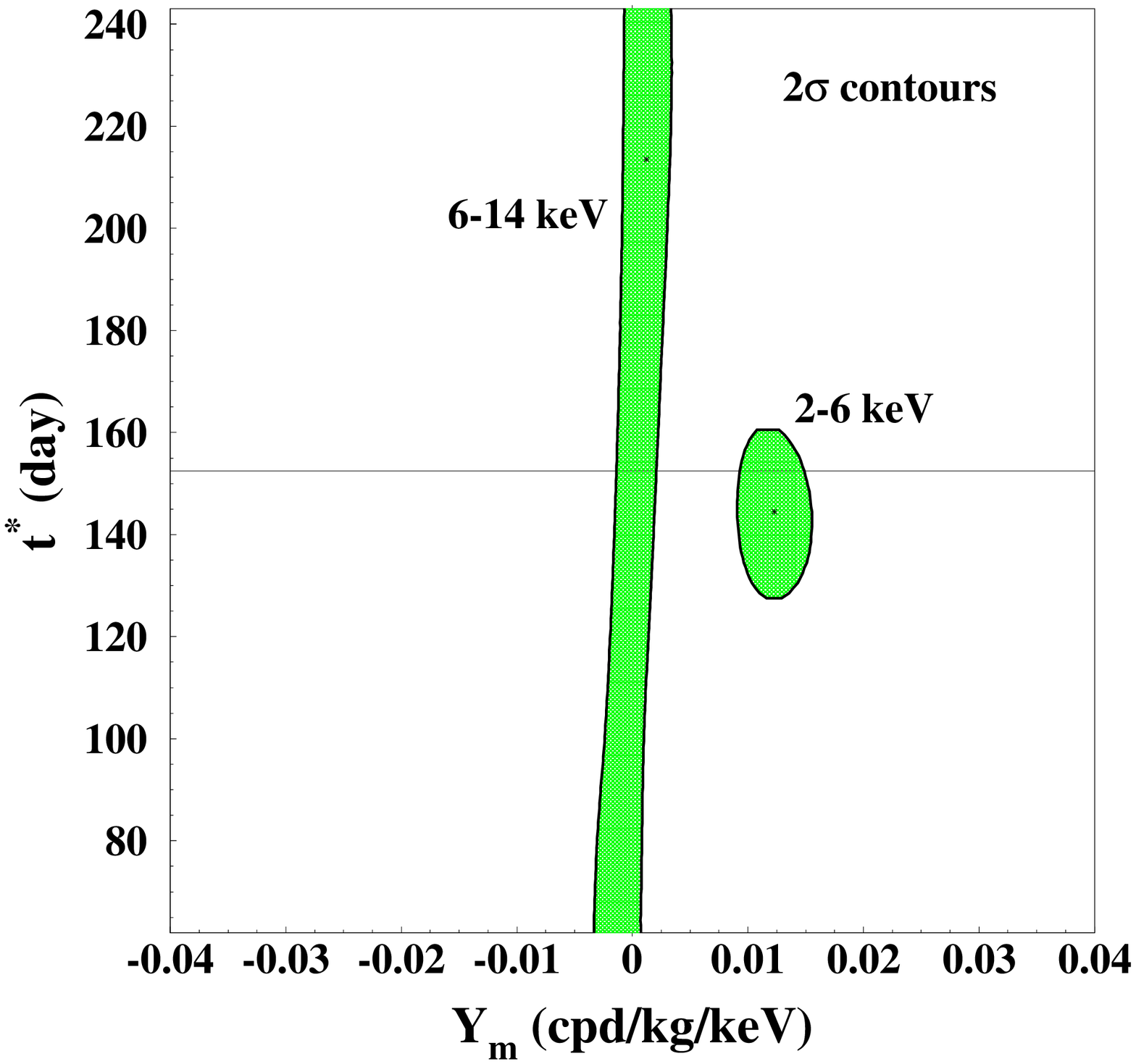}
\end{center}
\vspace{-0.2cm}
\caption{$2\sigma$ contours in the plane $(S_m , Z_m)$ ({\it left})
and in the plane $(Y_m , t^*)$ ({\it right})
for the (2--6) keV and (6--14) keV energy intervals.
The contours have been  
obtained by the maximum likelihood method, considering the
seven annual cycles of DAMA/NaI and the four annual cycles of DAMA/LIBRA all together.}
\label{fg:bid}
\end{figure}

Let us, finally, release the assumption of a phase $t_0=152.5$ day 
in the maximum likelihood procedure to evaluate the modulation amplitudes, as performed above; 
that is, let us alternatively write the expectation values, $\mu_{ijk}$, in eq. (\ref{eq:maxl}) as:
\begin{eqnarray}
\mu_{ijk} & = & \left[ b_{jk} + S_{0,k} + S_{m,k} \cos\omega(t_i-t_0) + Z_{m,k} \sin\omega(t_i-t_0)\right] 
                M_j \Delta t_i \Delta E \epsilon_{jk} \label{eq:mu1} \\
          & = & \left[ b_{jk} + S_{0,k} + Y_{m,k} \cos\omega(t_i-t^*)\right] M_j \Delta
                t_i \Delta E \epsilon_{jk} \;.
\label{eq:mu2}
\end{eqnarray}
Obviously, for signals induced by DM particles one would expect: 
i) $Z_{m,k} \sim 0$ (because of the orthogonality between the cosine and the sine functions); 
ii) $S_{m,k} \simeq Y_{m,k}$; iii) $t^* \simeq t_0=152.5$ day. 
In fact, these conditions hold for most of the dark halo models; however, it is worth noting that 
slight differences can be expected in case of possible contributions
from non-thermalized DM components, such as e.g. the SagDEG stream \cite{epj06} and the caustics \cite{caus}.
The analysis has been performed considering the data of the
seven annual cycles of DAMA/NaI and the four annual cycles of DAMA/LIBRA all together.
Fig. \ref{fg:bid}{\it --left} shows the 
$2\sigma$ contours in the plane $(S_m , Z_m)$ 
for the (2--6) keV and (6--14) keV energy intervals and 
Fig. \ref{fg:bid}{\it --right} shows, instead, those in the plane $(Y_m , t^*)$.
The best fit values for the (2--6) keV energy interval are ($1\sigma$ errors): 
$S_m= (0.0122 \pm 0.0016)$ cpd/kg/keV; 
$Z_m=-(0.0019 \pm 0.0017)$ cpd/kg/keV; 
$Y_m= (0.0123 \pm 0.0016)$ cpd/kg/keV; 
$t^*= (144.0  \pm 7.5)$ day;
while for the (6--14) keV energy interval are:
$S_m= (0.0005 \pm 0.0010)$ cpd/kg/keV;
$Z_m= (0.0011 \pm 0.0012)$ cpd/kg/keV;
$Y_m= (0.0012 \pm 0.0011)$ cpd/kg/keV
and $t^*$ obviously not determined (see Fig. \ref{fg:bid}). 
These results confirm those achieved above by analysing the residuals.
In particular, a modulation amplitude is present in the lower energy intervals and the period
and the phase agree with those expected for DM induced signals.
Finally, forcing to zero the contribution of the cosine function in eq. (\ref{eq:mu1}),
the $Z_{m,k}$ values as function of the energy have also been determined
by using the same procedure. Obviously, such values are expected to be zero in case of 
presence of a DM particles' signal with $t_0 = 152.5$ day.
From the data we get:
$-(0.0046 \pm 0.0041)$ cpd/kg/keV,
$-(0.0036 \pm 0.0042)$ cpd/kg/keV,
$ (0.0036 \pm 0.0034)$ cpd/kg/keV, and
$-(0.0015 \pm 0.0032)$ cpd/kg/keV,
for (2-3), (3-4), (4-5), (5-6) energy intervals, respectively.
Moreover, the $\chi^2$ test applied to the data in the (2-14) keV energy region
($\chi^2$ equal to 12.4 for 12 degrees of freedom) supports
the hypothesis that the $Z_{m,k}$ values are simply fluctuating around zero.

In conclusion, an annual modulation satysfying all the requirements of the DM annual 
modulation signature has been observed also in DAMA/LIBRA with high C.L.; it
credits cumulatively with DAMA/NaI an evidence 
for the presence of DM particles in the galactic halo at 8.2 $\sigma$ C.L..

\section{Further investigation on possible systematic effects and side reactions}
\label{sistse}

Also the data of the first four annual cycles of DAMA/LIBRA as previously those of DAMA/NaI
fullfill the requirements 
of the DM annual modulation signature and preliminar investigation on absence of
any significant systematics or side reaction effect is already present in some parts of 
the previous section; however, 
here the argument will be addressed in more details.

\begin{table}[!b]
\caption{Modulation amplitudes (1 $\sigma$ error) obtained by fitting the time behaviours of the main
running parameters including a possible annual modulation with phase and period
as for  DM particles. These running parameters, acquired with the production data, 
are: i) the operating temperature of the detectors;
ii) the HP Nitrogen flux in the inner Cu box housing the detectors; 
iii) the pressure of the HP Nitrogen atmosphere of the inner Cu box
housing the detectors; iv) the environmental Radon in the inner part of the
barrack from which the detectors are however excluded (see text and ref. \cite{perflibra} for details); 
v) the hardware rate above single photoelectron threshold. All the measured amplitudes are
compatible with zero.}
\vspace{0.4cm}
\centering
\resizebox{\textwidth}{!}{
\begin{tabular}{|c||c||c||c||c||} \hline
  & & & & \\
  & DAMA/LIBRA-1 & DAMA/LIBRA-2 & DAMA/LIBRA-3 & DAMA/LIBRA-4 \\
  & & & & \\
\hline
  & & & & \\
Temperature & $ -(0.0001 \pm 0.0061) ^{\circ}$C &
              $ (0.0026 \pm 0.0086) ^{\circ}$C  &
              $ (0.001 \pm 0.015) ^{\circ}$C &
              $ (0.0004 \pm 0.0047) ^{\circ}$C  \\
  & & & & \\
Flux  & $ (0.13 \pm 0.22)$ l/h &
        $ (0.10 \pm 0.25)$ l/h &
        $-(0.07 \pm 0.18)$ l/h &
        $-(0.05 \pm 0.24)$ l/h \\
  & & & & \\
Pressure   & $ (15 \pm 30) 10^{-3}$ mbar &
              $ -(13 \pm 25) 10^{-3}$ mbar &
              $ (22 \pm 27) 10^{-3}$ mbar &
              $ (1.8 \pm 7.4) 10^{-3}$ mbar \\
&  & & & \\
Radon       & $ -(0.029 \pm 0.029)$ Bq/m$^{3}$ &
              $ -(0.030 \pm 0.027)$ Bq/m$^{3}$ &
              $ (0.015 \pm 0.029)$ Bq/m$^{3}$ &
              $ -(0.052 \pm 0.039)$ Bq/m$^{3}$ \\
&  & & & \\
Hardware rate & $-(0.20 \pm 0.18) 10^{-2}$ Hz &
              $ (0.09 \pm 0.17) 10^{-2}$ Hz &
              $ -(0.03 \pm 0.20) 10^{-2}$ Hz &
              $ (0.15 \pm 0.15) 10^{-2}$ Hz \\
&  & & & \\
\hline\hline
\end{tabular}}
\label{tb:par1234}
\end{table}

Sometimes naive statements are put forwards as the fact that 
in nature several phenomena may show annual periodicity. 
It is worth noting that the point is whether they might  
mimic the annual modulation signature in DAMA/LIBRA, i.e. whether they 
might be not only quantitatively able to account for the observed 
modulation amplitude but also able to contemporaneously 
satisfy all the requirements of the DM annual modulation signature. 
This was deeply investigated  in the  
former DAMA/NaI experiment (see e.g. ref. \cite{RNC,ijmd} and refs. therein; no one has been 
either found or suggested by anyone so far)
and will be further addressed in the following for the present DAMA/LIBRA data.

Firstly, in order to continuously monitor the running conditions, several pieces of information 
are acquired with the production data 
and quantitatively analysed; note that information on technical aspects of DAMA/LIBRA has been  
discussed in ref. \cite{perflibra}, where the sources of possible residual radioactivity have also been 
analysed. 

In particular, all the time behaviours 
of the running parameters, acquired with the production data,
have been investigated. Table \ref{tb:par1234} shows the modulation amplitudes obtained for each  
annual cycle when fitting the time behaviours of the values of the main parameters including a cosine 
modulation with the same 
phase and period as for DM particles.
As can be seen, all the measured amplitudes are
well compatible with zero.

Let us now enter in more details.

\subsection{The temperature}

Since temperature at sea level varies along the year, sometimes it has been naively 
suggested that it can mimic the observed effect.

It is worth noting that the full experiment is placed underground and works in an air-conditioned
environment; moreover, the detectors have Cu housing in direct contact with the multitons
metallic passive shield whose huge heat capacity definitively assures 
a relevant stability of the detectors' operating temperature \cite{perflibra}.
Nevertheless the operating temperature 
is read out by a probe and stored with the production data, in order to offer 
the possibility of quantitative 
investigations (see also above).
 
\begin{figure}[!htb]
\begin{center}
\vspace{-0.4cm}
\includegraphics[width=4.cm] {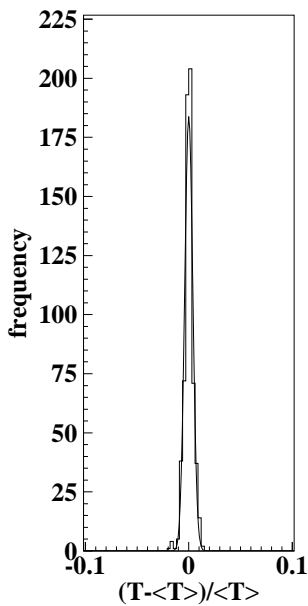}
\vspace{-.6cm}
\end{center}
\caption[]{Distribution of the relative variations of the operating temperature
measured during the DAMA/LIBRA-1 to -4 annual cycles (histogram); the superimposed curve is a gaussian 
fit. The standard deviation is 
0.4\%.}
\label{fig_temp}
\end{figure}

Specific information on the DAMA/LIBRA-1 to 4 annual cycles can be derived 
from Fig. \ref{fig_temp}; no evidence of any 
operating temperature modulation has been observed, as also 
quantitatively reported in Table \ref{tb:par1234}.

\begin{figure}[!ht]
\begin{center}
\vspace{-0.4cm}
\includegraphics[width=4.cm] {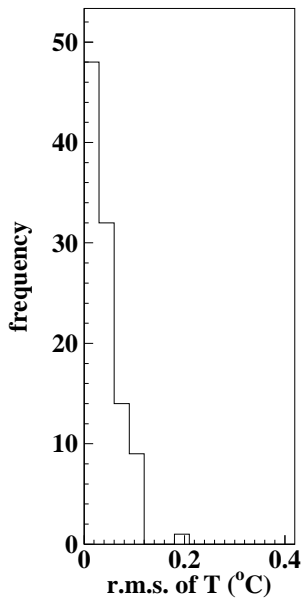}
\vspace{-.6cm}
\end{center}
\caption{Distribution of the root mean square (r.m.s.) 
detectors' operating temperature
variations within periods
with the same calibration factors (typically $\simeq$ 10 days)
during the DAMA/LIBRA-1 to -4 annual cycles.
The mean value is 0.04 $^o$C.}
\label{fig_rms_T}
\end{figure}

However, to properly evaluate the real effect of possible variations of the
detectors' operating temperature on the light output,
we consider the distribution of the root mean square temperature 
variations within periods with the same calibration factors (typically $\simeq$ 10 days); 
this is given in Fig.~\ref{fig_rms_T} cumulatively for the four-year data sets.

Considering the obtained mean value of the root mean square detectors' operating 
temperature variation: $\simeq$ 0.04 $^o$C,
and the known value of the slope of the light output around its value:
$\lsim$ -0.2\%/$^o$C, the 
relative light output variation is $\lsim 10^{-4}$, which would correspond
to a modulation amplitude
$\lsim 10^{-4}$ cpd/kg/keV (that is $\lsim 0.5\%$ of the observed modulation amplitude).

Thus, there is no significant effect from possible temperature variation in the DAMA/LIBRA 
installation; on the other 
hand, in that case some of the specific requirements of the DM annual modulation 
signature (such as e.g. the $4^{th}$ and the $5^{th}$) would fail, while they are instead 
satisfied by the DAMA/LIBRA 
production data.

In conclusion, all the arguments given above
quantitatively exclude any role of possible effects on the observed rate modulation 
directly correlated with temperature.

For the sake of completeness, we comment that 
sizeable temperature variations in principle might also induce variations in
the electronic noise, in the Radon release from the rocks and in
some environmental background;
these specific topics
will be further addressed in the following.

\subsection{The background}

As quantitatively discussed in previous section, 
in order to verify the absence of any significant background modulation, 
the energy distribution measured during the data taking periods
in energy regions not of interest for DM detection 
has been investigated.
The presence of background (of whatever nature) modulation  is already excluded 
by the results -- given in previous section -- on the measured 
rate integrated above 90 keV, R$_{90}$, as a function of the time; the latter one 
not only does not show any modulation, but  
allows one to exclude the presence of a background
modulation in the whole energy spectrum at a level some orders of 
magnitude lower than the annual modulation observed in the {\it single-hit} 
events in the (2 -- 6) keV energy region.

A further relevant support is given by the result of the analysis of the
{\it multiple-hits} events (see above) which independently proofs that there is no 
modulation at all in the background event in the same energy region where the
{\it single-hit} events present an annual modulation satifying all the requirements 
of the DM annual modulation signature.
  
The results given above obviously already account for whatever kind of background including 
that possibly induced by neutrons, by Radon and by side reactions. 
In the following we will focus our attention on the latter ones 
to offer independent cautious analyses to estimate their possible contribution,
as done in ref. \cite{Sist,RNC} for the DAMA/NaI data.

\subsubsection{... more on neutrons}

As regards the thermal neutrons, the neutron capture reactions 
$^{23}$Na(n,$\gamma$)$^{24}$Na and
$^{23}$Na(n,$\gamma$)$^{24m}$Na (cross sections to thermal neutrons
equal to 0.10 and 0.43 barn, respectively \cite{toi78}) have been investigated.
The $^{24}$Na isotope is a $\beta$-emitter (end point equal to 1.391 MeV) with two prompt
associated $\gamma$'s (2.754 and 1.369 MeV), while the $^{24m}$Na isotope decays 
100\% of the times in $^{24}$Na by internal transition with a $\gamma$ of 0.472 MeV.
Thus, the possible presence of $^{24}$Na isotopes in the NaI(Tl) crystals 
gives information about the thermal neutron flux surviving the DAMA/LIBRA shielding and 
impinging on the DAMA/LIBRA detectors;
hence, as reported in ref. \cite{perflibra}, this has been investigated with high sensitivity 
by looking for triple coincidences induced by
a $\beta$ in one detector and by the two $\gamma$'s in two adjacent ones. 
An upper limit on the thermal neutron flux surviving 
the multicomponent DAMA/LIBRA shield has been derived as \cite{perflibra}: 
$< 1.2 \times 10^{-7}$ cm$^{-2}$s$^{-1}$ (90\% C.L.)\footnote{We remind that
the thermal neutron flux has been measured in the LNGS to be 
$1.08 \times 10^{-6}$ neutrons cm$^{-2}$ s$^{-1}$ \cite{bel89} and
two consistent upper limits on the thermal neutron flux 
have been obtained with the DAMA/NaI
considering the same capture reactions and using different approaches \cite{Nim98,supclu}.}.
The corresponding capture rate is: $< 0.022$ captures/day/kg.
Assuming cautiously a 10\% modulation (of whatever origin) of the thermal neutrons flux, 
the corresponding modulation amplitude in the lowest energy region
has been calculated by MonteCarlo 
program to be $< 0.8 \times 10^{-6}$ cpd/kg/keV (that is
$<0.01\%$ of the observed modulation amplitude).
In addition, a similar contribution cannot anyhow 
mimic the annual modulation signature since 
possible modulation of thermal neutron captures would induce e.g.
variations in all the energy spectrum,
that is it would fail some of the six requirements of the
annual modulation signature (such as e.g. the $4^{th}$ and the $5^{th}$).

A similar analysis has also been carried out for the fast neutrons case, as already 
done for DAMA/NaI \cite{Sist,RNC}.
In particular, very safely, the effect of the about 1 m concrete (made from the Gran Sasso
rock material) which almost fully surrounds (outside the barrack) 
the DAMA/LIBRA passive shield -- acting as a further neutron moderator -- 
is not cautiously included here.
Thus, from the fast neutron flux measured at the Gran Sasso underground 
laboratory, $\simeq 10^{-7}$ fast neutrons cm$^{-2}$ s$^{-1}$ \cite{cri95} 
the differential counting rate above 2 keV has been estimated
by MonteCarlo code to be $\simeq 10^{-3}$ cpd/kg/keV. 
Therefore, assuming 
cautiously a 10\% modulation (of whatever origin) of the fast neutron flux,
the corresponding modulation amplitude is $<10^{-4}$ cpd/kg/keV.

Note that the use of other measurements of fast neutron flux at LNGS \cite{bel89,neutall}
does not change the given conclusions. 
Moreover, an independent evaluation of the fast neutron flux impinging on the
DAMA/LIBRA detectors has been performed by using the DAMA/LIBRA production data 
through the study of the inelastic
reaction $^{23}Na(n,n')^{23}Na^*(2076$ keV), which produces two $\gamma's$ in coincidence 
(1636 keV and 440 keV).
An upper limit -- limited by the sensitivity of the method -- has been 
found: $<2.2 \times 10^{-7}$ fast neutrons cm$^{-2}$ s$^{-1}$ (90\% C.L.),
well compatible with the measured values in the laboratory in ref. \cite{bel89,cri95,neutall} and 
estimated in ref. \cite{wul}. This further excludes 
any presence of a fast neutron flux in DAMA/LIBRA significantly larger than measured in 
ref. \cite{bel89,cri95,neutall}. 

\vspace{0.2cm}

It is worth noting that a possible neutron flux modulation as claimed in ref. 
\cite{icar} at LNGS ($\sim 5\%$) and in ref. \cite{giul} at the shallow deep Baradello mine
or suggested by phenomenological approaches \cite{wul} \footnote{In particular, it has been 
suggested that a possible origin of the variability of the neutron flux might be ascribed 
to possible variations of the water content of the environment.} 
cannot quantitatively contribute to the DAMA/LIBRA 
observed effect, even if the neutron flux at LNGS would be assumed two orders of magnitude larger than
measured.

\vspace{0.2cm}

Finally, a possible modulation in the fast neutron flux would induce
variation in all the energy spectrum and in the {\it multiple-hits} events at low energy,
that is some of the six requirements mentioned above would fail.

\subsubsection{... more on Radon}

The DAMA/LIBRA detectors are excluded from the air of the underground laboratory
by a 3-level sealing system \cite{perflibra}; in fact,
this air contains traces of the radioactive Radon gas ($^{222}$Rn
-- T$_{1/2}$ = 3.82 days -- and of $^{220}$Rn -- T$_{1/2}$ = 55 s -- isotopes, which belong to the
$^{238}$U and $^{232}$Th chains, respectively), whose 
daughters attach themselves to
surfaces by various processes. 
In particular: i) the walls, the floor and the top of the inner part of the installation are insulated 
by 
Supronyl (permeability: $2 \times 10^{-11}$
cm$^2$/s \cite{woj91}) and a large flux of HP Nitrogen is released
in the closed space of this inner part of the barrack housing the set-up.
An Oxygen level alarm informs the operator
before entering it, when necessary; ii) the whole passive shield is sealed in a 
plexiglas
box and maintained continuously in HP Nitrogen atmosphere in slight overpressure
with respect to the environment as well as 
the upper glove box for calibrating the detectors; iii)
the detectors are housed in an inner sealed 
Cu box also maintained continuously in HP Nitrogen atmosphere in slight 
overpressure
with respect to the environment; the Cu box can enter in contact only with 
the upper glove box -- during calibrations -- which is 
also continuously maintained in HP Nitrogen
atmosphere in slightly overpressure with respect to the external environment. 

\begin{figure}[!ht]
\centering
\includegraphics[width=6.cm] {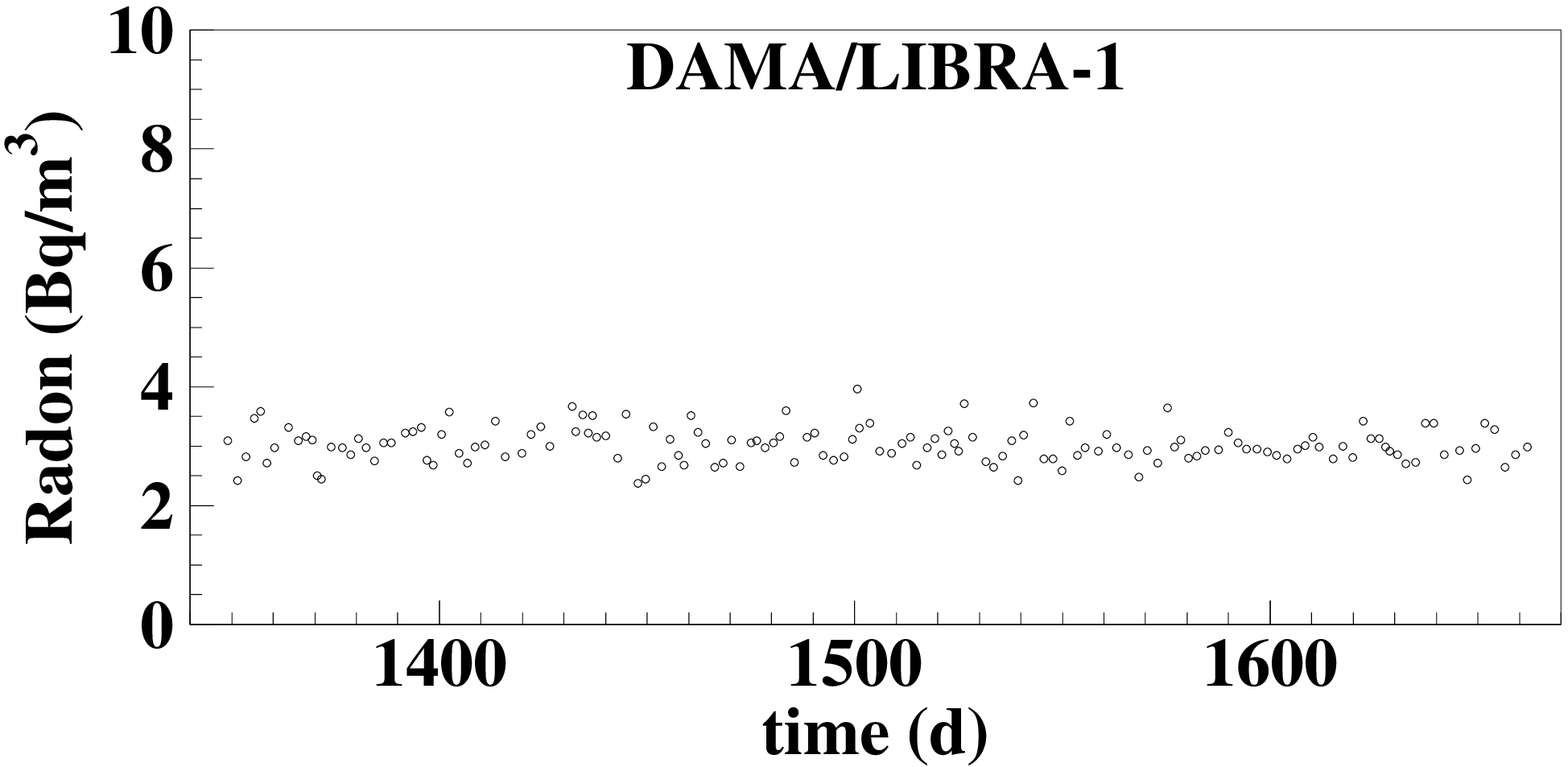}
\includegraphics[width=6.cm] {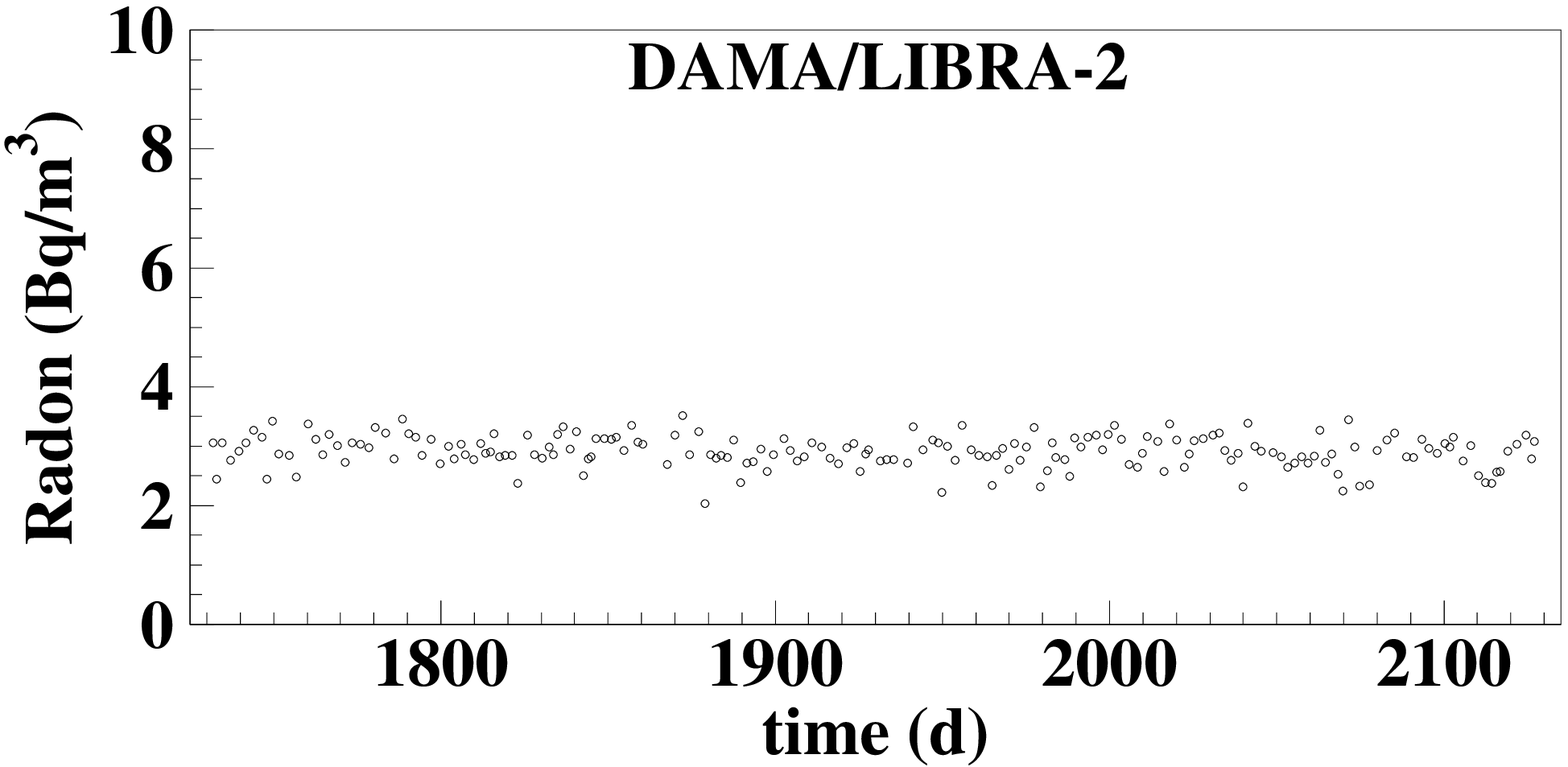}
\includegraphics[width=6.cm] {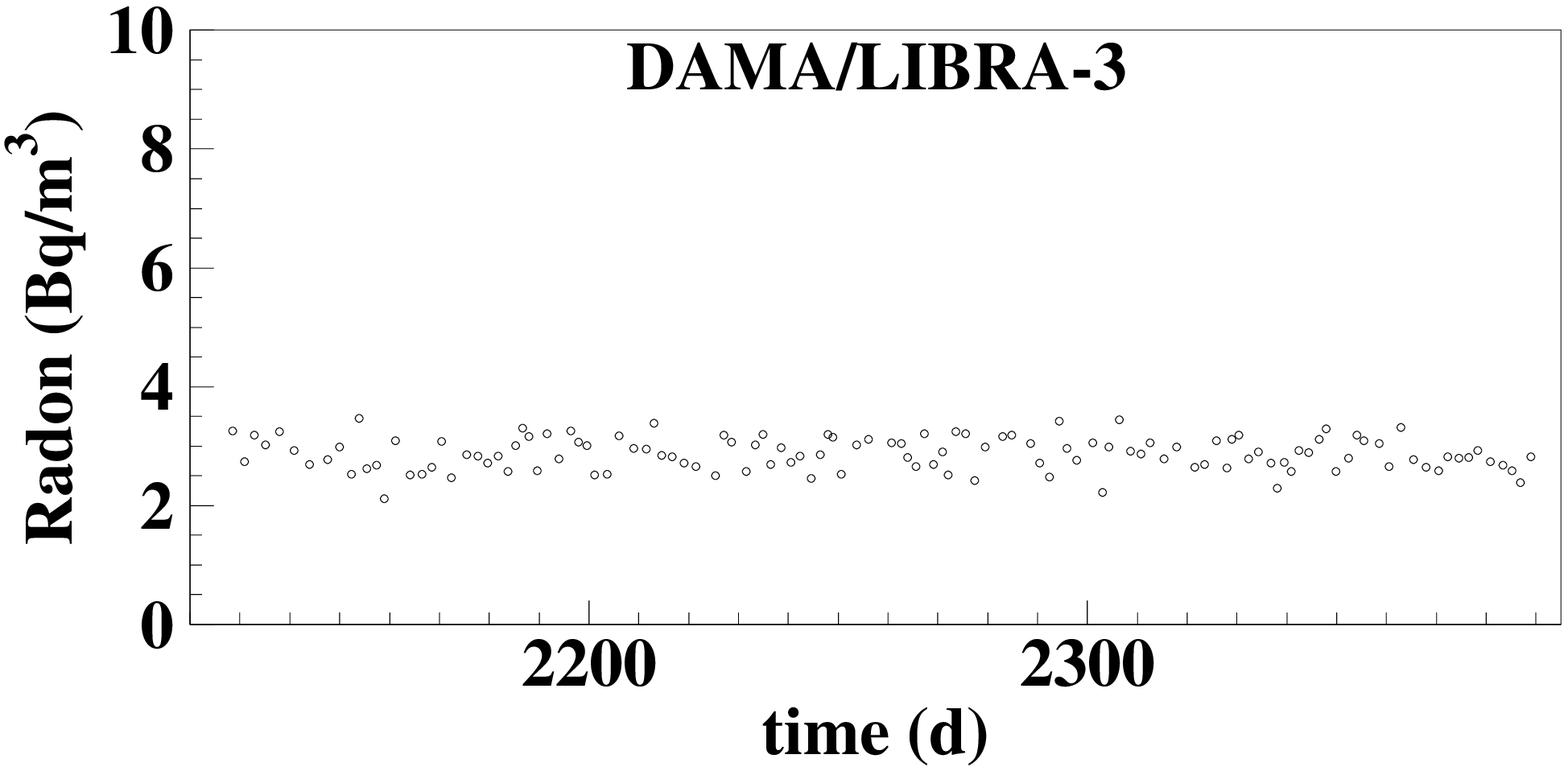}
\includegraphics[width=6.cm] {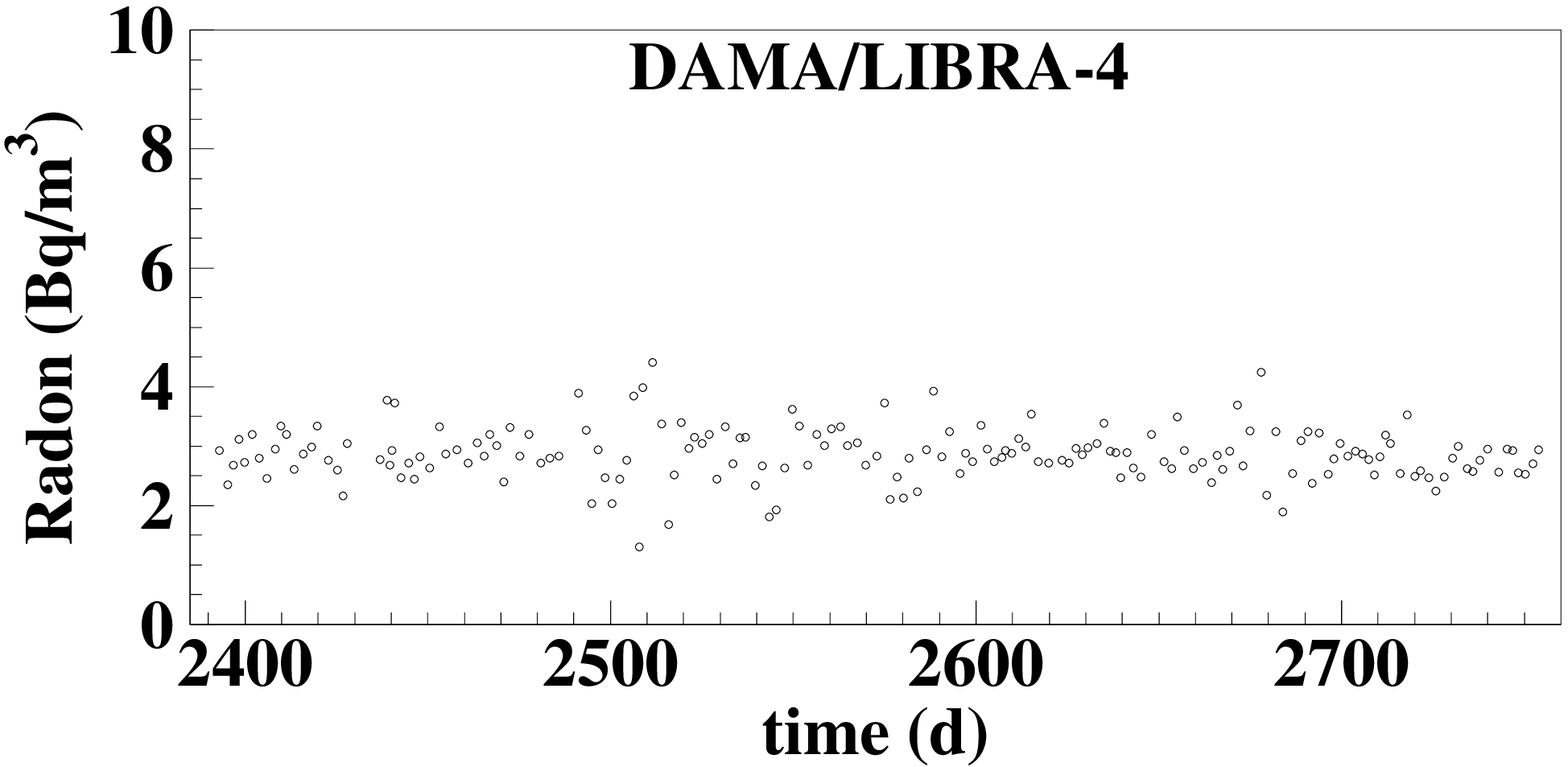}
\caption{Time behaviours of the Radon in the inner part of 
the barrack (from which -- in addition - the detectors are further isolated 
by other two levels of sealing \cite{perflibra}) 
during the DAMA/LIBRA-1 to -4 annual cycles, respectively. 
The measured values are at the level of sensitivity 
of the used radonmeter. The time scale has the origin at Jan. 1st, 2000.}
\label{fig_radon}
\end{figure}

Notwithstanding the above considerations, the Radon in the
installation outside the plexiglas box, containing the passive shield,
is continuously monitored; it is at level of sensitivity
of the used Radonmeter. The time behaviours for the DAMA/LIBRA-1, -2, -3, and -4
annual cycles are shown in Fig. \ref{fig_radon}. As quantitatively reported in Table 
\ref{tb:par1234}, 
no modulation of Radon is present in the environment of the set-up; moreover, 
the detectors are further isolated by the other two levels of sealing \cite{perflibra}.

\begin{figure}[!htb]
\begin{center}
\includegraphics[width=4.cm] {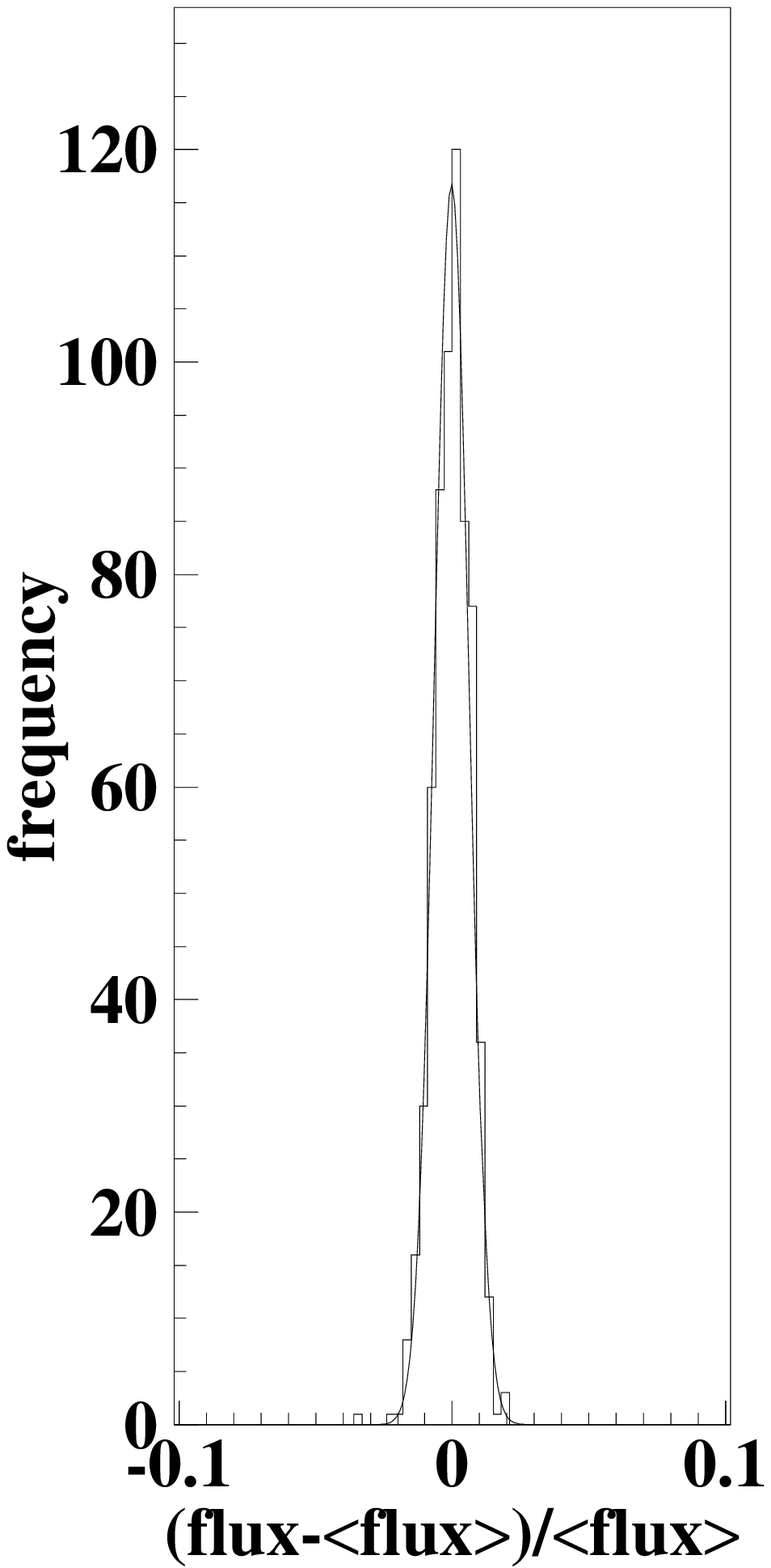}
\includegraphics[width=4.cm] {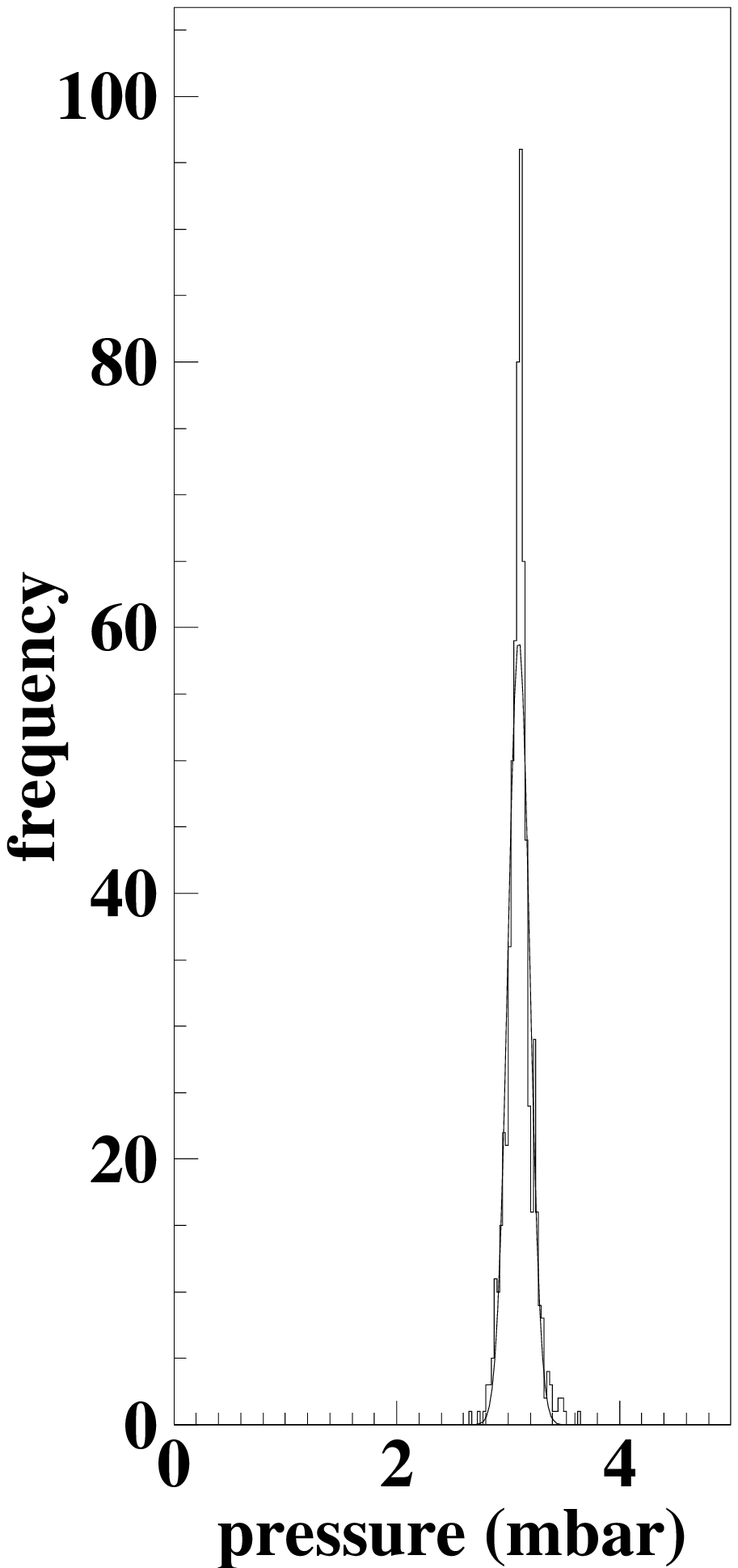}
\end{center}      
\vspace{-0.6cm}
\caption{Distributions of the HP Nitrogen flux in the inner Cu box 
housing the detectors and of its pressure as measured during the DAMA/LIBRA-1 to -4 annual 
cycles (histograms); the superimposed curves are gaussian fits. For clarity the HP Nitrogen flux has 
been given in terms of relative variations.}
\label{fig_pres}
\end{figure}

In Fig. \ref{fig_pres} the distributions of
the relative variations of the HP Nitrogen flux in the inner Cu box housing the detectors
and of its pressure as measured during the 
DAMA/LIBRA-1 to -4 annual cycles are shown (the typical flux mean value for each annual
cycle is of order of $\simeq$ 320 l/h and the typical overpressure 
mean value is of order of 3.1 mbar).

We have also investigated possible Radon trace in the HP Nitrogen atmosphere inside the Cu box, 
housing the detectors, by searching for the double coincidences of
the gamma-rays (609 and 1120 keV) from $^{214}$Bi
Radon daughter, obtaining an upper limit on 
the possible Radon concentration in the Cu box HP Nitrogen atmosphere:
$< 5.8 \times 10^{-2} $ Bq/m$^3$  (90\% C.L.); thus, 
roughly $< 2.5 \times 10^{-5}$ cpd/kg/keV can be expected from this source in the lowest energy 
bins and {\it single-hit} events in DAMA/LIBRA as of 
interest for DM detection. 
The improvement -- with respect to DAMA/NaI -- of the limit of the expected contribution to the {\it 
single-hit} events counting rate at low energy is due to the enlarged matrix of detectors and
to the better filling of the Cu box, housing the detectors, e.g. thanks to the new 
Cu shaped shield for PMTs and light guides \cite{perflibra}.

This shows that even an hypothetical, e.g. 10\%, modulation of possible Radon in 
the HP Nitrogen atmosphere of the Cu box, housing the detectors, would correspond to a modulation 
amplitude
$< 2.5 \times 10^{-6}$ cpd/kg/keV ($< 0.01\%$ of the observed modulation amplitude).

Moreover, it is worth noting that, while the possible presence of a sizeable quantity 
of Radon nearby a detector would forbid the investigation of the annual 
modulation signature (since every Radon variation would induce both the variation in the whole
energy distribution and the continuous pollution of the exposed surfaces by the non-volatile 
daughters), it cannot mimic the DM
annual modulation signature in experiments such as the former DAMA/NaI and the present DAMA/LIBRA
which record the whole energy distribution; in fact, possible presence of Radon 
variation can easily be identified in this case and 
some of the six requirements 
of the DM annual modulation signature would fail.

In conclusion, no significant effect is possible from the Radon.

\subsubsection{... more on side processes}

Finally, possible side reactions have also been carefully searched for. 
The only process which has been found 
as an hypothetical possibility is the muon flux modulation reported 
by the MACRO experiment \cite{Mac97}. In fact, MACRO has observed 
that the muon flux shows a nearly sinusoidal time behaviour 
with one year period and maximum in the summer with amplitude
of $\simeq$ 2 \%; this muon flux modulation is correlated
with the temperature of the Earth's atmosphere. 
A simple calculation to estimate the 
modulation amplitude expected 
from this process in the DAMA/LIBRA set-up can follow the analysis introduced in ref. 
\cite{Sist,RNC} and recalled 
in the following.
The muon flux ($\Phi_\mu$) and the
yield of neutrons produced by muons measured at the underground
Gran Sasso National Laboratory ($Y$) 
are:
$\Phi_\mu \simeq 20 $ muons m$^{-2}$d$^{-1}$ \cite{Mac97}
and 
$Y \simeq (1 - 7) \times 10^{-4}$ neutrons per muon per g/cm$^2$ \cite{Agl},
respectively. Thus,
the fast neutron rate produced by muons is given by: 
$R_n = \Phi_\mu \cdot Y \cdot M_{eff}$,
where $M_{eff}$ is the effective mass 
where muon interactions can give rise to events detected in the DAMA 
set-up.
Consequently, the annual modulation amplitude 
in the lowest energy region induced in DAMA/LIBRA by a muon flux 
modulation as measured by MACRO \cite{Mac97} can be estimated according to:
$S_m^{(\mu)} = R_n \cdot g \cdot \epsilon \cdot f_{\Delta E} \cdot 
f_{single} \cdot 2\% / (M_{set-up} \cdot \Delta E)$,
where $g$ is a geometrical factor, $\epsilon$ is the detection efficiency 
for elastic scattering interactions, $f_{\Delta E}$ is the acceptance
of the considered 
energy window (E $\ge$ 2 keV),
$f_{single}$ is the {\it single-hit} efficiency and 2\% is the MACRO 
measured effect. Since 
$M_{set-up} \simeq$ 250 kg and $\Delta E \simeq$  4 keV, assuming the very
cautious values $g \simeq \epsilon \simeq f_{\Delta E} \simeq f_{single} 
\simeq 0.5$ 
and $M_{eff}$ = 15 t, one obtains:
$S_m^{(\mu)} < (0.4 - 3) \times 10^{-5}$ cpd/kg/keV.
We stress that -- in addition --
the latter value has been 
overestimated of orders of magnitude both because of the extremely cautious 
values assumed in the
calculation and, as mentioned, of the omission of the effect of the
$\simeq$ 1 m concrete neutron moderation. Finally,
we remark that not only the modulation of the muon flux observed by MACRO 
would give rise in our set-up to a quantitatively 
negligible effect, but -- in addition -- 
some of the six requirements necessary to mimic 
the annual modulation signature (such as e.g. the $4^{th}$ and 
the $5^{th}$) would fail. 
Therefore, it can be safely ignored.

Just for the sake of completeness, we remind that the contribution of solar neutrinos,
whose flux is also expected to be modulated, is many orders of magnitude 
lower than the modulation amplitude measured by DAMA/LIBRA \cite{IDM96}.

\subsection{The noise}

Despite the good noise identification near energy threshold and the stringent noise
rejection procedure which is used  \cite{perflibra}, 
the role of a possible noise tail in the data after the noise rejection procedure
has been quantitatively investigated.

In particular, the hardware rate of each detector above a single photoelectron, 
$R_{Hj}$ ($j$ identifies the detector), has been considered. 
Indeed, this hardware rate is significantly determined by the noise. 

\begin{figure}[!ht]
\begin{center}
\includegraphics[width=6.0cm] {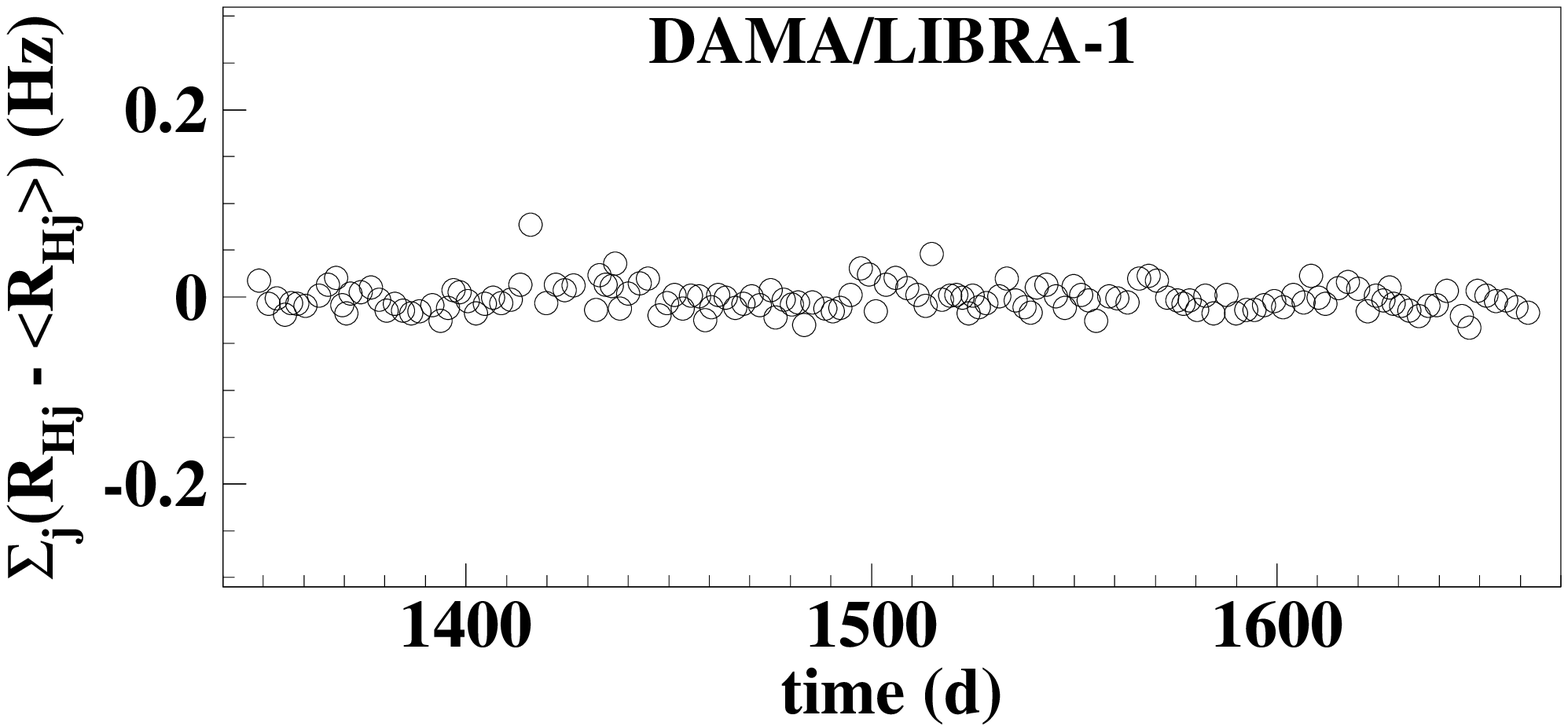}
\includegraphics[width=6.0cm] {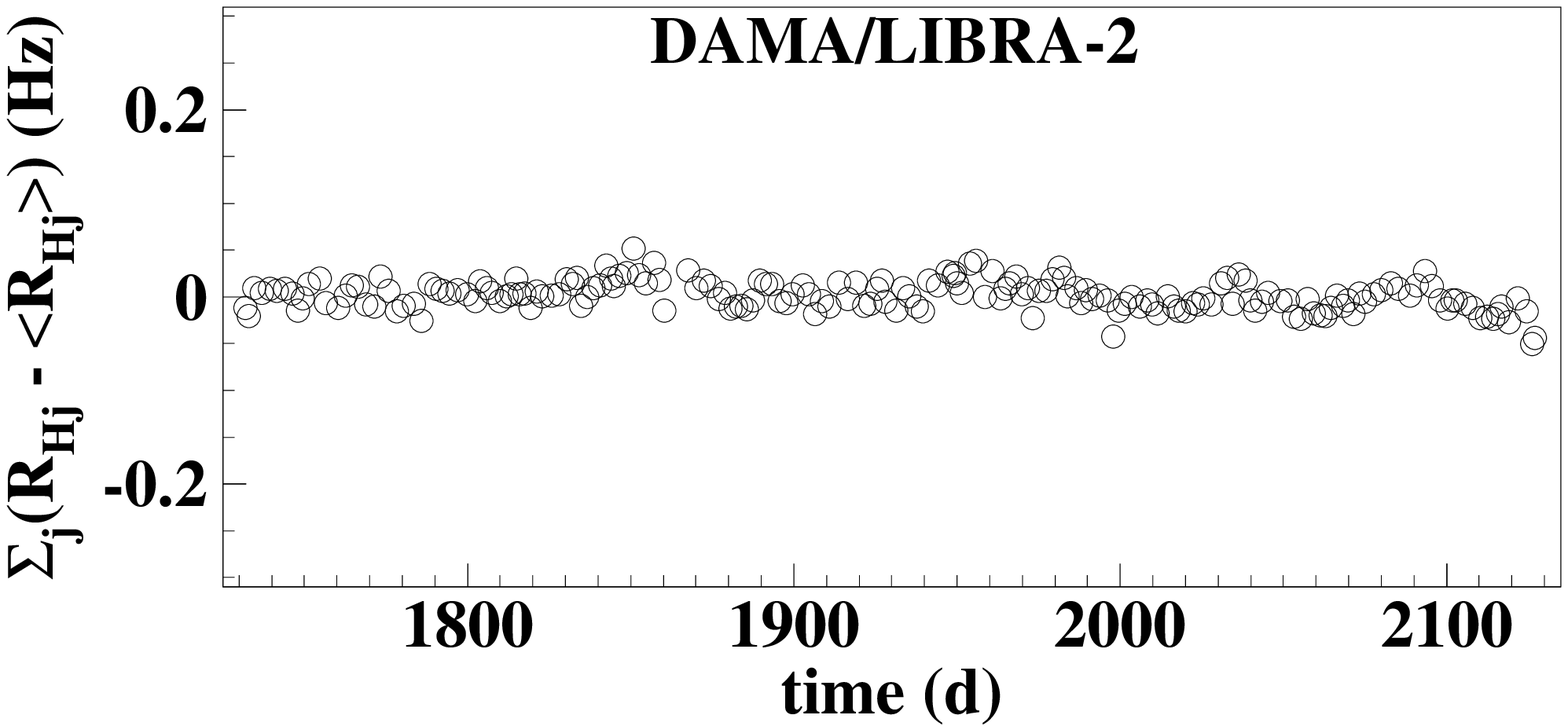}
\includegraphics[width=6.0cm] {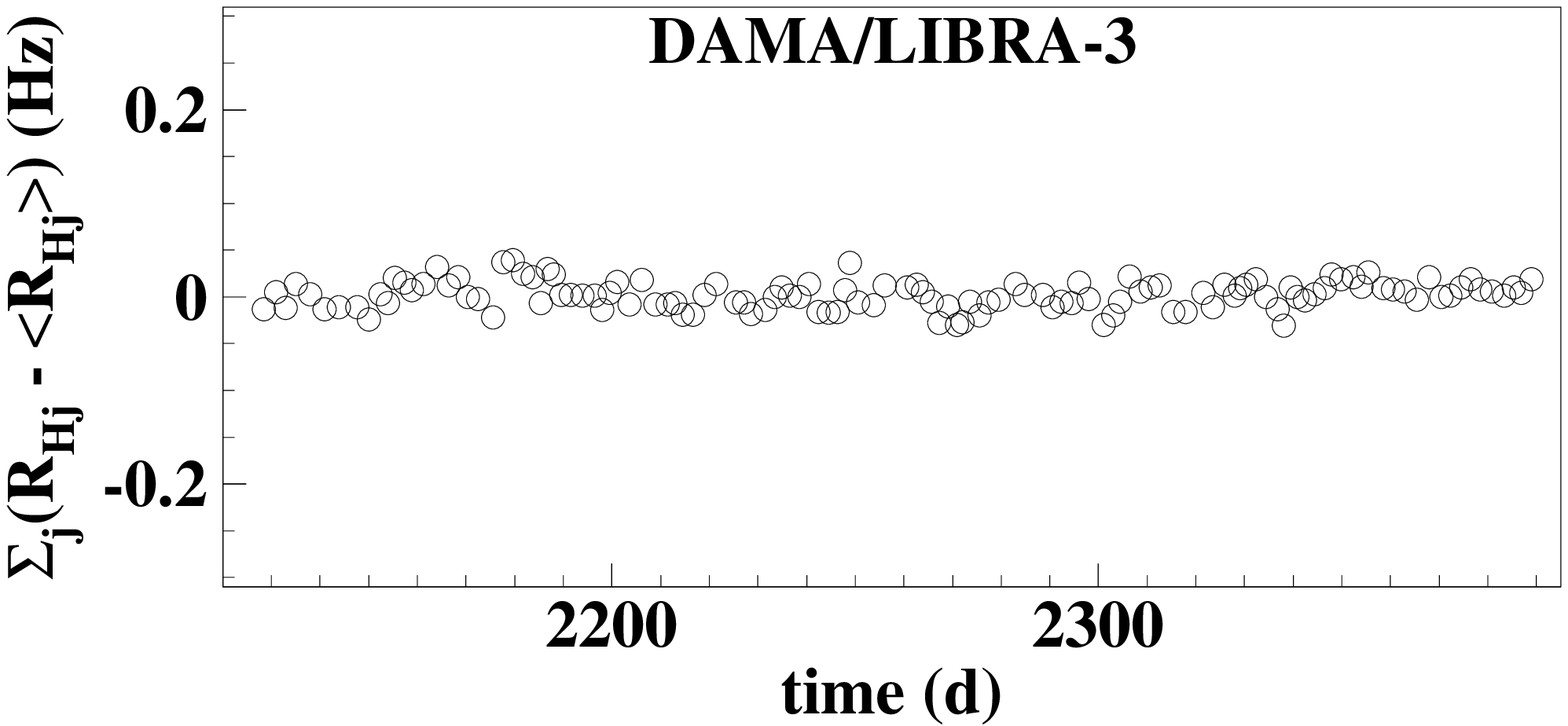}
\includegraphics[width=6.0cm] {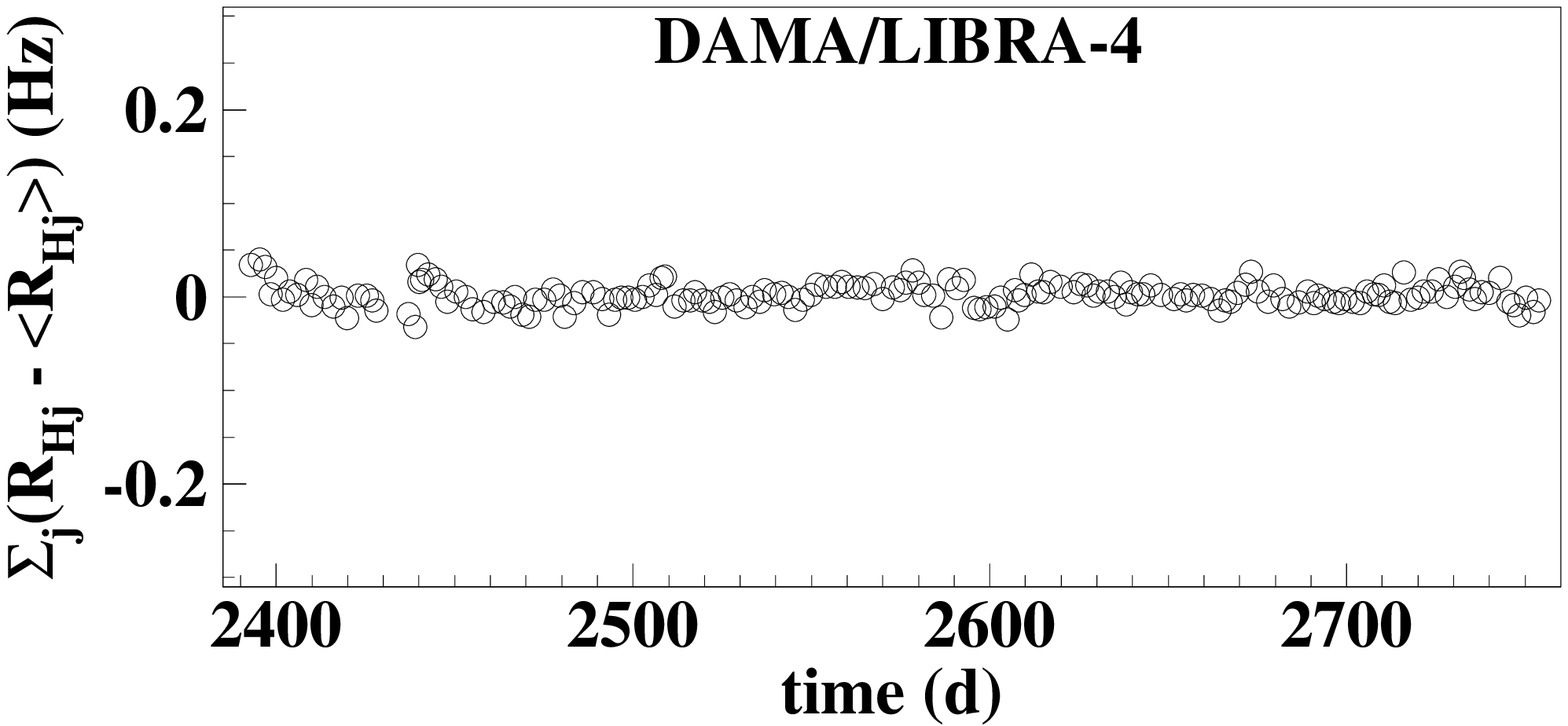}
\end{center}
\caption{Time behaviours of the variable $R_H = \Sigma_j (R_{Hj} - \langle R_{Hj} \rangle )$, where
$R_{Hj}$ is the hardware rate of each detector above single photoelectron threshold 
(that is including the noise), $j$ identifies the detector and $\langle R_{Hj} \rangle$ is the mean 
value of $R_{Hj}$ in the corresponding annual cycle.}
\label{fig_stab_rh}
\vspace{-0.1cm}
\end{figure}

For the proposed purpose the variable: $R_H = \Sigma_j (R_{Hj} - \langle R_{Hj} \rangle )$, can be built; 
in the present case $\langle R_{Hj} \rangle \lsim 0.2$ Hz. The time behaviours of $R_H$ 
during the DAMA/LIBRA-1 to -4 annual cycles are shown in Fig. \ref{fig_stab_rh}.

\begin{figure}[!ht]
\begin{center}
\vspace{-0.4cm}
\includegraphics[width=4.cm] {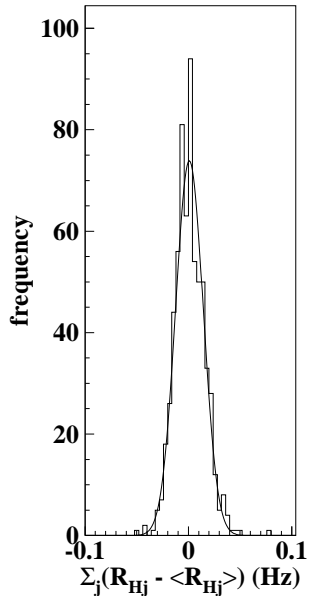}
\vspace{-0.2cm}
\caption{Distribution of R$_H$ during the DAMA/LIBRA-1 to -4 annual cycles (histogram); the 
superimposed curve is a gaussian fit. See text.}
\label{fg:fig_rh}
\end{center}
\end{figure}

As can be seen in Fig.\ref{fg:fig_rh}, the cumulative distribution of $R_H$ for 
the DAMA/LIBRA-1 to -4 annual cycles 
shows a gaussian behaviour with $\sigma$ = 0.3\%, value well in agreement with the one expected 
on the basis of simple statistical arguments. 

Moreover, by fitting the time behaviour of R$_H$ in the four data taking periods -- including a
modulation term as that for DM particles --
a modulation amplitude compatible with zero:
$(0.03 \pm 0.09) \times 10^{-2}$ Hz, is obtained. From this value 
the upper limit at 90\% C.L. on the modulation amplitude 
can be derived: $< 1.8 \times 10^{-3}$ Hz.
Since the typical noise contribution to the hardware rate 
of each detector is $\simeq$ 0.10 Hz, the upper limit on the noise relative 
modulation amplitude is given by:
$ \frac{1.8 \times 10^{-3} Hz} {2.5 Hz} 
\simeq 7.2 \times 10^{-4}$  (90\% C.L.).
Therefore, even in the worst hypothetical case of a 
10\% contamination of the residual noise -- after rejection -- in the 
counting rate, the noise contribution to the modulation   
amplitude in the lowest energy bins would be 
$< 7.2 \times 10^{-5}$ of the total counting rate.
This means that an hypothetical noise modulation could account at maximum
for absolute amplitudes less than $10^{-4}$ cpd/kg/keV.

In conclusion, there is no evidence for any role of an hypothetical tail of
residual noise after rejection.

\subsection{The calibration factor}

The performed calibrations have been discussed in ref. \cite{perflibra};
in particular, in long term running conditions 
periodical calibrations are performed every $\simeq$ 10 days with $^{241}$Am source.

Although it is highly unlikely that a variation of the calibration factor
(proportionality factor between the area of the recorded pulse  
and the energy), $tdcal$,
could play any role, 
a quantitative
investigation on that point has been carried out.

\begin{figure}[!t]
\begin{center}
\vspace{-0.6cm}
\includegraphics[width=4.cm]{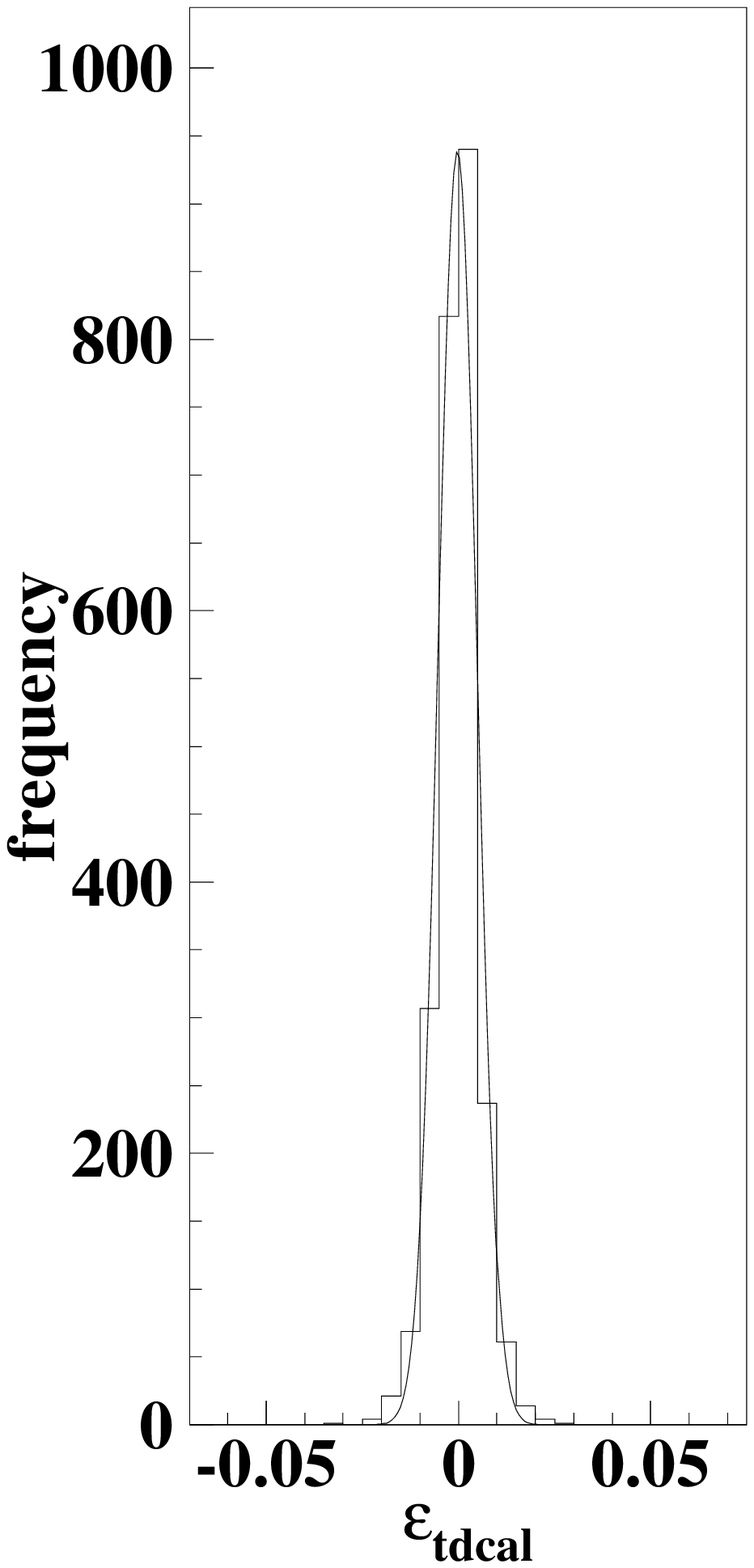}
\includegraphics[width=4.cm]{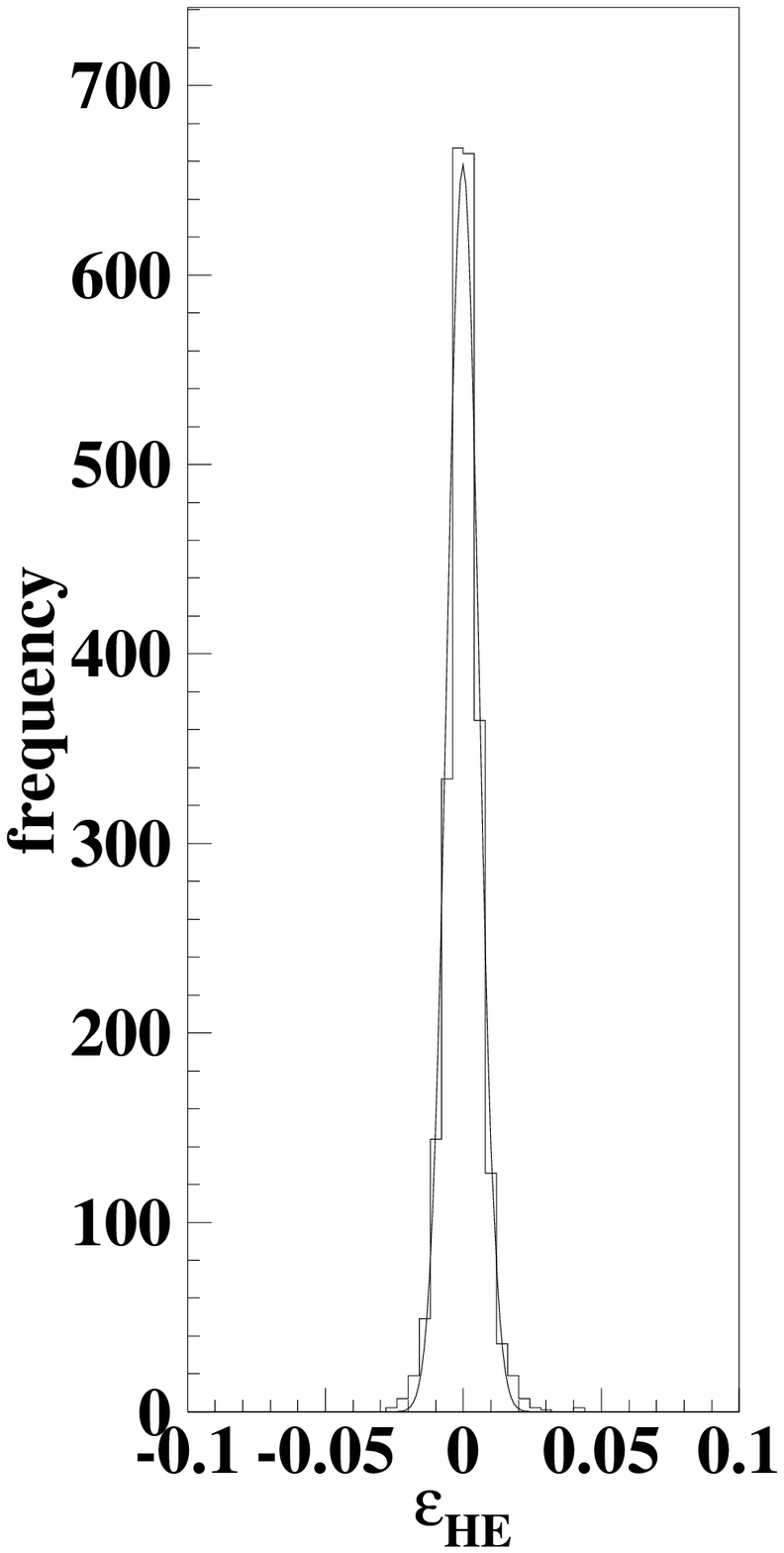}
\end{center}
\vspace{-0.4cm}
\caption{{\it Left:} Distribution of the percentage variations ($\epsilon_{tdcal}$)
of each energy scale factor ($tdcal$) with respect to the value measured
in the previous calibration for the DAMA/LIBRA-1 to -4 annual cycles
(histogram); the superimposed curve is a gaussian fit. 
The standard deviation is 0.5\%.
{\it Right:} Distribution of the percentage variations ($\epsilon_{HE}$)
of the high energy scale factor with respect to the mean values
for the DAMA/LIBRA-1 to -4 annual cycles (histogram); the superimposed curve is a gaussian fit.
The standard deviation is 0.6\%.}
\label{fig_tdcal}
\vspace{-0.2cm}
\end{figure}

For this purpose, we define the percentage variation 
of each energy scale factor ($tdcal$) with respect to the value measured
in the previous calibration: $\epsilon_{tdcal} = \frac{tdcal_k-tdcal_{k-1}}{tdcal_{k-1}}$
(here $tdcal_k$ is the value of the calibration factor in the $k$-th calibration). 
The distribution of $\epsilon_{tdcal}$
for all the detectors during the DAMA/LIBRA-1 to -4 
annual cycles is given in Fig.~\ref{fig_tdcal}{\it --Left}. 
This distribution shows a gaussian behaviour with
$\sigma \simeq 0.5\%$. Since the results of the routine calibrations
are properly taken into account in the data analysis,  
such a result allows us to conclude that
the energy calibration factor for each detector
is known with an uncertainty $\ll 1\%$ during the data taking periods.

\vspace{0.3cm}

Moreover, the distribution of the percentage variations ($\epsilon_{HE}$)
of the high energy scale factor with respect to the mean values
for all the detectors and for the DAMA/LIBRA-1 to -4 annual cycles
is reported in Fig.~\ref{fig_tdcal}{\it --right}. 
Also this distribution shows a gaussian behaviour with
$\sigma \simeq 0.6\%$. 

\vspace{0.3cm}

As discussed also in ref. \cite{Sist,RNC},
the possible variation of the calibration factor for each detector
during the data taking would give rise to an additional energy
spread ($\sigma_{cal}$) besides the detector energy resolution
($\sigma_{res}$). The total
energy spread can be, therefore, written as: $\sigma = \sqrt{\sigma^2_{res} +
\sigma^2_{cal}} \simeq \sigma_{res} \cdot
[1+\frac{1}{2} \cdot (\frac{\sigma_{cal}}{\sigma_{res}})^2]$; 
clearly the contribution due to the calibration factor variation
is negligible since
$\frac{1}{2} \cdot (\frac{\sigma_{cal}/E}{\sigma_{res}/E})^2 \lsim   
7.5 \times 10^{-4} \frac{E}{20 keV} $ (where the adimensional ratio
$\frac{E}{20 keV}$ accounts for the energy dependence of this limit value).
This order of magnitude is confirmed by a MonteCarlo calculation,
which credits -- as already reported in ref. \cite{Sist,RNC} -- 
a maximum value of the effect of similar variations of $tdcal$ on the
modulation amplitude equal to $1-2 \times 10^{-4}$ cpd/kg/keV.
Thus, also the unlikely idea that the calibration factor could 
play a role can be safely ruled out.

\vspace{0.3cm}
\subsection{The efficiencies}

The behaviour of the used overall efficiencies   during the 
whole data taking periods has even been investigated. 
Their possible time variation depends essentially on the stability
of the efficiencies related to the 
adopted acceptance windows; they are regularly measured by dedicated 
calibrations \cite{perflibra}.
\begin{figure}[!ht]
\begin{center}
\vspace{-0.6cm}
\includegraphics[width=4.cm] {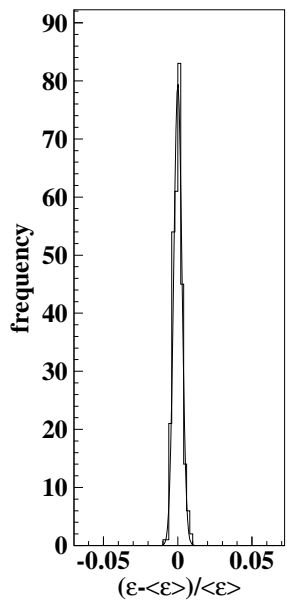}
\end{center}
\vspace{-0.4cm}
\caption{Distribution of the percentage variations
of the overall efficiency values with the respect to their mean values during the DAMA/LIBRA-1 
to -4 annual cycles (histogram); the superimposed curve is a gaussian fit. See text.}
\label{fig_eff}
\end{figure}   
In particular, Fig.~\ref{fig_eff} shows
the percentage variations of the efficiency values
in the (2-8) keV energy interval considering 2 keV bins. They
show a gaussian distribution with $\sigma$ = 0.3\%
for DAMA/LIBRA-1 to -4, cumulatively. Moreover, we have verified
that the time behaviour of these percentage variations
does not show any modulation with period
and phase expected for a possible DM signal.
In Table \ref{tb:eff1234}
the modulation amplitudes of the efficiencies in each energy
bin between 2 and 10 keV are reported, showing that they are all consistent with zero.
In particular, modulation amplitudes -- considering the four DAMA/LIBRA annual cycles
all together -- equal to $(0.1 \pm 0.4) \times 10^{-3}$ and
$-(0.4 \pm 0.4) \times 10^{-3}$ are found for the (2-4) keV and
(4-6) keV energy bins, respectively; both consistent with zero. 
Thus, also the unlikely idea of a possible role played by the 
efficiency values is ruled out.

\begin{table}[!ht]
\caption{Modulation amplitudes
obtained by fitting the time behaviour of the
efficiencies including a cosine
modulation with phase and period as for DM particles for
the DAMA/LIBRA-1 to -4 annual cycles.}
\vspace{0.4cm}
\centering
\resizebox{\textwidth}{!}{
\begin{tabular}{|c|r|r|r|r|} \hline
  & \multicolumn{4}{c|}{Amplitudes ($\times 10^{-3}$)}  \\
  & & & & \\
Energy  & DAMA/LIBRA-1 & DAMA/LIBRA-2 & DAMA/LIBRA-3 & DAMA/LIBRA-4\\
  & & & & \\
\hline
  & &  & & \\
2-4 keV   & $ (0.3 \pm 0.6)$ & $ (0.1 \pm 0.6)$ & $-(0.4 \pm 1.1)$ & $ -(0.4 \pm 1.0)$ \\
4-6 keV   & $ (0.0 \pm 0.6)$ & $-(0.7 \pm 0.6)$ & $-(0.3 \pm 1.0)$ & $ -(0.7 \pm 1.0)$ \\
6-8 keV   & $-(0.3 \pm 0.6)$ & $-(1.0 \pm 0.7)$ & $-(0.2 \pm 0.8)$ & $ -(1.0 \pm 0.8)$ \\
8-10 keV  & $-(0.5 \pm 0.5)$ & $-(0.5 \pm 0.5)$ & $-(0.2 \pm 0.6)$ & $  (0.7 \pm 0.6)$ \\
 & & &  &\\
\hline\hline
\end{tabular}
}
\label{tb:eff1234}
\end{table}

\vspace{-0.5cm}
\subsection{Conclusions on possible systematics effects}

No modulation has been found in any  
\begin{table}[!ht]
\vspace{-0.1cm}
\caption{Summary of the results obtained by investigating possible sources 
of systematics or of side reactions
in the data of the DAMA/LIBRA-1 to -4 annual cycles.
None able to give a modulation amplitude different from zero has been 
found;
thus cautious upper limits (90\% C.L.)
on the possible contributions to the measured modulation amplitude
have been calculated and are shown here.}
\vspace{-0.1cm}
\begin{center}
\begin{tabular}{|c|c|c|}
\hline \hline
 Source      & Main comment                       &  Cautious upper limit \\
             & (see also ref. \cite{perflibra})   &       (90\%C.L.) \\
\hline\hline
             & Sealed Cu Box in         &  \\
  Radon      & HP Nitrogen atmosphere,  &  $<2.5 \times 10^{-6}$ cpd/kg/keV \\
             & 3-level of sealing       &  \\
\hline
Temperature  & Air conditioning         &  $<10^{-4}$ cpd/kg/keV \\
             & + huge heat capacity     &                        \\
\hline
Noise        & Efficient rejection      &  $<10^{-4}$ cpd/kg/keV \\
\hline
Energy scale & Routine                  &  $<1 - 2 \times 10^{-4}$ cpd/kg/keV \\
             & + intrinsic calibrations & \\
\hline
Efficiencies & Regularly measured       & $<10^{-4}$ cpd/kg/keV \\
\hline
             &  No modulation above 6 keV;       & \\
             &  no modulation in the (2 -- 6) keV  & \\
 Background  &  {\it multiple-hit} events;       & $<10^{-4}$ cpd/kg/keV \\
             &  this limit includes all possible & \\
             &  sources of background            & \\
\hline
Side reactions & From muon flux variation& $<3 \times 10^{-5}$ cpd/kg/keV \\
               & measured by MACRO  & \\
\hline
\multicolumn{3}{|c|} {In addition: no effect can mimic the signature} \\
\hline \hline
\end{tabular}
\end{center}
\label{tb:sist}
\vspace{-0.1cm}
\end{table}
possible source of systematics or side reactions; thus, upper limits (90\% C.L.)  
on the possible contributions to the DAMA/LIBRA measured modulation amplitude
are summarized in Table \ref{tb:sist}.
In particular, they cannot account for the
measured modulation both because quantitatively not relevant and
unable to mimic the observed effect.

\section{Conclusions}

The model independent results achieved by the second generation DAMA/LIBRA set-up
in operation at the Gran Sasso National Laboratory confirms evidence of Dark Matter particles 
in the galactic halo with high confidence level; a cumulative C.L. of 8.2 $\sigma$ 
is reached when considering the data of the former DAMA/NaI experiment and the present ones 
of DAMA/LIBRA all together. In particular, deep quantitative analyses 
exclude any effect either from systematics or from side processes (temperature, noise, 
hardware or software procedures, background of whatever nature including also radon, neutrons and 
cosmic rays). We note that no experiment exists whose result can be directly 
compared with those presented here.

The wide sensitivity of the used target-detector
material to many of the possible DM candidates and of the 
possible astrophysical, nuclear and particle Physics scenarios, 
the reached intrinsic radiopurity, the used approach, the 
specific performances and operating conditions, the large collected exposures 
of the former DAMA/NaI and of the present
DAMA/LIBRA set-ups, have offered an unique possibility of an effective model independent 
investigation.

Model dependent considerations will be presented in later publication specifically devoted 
to this aspect. Just few arguments for some illustrative purposes are given in Appendix A.

The collection of a larger exposure with DAMA/LIBRA (and with the possible DAMA/1ton, which is
at R\&D stage) will also allow the improvement of corollary information which can be derived  
on the nature of the candidate particle(s) and on the various related astrophysical, nuclear and 
particle Physics scenarios, and the investigation with very high sensitivity of the
other DM features and second order effects as well as of several rare processes other than DM.

\vspace{0.8cm}
\section{Appendix A}
\label{impl}

As in the past (see e.g. \cite{RNC,ijmd,ijma,epj06,ijma07,chan,wimpele,ldm}),
corollary investigations can also be pursued 
-- on the basis of the cumulative 
8.2 $\sigma$ C.L. model-independent result by DAMA/NaI and DAMA/LIBRA -- 
on the nature of the DM 
candidate particle and on related astrophysical, nuclear and particle Physics scenarios.  
As widely discussed elsewhere, these investigations are instead model-dependent and -- considering
the large uncertainties which exist on the astrophysical, nuclear and particle
physics assumptions and on the theoretical and experimental parameters needed in the calculations 
-- have no general meaning (as it is also the case of exclusion plots and of the DM 
particle parameters evaluated in indirect detection experiments). 

Complete model dependent analyses, to update the allowed regions
in various scenarios and to enlarge the investigations to other ones, will be presented elsewhere.
Here, we just remind that many astrophysical, nuclear and particle physics scenarios 
and many DM particle candidates exist.
Just to offer some naive feeling on the complexity of the
argument, we show in Fig. \ref{fg:tem} 
the experimental $S_m$ values of Fig. \ref{sme}
with superimposed the expected behaviours for some DM candidates in few of the many 
possible scenarios and parameters values (see Table \ref{tb:legenda}). 
In fact, despite the behaviour of the $S_m$ values can be effective a posteriori for template purpose,  
very accurate results on corollary model dependent quests at given C. L.  
should be evaluated by applying the maximum likelihood analysis
in time and energy of all the events (as described elsewhere),
which offers 
efficient and complete data analyses accounting for all the experimental 
information carried out by the data and, when of interest, for priors.

\begin{figure}[!p]
\begin{center}
\vspace{-0.6cm}
\includegraphics[width=10.cm] {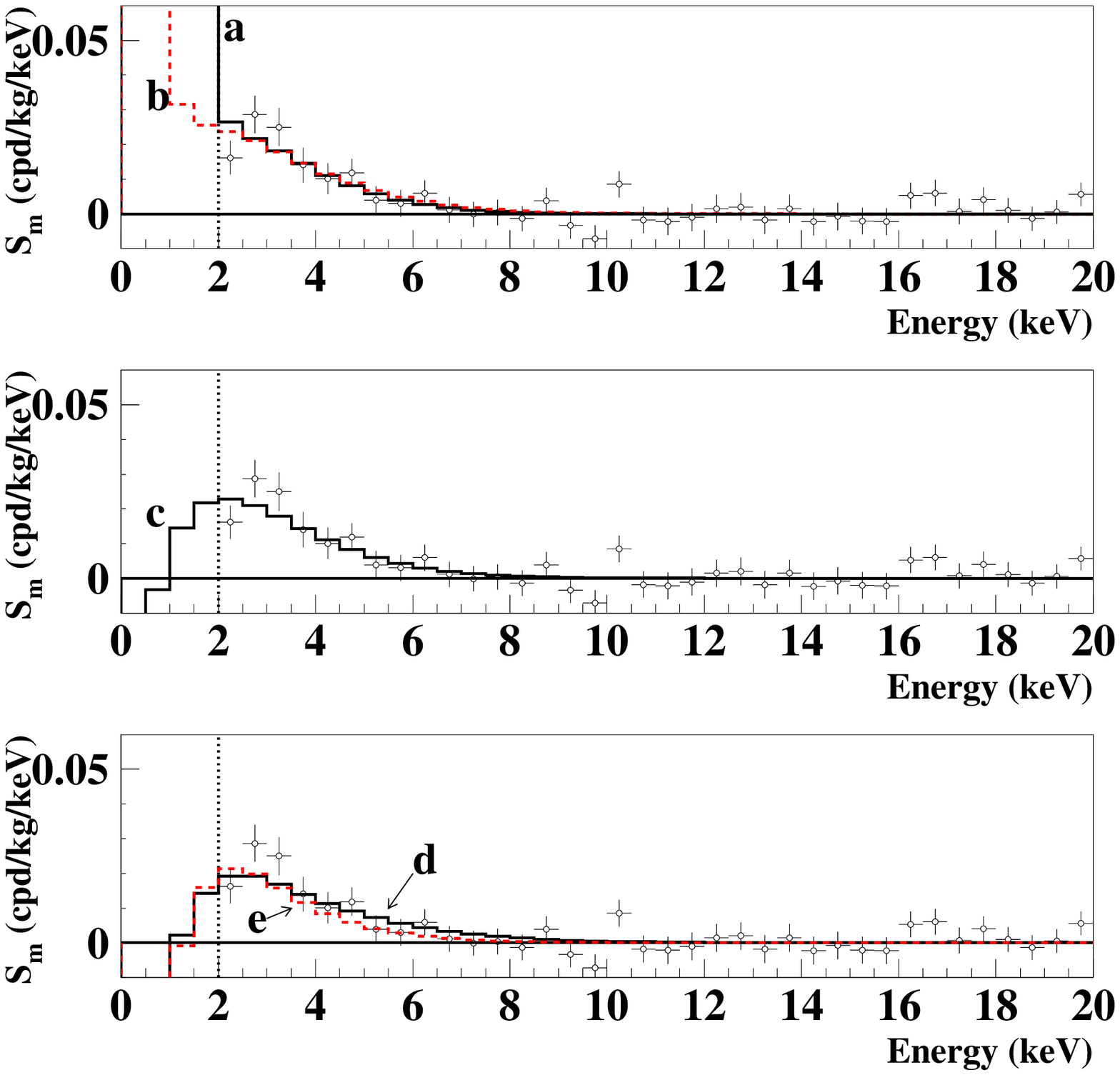}
\includegraphics[width=10.cm] {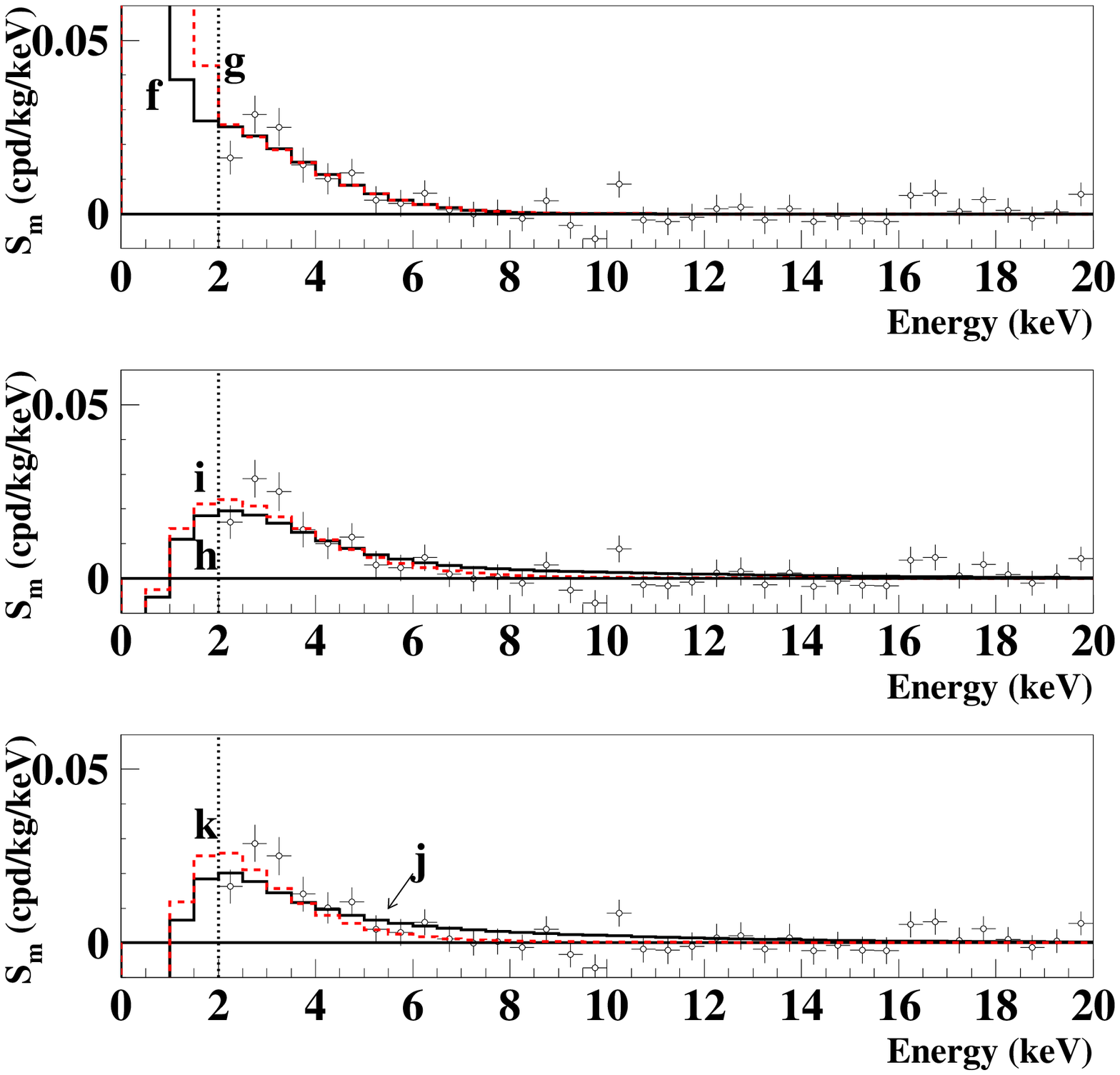}
\end{center}
\end{figure}   

\begin{figure}[!t]
\begin{center}
\vspace{-0.6cm}
\includegraphics[width=10.cm] {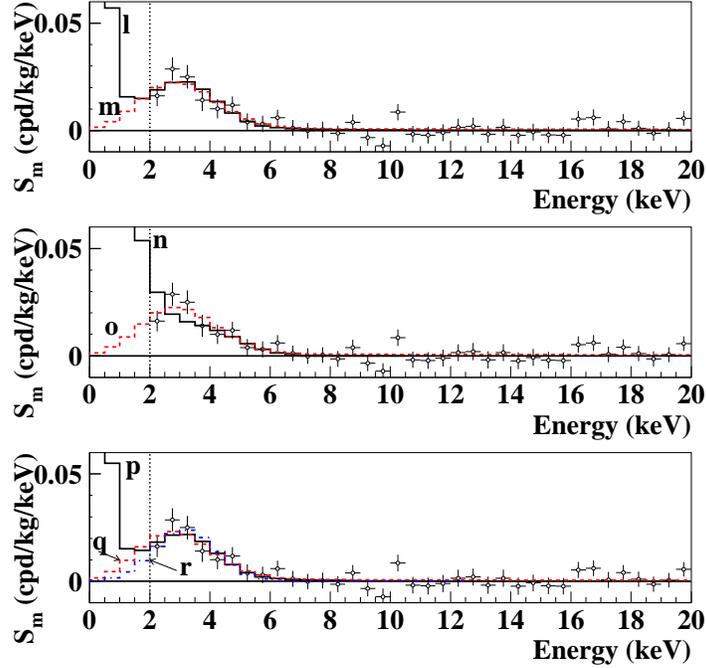}
\end{center}
\vspace{-0.4cm}
\caption{Few examples of expected behaviours for few of the many possible scenarios 
(see Table \ref{tb:legenda}), superimposed to the measured $S_{m,k}$ values of Fig. 
\ref{sme}. The shown behaviours have not been obtained by the maximum likelihood method (see
in our quoted literature) and are shown just for illustrative purpose; they all give practically 
about the same C.L.. 
As mentioned in the text, the full treatment of
the data by maximum likelihood method to update the volumes/regions allowed at given C.L. 
by the cumulative DAMA/NaI and DAMA/LIBRA data
for the considered scenarios 
will be presented elsewhere following the full analysis method of refs. 
\cite{RNC,ijmd,ijma,epj06,ijma07,chan,wimpele,ldm}.}
\label{fg:tem}
\end{figure}   

\begin{table}[!ht]
\caption{Scenarios and parameters values used in Fig. \ref{fg:tem}; they have been chosen 
among the many possible ones \cite{RNC,ijmd,ijma,epj06,ijma07,chan,wimpele,ldm}.
In the fourth column the considered Set -- as in ref. \cite{RNC} -- 
of nuclear form factors and/or of nuclear quenching factors is reported. 
Here: 
i)    $\sigma_{SI}$ is the spin independent point-like cross section; 
ii)   $\sigma_{SD}$ is the spin dependent point-like cross section;
iii)  $\theta$ is an angle defined in the [0,$\pi$) interval, whose tangent is 
      the ratio between the WIMP-neutron and the WIMP-proton effective SD coupling strengths, 
      respectively \cite{RNC};
iv)   $m_H$ is the mass of the LDM particle;
v)    $\Delta$ is the mass splitting \cite{ldm};
vi)   $g_{aee}$ is the bosonic axion-like particle coupling to electrons.
For the cross sections of the LDM particle see ref. \cite{ldm}
and for the channeling effect see ref. \cite{chan}.}
\begin{center}
\resizebox{\textwidth}{!}{
\begin{tabular}{|c|c|c|c|c|c|c|c|c|}
\hline
\multicolumn{9}{|c|}{DM particle elastic scattering on nuclei, spin-independent (SI) and spin-dependent (SD) couplings,} \\
\multicolumn{9}{|c|}{local velocity = 170 km/s and nuclear cross section scaling laws as in \cite{RNC}} \\
\hline
 Curve & Halo model & Local density  & Set as & DM particle & $\xi\sigma_{SI}$&$\xi\sigma_{SD}$& $\theta$ & Channeling \\
 label & (see ref. \cite{RNC,Hep}) & (GeV/cm$^{3}$) & in \cite{RNC} & mass & (pb) & (pb)     & (rad)    & \cite{chan} \\
\hline
\hline
 $a$ & A5 (NFW)          & 0.2  & A &  15 GeV & $ 3.1 \times 10^{-4}$ & 0 & -- & no \\ 
 $b$ & A5 (NFW)          & 0.2  & A &  15 GeV & $ 1.3 \times 10^{-5}$ & 0 & -- & yes \\ 
 $c$ & A5 (NFW)          & 0.2  & B &  60 GeV & $ 5.5 \times 10^{-6}$ & 0 & -- & no \\ 
 $d$ & B3 (Evans         & 0.17 & B & 100 GeV & $ 6.5 \times 10^{-6}$ & 0 & -- & no \\ 
     & power law)        &      &   &         &                       &   &    &    \\ 
 $e$ & B3 (Evans         & 0.17 & A & 120 GeV & $ 1.3 \times 10^{-5}$ & 0 & -- & no \\ 
     & power law)        &      &   &         &                       &   &    &    \\ 
\hline
 $f$ & A5 (NFW)          & 0.2  & A &  15 GeV & $ 10^{-7}$            & 2.6 & 2.435 & no \\ 
 $g$ & A5 (NFW)          & 0.2  & A &  15 GeV & $ 1.4 \times 10^{-4}$ & 1.4 & 2.435 & no \\ 
 $h$ & A5 (NFW)          & 0.2  & B &  60 GeV & $ 10^{-7}$            & 1.4 & 2.435 & no \\ 
 $i$ & A5 (NFW)          & 0.2  & B &  60 GeV & $ 8.7 \times 10^{-6}$ & $8.7 \times 10 ^{-2}$ & 2.435 & no \\ 
 $j$ & B3 (Evans         & 0.17 & A & 100 GeV & $ 10^{-7}$            & 1.7 & 2.435 & no \\ 
     & power law)        &      &   &         &                       &     &       &    \\ 
 $k$ & B3 (Evans         & 0.17 & A & 100 GeV & $ 1.1 \times 10^{-5}$ & 0.11& 2.435 & no \\ 
     & power law) &      &      &             &                       &     &       &    \\ 
\hline
\hline
\end{tabular}}
\resizebox{\textwidth}{!}{
\begin{tabular}{|c|c|c|c|c|c|c|c|}
\hline
\multicolumn{8}{|c|}{Light Dark Matter (LDM) inelastic scattering and bosonic axion-like interaction as in \cite{ijma,ldm},} \\
\multicolumn{8}{|c|}{A5 (NFW) halo model as in \cite{RNC,Hep}, local density = 0.17 GeV/cm$^{3}$, local 
velocity = 170 km/s } \\
\hline
 Curve & DM particle & Interaction  & Set as        & $m_H$ & $\Delta$ & Cross & Channeling \\
 label &             &              & in \cite{RNC} &       &          & section (pb) & \cite{chan} \\
\hline
\hline
 $l$ & LDM & coherent   & A &  30 MeV & 18 MeV & $\xi\sigma_m^{coh} = 1.8 \times 10^{-6}$ & yes \\ 
     &     & on nuclei  &   &         &        &                               &     \\ 
 $m$ & LDM & coherent   & A & 100 MeV & 55 MeV & $\xi\sigma_m^{coh} = 2.8 \times 10^{-6}$ & yes \\ 
     &     & on nuclei  &   &         &        &                               &     \\ 
 $n$ & LDM & incoherent & A &  30 MeV &  3 MeV & $\xi\sigma_m^{inc} = 2.2 \times 10^{-2}$ & yes \\ 
     &     & on nuclei  &   &         &        &                               &     \\ 
 $o$ & LDM & incoherent & A & 100 MeV & 55 MeV & $\xi\sigma_m^{inc} = 4.6 \times 10^{-2}$ & yes \\ 
     &     & on nuclei  &   &         &        &                               &     \\ 
 $p$ & LDM & coherent   & A &  28 MeV & 28 MeV & $\xi\sigma_m^{coh} = 1.6 \times 10^{-6}$ & yes \\ 
     &     & on nuclei  &   &         &        &                               &     \\ 
 $q$ & LDM & incoherent & A &  88 MeV & 88 MeV & $\xi\sigma_m^{inc} = 4.1 \times 10^{-2}$ & yes \\ 
     &     & on nuclei  &   &         &        &                               &     \\ 
 $r$ & LDM & on electrons & -- & 60 keV & 60 keV & $\xi\sigma_m^{e} = 0.3 \times 10^{-6}$ & --  \\ 
\hline
 $r$ & pseudoscalar & see ref. \cite{ijma} & -- & \multicolumn{2}{|c|}{Mass = 3.2 keV} & 
$g_{aee} = 3.9 \times 10^{-11}$ & -- \\
     & axion-like   &                      &    & \multicolumn{2}{|c|}{}               &   & \\ 
\hline
\hline
\end{tabular}}
\end{center}
\label{tb:legenda}
\end{table}

It is worth noting that an increase of the exposure and a possible lowering 
of the used 2 keV threshold will improve the discrimination capability among different
astrophysical, nuclear and particle Physics scenarios. 

\vspace{0.3cm}

Let us, finally, note that results obtained with different target materials and/or different
approaches cannot intrinsically 
be directly compared among them even when considering the same kind of candidate and of coupling, 
although apparently all the presentations generally refer to cross section on nucleon. 

For completeness, we also further note that no experiment exists 
whose result can be directly compared in a model independent way with the ones by DAMA/NaI and 
DAMA/LIBRA. Thus claims for contradictions are arbitrary, in fact, e.g.: 
1)  the others are insensitive to the annual modulation signature; 
2)  the others use different target materials;
3)  the plots they show and those they attribute to others are
    built at a ``single cooking'' without accounting at all for the existing experimental,
    theoretical and phenomenological uncertainties and for the existing alternative choices;
4)  DAMA/NaI and DAMA/LIBRA have a favoured sensitivity with respect to others in 
    several scenarios; moreover, scenarios exist (see literature) to which the others are 
    not only disfavoured with respect to the DAMA experiments, but even blind.
Furthermore, additional realistic limitations in those claimed model dependent sensitivities 
(just for ``nuclear recoils'' and 
a single assumed scenario and parameters set) arise so far e.g. from: 1) the
unproved physical threshold with suitable keV source calibrations; 2) 
the energy scale extrapolated from  
higher energy; 3) unproved stability of the running parameters and of all the used ``rejection'' windows
over long term at the needed precision; 4) insensitivity to candidates giving part (even WIMPs) or all 
the signals in electromagnetic form; 
5) marginal exposures; 6) unproved determination of the 
efficiencies in each one of the many data handlings, they apply, at the needed level of claimed 
precision (a control of systematics at level of $10^{-4}$ -- $10^{-8}$ is required); 7) 
disuniformity in the detector response e.g. in two-phase liquid Xenon detectors; etc. Moreover, they 
generally quoted in an uncorrect, partial and unupdated way the implications of the DAMA/NaI model 
independent result. Some arguments have been addressed e.g. in ref. \cite{RNC,ijmd,ijma,epj06,ijma07,chan,wimpele,ldm,nat} and 
in some literature. On the other hand, whenever there might be 
in future some correct claim for exclusion in one or more particular astrophysical, nuclear and 
particle physics model framework(s) and assumed parameters set (but correctly accounting both for 
experimental and theoretical uncertainties and for the implications of the DAMA model independent 
results in the considered scenario and assumptions), there will be still many other scenarios, 
parameters sets and DM candidates which can explain 
the DAMA/NaI and DAMA/LIBRA model independent results to which other target materials and approaches are 
disfavoured or even 
blind; thus, this never will exclude the model independent results of the DAMA experiments.
In addition, whenever an experiment using the same identical target material and methodological 
approach would be available, as usual in whatever field of Physics 
a serious comparison would require -- in every case -- e.g. a deep investigation of the radiopurity of 
all the part of the different set-ups, of their specific performances in all the aspects, of the detailed 
procedures used by each one, etc.


\end{document}